\documentclass[12pt]{article}
\usepackage{amssymb}
\def\a{\alpha}
\def\b{\beta}

\def\e{\epsilon}
\def\g{\gamma}
\def\k{\kappa}
\def\l{\lambda}
\def\m{\mu}

\def\O{\Omega}

\def\s{\sigma}
\def\t{\tilde}
\def\x{\xi}
\def\up{\uparrow}
\def\down{\downarrow}
\def\str{\mbox{str}}

\newtheorem{Theorem}{Theorem}
\newtheorem{Definition}{Definition}
\newtheorem{Proposition}{Proposition}

\makeatletter
\renewcommand\thesection{\@arabic\c@section}
\renewcommand\thesubsection{\thesection.\@arabic\c@subsection}
\makeatother

\newcommand{\sect}[1]{\setcounter{equation}{0}\section{#1}}

\begin{document}

%************************** Text Begins here ******************************

%  Greek letters

\def\R{\overline{R}}
% Shorthands for \begin{equation} and the like

%\newcommand{\sect}[1]{\setcounter{equation}{0}\section{#1}}
%\renewcommand{\theequation}{\thesection.\arabic{equation}}

\baselineskip=20pt

%%%%%%%%%%%%%%%%%%%%%%%%%%%%%%%%%%%%%%%%%%%%%%%%%%%%%%%%%%%%
%                                                          %
%  Title page                                              %
%                                                          %
%%%%%%%%%%%%%%%%%%%%%%%%%%%%%%%%%%%%%%%%%%%%%%%%%%%%%%%%%%%%
\newfont{\elevenmib}{cmmib10 scaled\magstep1}
\newcommand{\preprint}{
   %\begin{flushleft}
    % \elevenmib Yukawa\, Institute\, Kyoto\\
   %\end{flushleft}\vspace{-1.3cm}
   \begin{flushright}\normalsize
     {\tt hep-th/0511028} \\ November 2005
   \end{flushright}}
\newcommand{\Title}[1]{{\baselineskip=26pt
   \begin{center} \Large \bf #1 \\ \ \\ \end{center}}}
\newcommand{\Author}{\begin{center}
   \large \bf
Shao-You Zhao${}^{a,b}$, Wen-Li Yang${}^{a,c}$ and ~Yao-Zhong
Zhang${}^a$
 \end{center}}
\newcommand{\Address}{\begin{center}

     ${}^a$ Department of Mathematics, University of Queensland,
            Brisbane, QLD 4072,
     Australia\\
     ${}^b$ Department of Physics, Beijing Institute of
     Technology, Beijing 100081, China\\
     ${}^c$ Institute of Modern Physics, Northwest University,
     Xian 710069, P.R. China\\\vskip.1in

% E-mail: syz@maths.uq.edu.au, wenli@maths.uq.edu.au, yzz@maths.uq.edu.au

 \end{center}}
   \newcommand{\Accepted}[1]{\begin{center}
   {\large \sf #1}\\ \vspace{1mm}{\small \sf Accepted for Publication}
   \end{center}}
\preprint \thispagestyle{empty}
\bigskip\bigskip\bigskip

\Title{Determinant Representations of Correlation Functions for
the Supersymmetric $t$-$J$ Model }  \Author

\Address \vspace{1cm}

\begin{abstract}

Working in the $F$-basis provided by the factorizing $F$-matrix,
the scalar products of Bethe states for the supersymmetric $t$-$J$
model are represented by determinants. By means of these results,
we obtain determinant representations of correlation functions for
the model.

\vspace{1truecm}

%\noindent {\it PACS:} 03.65.Fd; 04.20.Jb; 05.30.-d; 75.10.Jm

%\noindent {\it Keywords}: Correlation function; Drinfeld twist;
%Supersymmetric $t$-$J$ model.
\end{abstract}

\newpage
%%%%%%%%%%%%%%%%%%%%%%%%%%%%%%%%%%%%%%%%%%%%%%%%%%%%%%%%%%%%%%%
%                                                             %
%  1. Introduction                                            %
%                                                             %
%%%%%%%%%%%%%%%%%%%%%%%%%%%%%%%%%%%%%%%%%%%%%%%%%%%%%%%%%%%%%%%
\sect{Introduction}

The computation of correlation functions is one of the major
challenging problems in the theory of quantum integrable models
\cite{Smirnov92,Korepin93}. There are currently two approaches for
computing the correlation functions of a quantum integrable model.
One is the vertex operator method (see e.g. [3-8]), and another
one is based on the detailed analysis of the structure of the Beth
states \cite{Korepin82,Izergin87}.

Progress has recently been made in the literature on the second
approach with the help of the Drinfeld twists. Working in the
$F$-bases provided by the $F$-matrices (Drinfeld twists), the
authors in \cite{Maillet96,Kitanine98} managed to compute the form
factors and correlation functions of the XXX and XXZ models
analytically and expressed them in compact determinant forms.

Recently we have constructed the Drinfeld twists for both the
rational $gl(m|n)$ and the quantum $U_q(gl(m|n))$ supersymmetric
models and resolved the hierarchy of their nested Bethe vectors in
the $F$-basis \cite{zsy04,zsy0502,zsy0503}.  In \cite{zsy0506}, we
obtained the determinant representation of the scalar products and
the correlation functions for the $U_q(gl(1|1))$ free fermion
model.

Quantum integrable models associated with Lie superalgebras
\cite{Per81,Kul82,Kul86} are physically important because they
give strongly correlated fermion models of superconductivity.
Among them, the $t$-$J$ model with the Hamiltonian given by
\begin{eqnarray}
H&=&-t\sum _{j=1}^N\sum _{\sigma =\uparrow,\downarrow }
[c_{j,\sigma }^{\dagger}(1-n_{j,-\sigma })c_{j+1,\sigma }
(1-n_{j+1,-\sigma}) +c_{j+1,\sigma }^{\dagger}(1-n_{j+1,-\sigma
})c_{j+1,\sigma } (1-n_{j,-\sigma})]
\nonumber \\
&&+J\sum _{j=1}^N[S_j^zS_{j+1}^z+{1\over 2}(S_j^{\dagger }S_{j+1}
+S_jS_{j+1}^{\dagger })-{1\over 4}n_jn_{j+1}]
\label{eq:Hamiltonian}
\end{eqnarray}
was proposed in an attempt to understand high-$T_c$
superconductivity \cite{Suth75,Schultz83,Wieg88,And90}. It is a
strongly correlated electron system with nearest-neighbor hopping
($t$) and anti-ferromagnetic exchange ($J$) of electrons. When
$J=2t$, the $t$-$J$ model becomes $gl(2|1)$ invariant (i.e.
supersymmetric). Using the nested algebraic Bethe ansatz method,
Essler and Korepin obtained the eigenvalues of the supersymmetric
$t$-$J$ model \cite{Korepin92}. The algebraic structure and
physical properties of the model were investigated in
\cite{Foer931,Ambjorn01}.

In this paper, using our previous results in \cite{zsy04} and
\cite{zsy0503}, we give the determinant representation of the
scalar products and the correlation functions of the
supersymmetric $t$-$J$ model. In section 2, we review the
background of the supersymmetric $t$-$J$ model and its algebraic
Bethe ansatz. In section 3, we apply our results on the Drinfeld
twists to construct the determinant representations of the
components of the Bethe states. In section 4, we obtain the
determinant representation of the scalar products of the Bethe
states. Then in section 5, we compute the correlation functions of
the local fermion operators of the model. We conclude the paper by
offering some discussions in section 6.

%%%%%%%%%%%%%%%%%%%%%%%%%%%%%%%%%%%%%%%%%%%%%%%%%%%%%%%%%%%%%%%
%                                                             %
%  2. The supersymmetry $t$-$J$ model                         %
%                                                             %
%%%%%%%%%%%%%%%%%%%%%%%%%%%%%%%%%%%%%%%%%%%%%%%%%%%%%%%%%%%%%%%
\sect{The supersymmetry $t$-$J$ model}

\subsection{Some background of the model }

Let $V$ be the 3-dimensional $gl(2|1)$-module and $R\in
End(V\otimes V)$ the $R$-matrix associated with this module. $V$
is $Z_2$-graded, and in the following we choose the FFB grading
for $V$, i.e. $[1]=[2]=1,[3]=0$. The $R$-matrix depends on the
difference of two spectral parameters $\l_1$ and $\l_2$ associated
with the two copies of $V$, and is, in the FFB grading, given by
\begin{eqnarray}
 R_{12}(\l_1,\l_2)\equiv R_{12}(\l_1-\l_2)%\nonumber\\
% &=& \nonumber\\
          &=&\left(\begin{array}{ccccccccc}
 c_{12}&0&0 & 0&0&0 & 0&0&0\\
 0&a_{12}&0 & -b_{12}&0&0 & 0&0&0 \\
 0&0&a_{12} & 0&0&0  & b_{12}&0&0\\
 0&-b_{12}&0 & a_{12}&0&0  & 0&0&0\\
 0&0&0      & 0&c_{12}&0  & 0&0&0\\
 0&0&0      & 0&0&a_{12}  & 0&b_{12}&0\\
 0&0&b_{12} & 0&0&0       & a_{12}&0&0\\
 0&0&0      & 0&0&b_{12}  & 0&a_{12}&0\\
 0&0&0      & 0&0&0       & 0&0&1
 \end{array} \right), \label{de:R} \nonumber\\
\end{eqnarray}
where
\begin{eqnarray}
&& a_{12}=a(\l_1,\l_2)\equiv
  {\l_1-\l_2\over \l_1-\l_2+\eta},\quad \quad
b_{12}=b(\l_1,\l_2)\equiv
  {\eta\over \l_1-\l_2+\eta},\quad\quad
\nonumber\\
&&c_{12}=c(\l_1,\l_2)\equiv {\l_1-\l_2-\eta\over \l_1-\l_2+\eta}
\end{eqnarray}
with $\eta\in C$ being the crossing parameter. One can easily
check that the $R$-matrix satisfies the unitary relation
\begin{equation}
R_{21}R_{12}=1.
\end{equation}
Here and throughout $R_{ij}\equiv R_{ij}(\l_i,\l_j)$. The
$R$-matrix satisfies the graded Yang-Baxter equation (GYBE)
\begin{equation}
R_{12}R_{13}R_{23}=R_{23}R_{13}R_{12}.
\end{equation}
In terms of the matrix elements defined by
\begin{equation}
R(\l)(v^{i'}\otimes
v^{j'})=\sum_{i,j}R(\l)^{i'j'}_{ij}(v^{i}\otimes v^{j}),
\end{equation}
the GYBE reads
\begin{eqnarray}
&&
\sum_{i',j',k'}R(\l_1-\l_2)^{i'j'}_{ij}R(\l_1-\l_3)^{i''k'}_{i'k}
R(\l_2-\l_3)^{j''k''}_{j'k'}
    (-1)^{[j']([i']+[i''])}\nonumber\\
&=&\sum_{i',j',k'}R(\l_2-\l_3)^{j'k'}_{jk}R(\l_1-\l_3)^{i'k''}_{ik'}
R(\l_1-\l_2)^{i''j''}_{i'j'}
    (-1)^{[j']([i]+[i'])}.
\end{eqnarray}

The quantum monodromy matrix $T(\l)$ of the supersymmetric $t$-$J$
model on a lattice of length $N$ is defined as
\begin{eqnarray}
T(\l)=R_{0N}(\l,\x_N)R_{0N-1}(\l,\x_{N-1})_{\ldots}
R_{01}(\l,\x_1), \label{de:T}
\end{eqnarray}
where the index 0 refers to the auxiliary space  and $\{\x_i\}$
are arbitrary inhomogeneous parameters depending on site $i$.
$T(\l)$ can be represented in the auxiliary space as the $3\times
3$ matrix whose elements are operators acting on the quantum space
$V^{\otimes N}$:
\begin{eqnarray}
 {T}(\l)=\left(\begin{array}{ccc}
{ A}_{11}(\l)& {A}_{12}(\l)&{ B}_1(\l)\\
{ A}_{21}(\l)& { A}_{22}(\l)&{ B}_2(\l)\\
{ C}_{1}(\l)& { C}_{2}(\l)&{ D}(\l)
 \end{array}\right)_{(0)}. \label{de:T-marix}
 \end{eqnarray}
 By using the GYBE, one may prove that the monodromy matrix
satisfies the GYBE
\begin{eqnarray}
R_{12}(\l-\m)T_1(\l)T_2(\m)=T_2(\m)T_1(\l)R_{12}(\l-\m).\label{eq:GYBE}
\end{eqnarray}
or in matrix form,
\begin{eqnarray}
&&\sum_{i',j'}R(\l-\m)^{i'j'}_{ij}
 T(\l)^{i''}_{i'}T(\m)^{j''}_{j'}(-1)^{[j']([i']+[i''])}
    \nonumber\\ && \mbox{} \quad\quad
=\sum_{i',j'}T(\m)^{j'}_{j}T(\l)^{i'}_{i}
R(\l-\m)^{i''j''}_{i'j'}(-1)^{[j']([i]+[i'])}.
\end{eqnarray}

Define the transfer matrix $t(\l)$
\begin{eqnarray}
t(\l)=str_0T(\l),\label{de:t}
\end{eqnarray}
where $str_0$ denotes the supertrace over the auxiliary space.
With the help of the GYBE, one may check that the transfer matrix
satisfies the commutation relation $[t(\l),t(\m)]=0,$ ensuring the
integrability of the system.

The transfer matrix gives the Hamiltonian of the system:
\begin{eqnarray}
H&=&{\partial \ln t(\l)\over \partial \l}|_{\l=0}\nonumber\\
  &=&-\sum _{j=1}^N
\left\{ \sum _{\sigma=\uparrow,\downarrow } (Q_{j+1,\sigma
}^{\dagger }Q_{j,\sigma } +Q_{j,\sigma }^{\dagger }Q_{j+1,\sigma
}) -2S_j^zS_{j+1}^z-S_jS_{j+1}^{\dagger
}-S_{j+1}S_j^{\dagger } +2T_jT_{j+1}\right\},\nonumber\\
\label{eq:Ham-QQ}
\end{eqnarray}
where $Q_{\sigma},Q_{\sigma}^\dag,S,S^\dag,S^z,T$ are generators
of the superalgebra $gl(2|1)$. The fundamental representations of
these operators take the following form
\begin{eqnarray}
&& S_j^z=\left( \begin{array}{ccc} -{1\over 2}&0&0\\ 0&{1\over 2}&0\\
0&0&0 \end{array}\right),\quad T_j=\left( \begin{array}{ccc}
{1\over 2}&0&0\\ 0&{1\over 2}&0\\ 0&0&1 \end{array}\right),\quad
 S_k=e^k_{21}, ~~~S_k^{\dagger }=e^k_{12},
 \nonumber \\ &&
 Q_{k,\uparrow}=e^k_{32},\quad
%\nonumber \\
Q_{k,\uparrow}^{\dagger }=e^k_{23},
~~~~Q_{k,\downarrow}=e^k_{31},~~~~Q_{k,\downarrow}^{\dagger
}=e^k_{13},
\end{eqnarray}
where $e_{ij}^k$ is a $3\times 3$ matrix acting on the $k$-th
space with elements $(e_{ij}^k)_{\alpha \beta }=\delta _{i\alpha }
\delta _{j\beta }$. Using the standard fermion representation
\begin{eqnarray}
 &&S_j=c_{j,\uparrow}^{\dagger }c_{j,\downarrow}, ~~~~S_j^{\dagger
}=c_{j,\downarrow}^{\dagger }c_{j,\uparrow}, ~~~~S_j^z={1\over
2}(n_{j,\uparrow}-n_{j,\downarrow}) ~~~~
%c_{j,\s},
 \nonumber\\
 && Q_{j,\s }=(1-n_{j,-\s
})c_{j,\s }, ~~~~Q_{j,\s }^{\dagger }=(1-n_{j,-\s })c_{j,\s
}^{\dagger }, ~~~~T_j=1-{1\over
2}n_j,\nonumber\\
 &&n_{j,\s}=c^\dag_{j,\s}c_{j,\s},
 ~~~~n_j=n_{j,\uparrow}+n_{j,\downarrow},
\label{de:repre-fermi}
\end{eqnarray}
one finds that (\ref{eq:Ham-QQ}) gives the Hamiltonian
(\ref{eq:Hamiltonian}) at the supersymmetric point $J=2t$.

\subsection{Algebraic Bethe ansatz}

The transfer matrix (\ref{de:t}) can be diagonalized by using the
nested algebraic Bethe ansatz. The Bethe state of the
supersymmetric $t$-$J$ model is defined as follows.
\begin{Definition}
Let $|0\rangle$ be the pseudo-vacuum state of the quantum tensor
space $V^{\otimes N}$, and $|0\rangle^{(1)}$ be the
pseudo-nested-vacuum state of the nested quantum tensor space
$(V^{(1)})^{\otimes n}$, i.e.,
\begin{equation}
 |0\rangle=\otimes_{i=1}^N\left(\begin{array}{c} 0\\0\\1\end{array}
 \right)_{(i)}, \quad\quad
 |0\rangle^{(1)}=\otimes_{j=1}^n\left(\begin{array}{c}
 0\\1\end{array} \right)_{(j)}.
 \label{de:vacuum-tj}
\end{equation}
The Bethe state of the supersymmetric $t$-$J$ model is then
defined by
\begin{eqnarray}
|\Omega_N(\{\l_j\})\rangle=\sum_{d_1\ldots d_n}
  (\Omega^{(1)}_n)^{d_1\ldots d_n}C_{d_1}(\l_1)\ldots
  C_{d_n}(\l_n)|0\rangle \quad\quad(\l_1\ne\ldots\ne\l_n),
 \label{de:Omega}
\end{eqnarray}
where $d_i=1,2$, $(\Omega^{(1)}_n)^{d_1\ldots d_n}$ is a component
of the nested Bethe state $|\O\rangle^{(1)}$ via
\begin{equation}
|\Omega_n(\{\l_j^{(1)}\})\rangle^{(1)}=C^{(1)}(\l^{(1)}_1)
 \cdots C^{(1)}(\l^{(1)}_m)|0\rangle^{(1)}
 \quad\quad(\l^{(1)}_1\ne\ldots\ne \l^{(1)}_m),
 \label{de:Bethe-vector-nested}
\end{equation}
and $C^{(1)}$, the creation operator of the nested $gl(2)$ system,
is the lower-triangular entry of the nested monodromy matrix
$T^{(1)}$
\begin{eqnarray}
T^{(1)}(\l^{(1)})&=&r_{0n}(\l^{(1)}-\l_n)r_{0n-1}(\l^{(1)}-\l_{n-1})
 \ldots r_{01}(\l^{(1)}-\l_1) \label{de:T-nested}
\nonumber\\
&\equiv& \left(\begin{array}{ccc}
 A^{(1)}(\l^{(1)})& B^{(1)}(\l^{(1)})\\
 C^{(1)}(\l^{(1)})& D^{(1)}(\l^{(1)})\end{array}\right)_{(0)}
\end{eqnarray}
with
\begin{equation}
r_{12}(\l_1,\l_2)\equiv r_{12}(\l_1-\l_2)= \left(
 \begin{array}{cccc}
 c_{12}&0&0&0\\ 0&a_{12}&-b_{12}&0\\
 0&-b_{12}&a_{12}&0\\ 0&0&0&c_{12} \end{array}\right).
 \label{de:r-nested}
\end{equation}
\end{Definition}
Similarly, we can also define the dual Bethe state
$\langle\Omega_N|$.
\begin{Definition}
With the help of the dual pseudo-vacuum state $\langle 0|$ and the
dual pseudo-nested-vacuum state $\langle 0|^{(1)}$, the dual Bethe
state is defined by
\begin{eqnarray}
\langle \O_N(\{\m_j\})|=\sum_{f_n,\ldots,f_1}(\O^{(1)})^{f_n\ldots
f_1}
 \langle0| B_{f_n}(\m_n)\ldots B_{f_1}(\m_1)
 \quad\quad(\m_n\ne\ldots\ne \m_1),\label{de:dual-state}
\end{eqnarray}
where $(\O^{(1)})^{f_n\ldots f_1}$ is a component of the dual
nested Bethe state $\langle\O|^{(1)}$
\begin{eqnarray}
 \langle\O_n(\{\m_j^{(1)}\})|^{(1)}=\langle 0|^{(1)}B^{(1)}(\m_m^{(1)})
  \ldots B^{(1)}(\m_1^{(1)})\quad\quad (\m_m^{(1)}\ne\ldots\ne\m_1^{(1)}).
\end{eqnarray}
\end{Definition}

The diagonalization of the transfer matrix $t(\l)$ (\ref{de:t})
leads to the following theorem \cite{Korepin92}:
\begin{Theorem}
The Bethe states $|\O_N(\{\l_j\})\rangle$ defined by
(\ref{de:Omega}) are eigenstates of the transfer matrix $t(\l)$ if
the spectral parameters $\l_j$ $(j=1,\ldots, n)$ satisfy the Bethe
ansatz equations (BAE)
\begin{eqnarray}
&&\prod_{k=1}^N a(\l_j,\x_k)
\prod_{l=1}^ma^{-1}(\l_j,\l_l^{(1)})=1
 \quad\quad (j=1,\ldots,n)
\label{eq:BAE}
\end{eqnarray}
and the nested Bethe ansatz equations (NBAE)
\begin{eqnarray}
&&\prod_{j=1}^n a(\l_j,\l^{(1)}_l)\prod_{k=1,\ne l}^m
  {\l^{(1)}_k-\l^{(1)}_l+\eta\over \l^{(1)}_k-\l^{(1)}_l-\eta }=1
\quad\quad (l=1,\ldots,m). \label{eq:BAE-nested}
\end{eqnarray}
The eigenvalues $\Lambda(\l,\{\l_k\},\{\l^{(1)}_j\})$ of the
transfer matrix $t(\l)$ are given by
\begin{eqnarray}
&&\Lambda(\l,\{\l_k\},\{\l^{(1)}_j\})\nonumber\\
 &&=\prod_{i=1}^Na(\l,\x_i)
    \prod_{j=1}^n{1\over a(\l,\l_j)}\Lambda^{(1)}(\l)
    +\prod_{j=1}^n{1\over a(\l_j,\l)},
    \label{eq:eigenvalue}
\end{eqnarray}
where $\Lambda^{(1)}(\l)$ is the eigenvalues of the nested
transfer matrix $t^{(1)}(\l)=\str_0T^{(1)}(\l)$
\begin{eqnarray}
 \Lambda^{(1)}(\l)=-\prod_{j=1}^n{a(\l,\l_j)\over a(\l_j,\l)}
 \prod_{k=1}^m{1\over a(\l,\l^{(1)}_k)}
 -\prod_{j=1}^n a(\l,\l_j)
 \prod_{k=1}^m{1\over a(\l^{(1)},\l)}.   \label{eq:eigenvalue-nested}
\end{eqnarray}
\end{Theorem}
One easily checks that this theorem also holds for the dual Bethe
state $\langle \O_N(\{\m_j\})|$ defined by (\ref{de:dual-state})
if we change the spectral parameters $\l_j$ and $\l_j^{(1)}$ in
(\ref{eq:BAE})-(\ref{eq:eigenvalue-nested}) to $\m_j$ and
$\m_j^{(1)}$, respectively.

%%%%%%%%%%%%%%%%%%%%%%%%%%%%%%%%%%%%%%%%%%%%%%%%%%%%%%%%%%%%%%%%%%%%%%%%
%                                                                      %
%     3. Symmetric representations of the Bethe state                  %
%                                                                      %
%%%%%%%%%%%%%%%%%%%%%%%%%%%%%%%%%%%%%%%%%%%%%%%%%%%%%%%%%%%%%%%%%%%%%%%%

\sect{Symmetric representations of the Bethe state}
\subsection{Factorizing $F$-matrix and its inverse}

With the help of the permutation group $\s\in {{\cal S}_N}$, one
may introduce the $R$-matrix $R^\sigma_{1\ldots N}$
\cite{Maillet96,Albert00} which can be expressed in terms of the
elementary $R$-matrix $R_{i\,i+1}=R^{\sigma_i}_{1\ldots N}$ for
any elementary permutation $\s_i(i,i+1)=(i+1,i)$ through the
decomposition law $
 R^{\sigma'\sigma}_{1\ldots N}
 =R^{\sigma}_{\sigma'(1\ldots N)}
  R^{\sigma'}_{1\ldots N}.  % \label{eq:R-RR}
$
We proved in \cite{zsy04,zsy0502,zsy0503} that for the $R$-matrix
$R^\sigma_{1\ldots N}$, there exists a non-degenerate
lower-diagonal $F$-matrix satisfying the relation
\begin{eqnarray}
F_{\sigma(1\ldots N)}(\x_{\sigma(1)},\ldots,\x_{\sigma(N)})
  R_{1\ldots N}^\sigma(\x_1,\ldots,\x_N)
 =F_{1\ldots N}(\x_1,\ldots,\x_N). \label{eq:R-F-N}
\end{eqnarray}
Explicitly,
\begin{eqnarray}
F_{1\ldots N}=\sum_{\sigma\in {\cal S}_N}
   \sum_{\alpha_{\sigma(1)}\ldots\alpha_{\sigma(N)}}^{\quad\quad *}
   \prod_{j=1}^N P_{\sigma(j)}^{\alpha_{\sigma(j)}}
   S(c,\sigma,\alpha_\sigma)R_{1\ldots N}^\sigma, \label{de:F}
\end{eqnarray}
where the sum $\sum^*$  is over all non-decreasing sequences of
the labels $\alpha_{\sigma(i)}$:
\begin{eqnarray}
&& \alpha_{\sigma(i+1)}\geq \alpha_{\sigma(i)},
 \quad\mbox{if}\quad
              \sigma(i+1)>\sigma(i), \nonumber\\
&& \alpha_{\sigma(i+1)}> \alpha_{\sigma(i)},
 \quad \mbox{if}\quad
              \sigma(i+1)<\sigma(i) \label{cond:F}
\end{eqnarray}
and the c-number function $S(c,\sigma,\alpha_\sigma)$ is given by
\begin{eqnarray}
S(c,\sigma,\alpha_\sigma)\equiv \exp\left\{{1\over2}
 \sum_{l>k=1}^N\left(1-(-1)^{[\alpha_{\sigma(k)}]}\right)
 \delta_{\alpha_{\sigma(k)},\alpha_{\sigma(l)}}
    \ln(1+c_{\sigma(k)\sigma(l)})\right\}.
\end{eqnarray}
The inverse of the $F$-matrix is given by
\begin{equation}
F^{-1}_{1\ldots N}=F^*_{1\ldots N}\prod_{i<j}\Delta_{ij}^{-1}
\label{eq:F-inverse}
\end{equation}
with
\begin{eqnarray}
[\Delta_{ij}]^{\beta_i\beta_j}_{\alpha_i\alpha_j} =
 \delta_{\alpha_i\beta_i}\delta_{\alpha_j\beta_j}\left\{
 \begin{array}{cl}
  a_{ij}& \mbox{if} \ \alpha_i>\alpha_j\\
  a_{ji}& \mbox{if} \ \alpha_i<\alpha_j\\
  1& \mbox{if}\ \alpha_i=\alpha_j=3 \\
  4a_{ij}a_{ji}& \mbox{if}\ \alpha_i=\alpha_j=1,2
  \end{array}\right.. \label{de:F-Inv}
\end{eqnarray}
and
\begin{eqnarray}
F^*_{1\ldots N}&=&\sum_{\sigma\in {\cal S}_N}
   \sum_{\alpha_{\sigma(1)}\ldots\alpha_{\sigma(N)}}^{\quad\quad **}
   S(c,\sigma,\alpha_\sigma)R_{\sigma(1\ldots N)}^{\sigma^{-1}}
   \prod_{j=1}^N P_{\sigma(j)}^{\alpha_{\sigma(j)}},
    \label{de:F*}  \nonumber\\
\end{eqnarray}
where the sum $\sum^{**}$ is taken over all possible $\alpha_i$
which satisfies the following non-increasing constraints:
\begin{eqnarray}
&& \alpha_{\sigma(i+1)}\leq \alpha_{\sigma(i)},
 \quad\mbox{if}\quad
              \sigma(i+1)<\sigma(i), \nonumber\\
&& \alpha_{\sigma(i+1)}< \alpha_{\sigma(i)},\quad \mbox{if}\quad
              \sigma(i+1)>\sigma(i). \label{cond:F*}
\end{eqnarray}

\subsection{Bethe state in the $F$-basis}
The non-degeneracy of the $F$-matrix means that its column vectors
form a complete basis, which is called the $F$-basis. In
\cite{zsy0503,zsy04}, we found that in the $F$-basis, the creation
and annihilation operators  $C_i(\l)$ and $B_i(\l)$ $(i=1,2)$ of
the supersymmetrix $t$-$J$ model have the symmetric form:
\begin{eqnarray}
\tilde C_2(\l)&=&F_{1\ldots N} C_2(\l)F^{-1}_{1\ldots N}
 =\sum_{i=1}^N b_{0i}E^{23}_{(i)}\otimes_{j\ne i}
 \mbox{diag}\left(a_{0j},2a_{0j},1\right)_{(j)},\label{eq:C2-tilde}\\
\tilde C_1(u)&=&F_{1\ldots N} C_1(\l)F^{-1}_{1\ldots N}
 =\sum_{i=1}^N b_{0i}E^{23}_{(i)}\otimes_{j\ne i}
 \mbox{diag}\left(2a_{0j},a_{0j}a^{-1}_{ij},1\right)_{(j)}
 \nonumber\\ &&\mbox{}
  -\sum_{i\ne j}^N{\eta b_{0j}a_{0i}\over \x_i-\x_j}
     E^{12}_{(i)}\otimes E^{23}_{(j)}\otimes_{k\ne i,j}
     \mbox{diag}\left(2a_{0k},a_{0k}a_{ik}^{-1},
     1\right)_{(k)},\label{eq:C1-tilde}\\
\tilde B_2(\l)&=&F_{1\ldots N} B_2(\l)F^{-1}_{1\ldots N}
 =-\sum_{i=1}^N b_{0i}E^{32}_{(i)}\otimes_{j\ne i}
  \mbox{diag}\left(a_{0j},a_{0j}(2a_{ji})^{-1},
     a_{ji}^{-1}\right)_{(j)}, \label{eq:B2-tilde}\\
\tilde B_1(\l)&=&F_{1\ldots N} B_1(\l)F^{-1}_{1\ldots N}
 =-\sum_{i=1}^N b_{0i}E^{31}_{(i)}\otimes_{j\ne i}
  \mbox{diag}\left(a_{0j}(2a_{ji})^{-1},a_{0j}a_{ji}^{-1},
     a_{ji}^{-1}\right)_{(j)}\nonumber\\ &&\mbox{}
  -\sum_{i\ne j}^N{\eta b_{0i}a_{0j}\over \x_j-\x_i}
     E^{32}_{(i)}\otimes E^{21}_{(j)}\otimes_{k\ne i,j}
     \mbox{diag}\left(a_{0k}(2a_{kj})^{-1},a_{0k}a_{ki}^{-1},
     a_{ki}^{-1}\right)_{(k)}, \label{eq:B1-tilde}
\end{eqnarray}
where $a_{0j}\equiv a(\l,\x_j)$  $b_{0j}\equiv b(\l,\x_j)$.

Acting the associated $F$-matrix on the pseudo-vacuum state
$|0\rangle$, one finds that the pseudo-vacuum state is invariant.
It is due to the fact that only the term with all roots equal to 3
will produce  non-zero results. Therefore, the $gl(2|1)$ Bethe
state (\ref{de:Omega}) in the $F$-basis can be written as
\begin{eqnarray}
   |\tilde\O_N(\{\l_j\})\rangle
 &\equiv& F_{1\ldots N}|\O_N(\{\l_j\})\rangle\nonumber\\
 &=& \sum_{d_1\ldots d_{n}}
  (\Omega^{(1)}_{n})^{d_1\ldots d_{n}}\tilde C_{d_1}(\l_1)\ldots \tilde
  C_{d_{n}}(\l_{n})|0\rangle. \label{eq:Omega-F}
\end{eqnarray}
Without loss of generality, we will only concentrate on the Bethe
state with the {\bf quantum number $p$} which indicates the number
of $d_i=2$, and will use the notation
$|\tilde\O_N(\{\l_j\}_{(p,n)})\rangle$ with the subscript pair
$(p,n)$ to denote a Bethe state which has quantum number $p$ and
has $n$ spectral parameters.
\begin{Proposition}
The Bethe state of the supersymmetric t-J model can be represented
in the $F$-basis  by
\begin{eqnarray}
 && |\tilde\O_N(\{\l_j\}_{(p,n)})\rangle \nonumber\\
 &&=\sum_{\s\in{\cal S}_N}Y_R(\{\l_{\s(i)}\},\{\l_{\s(j)}^{(1)}\})
 {\tilde C}_{2}(\l_{\sigma(1)})\ldots {\tilde C}_{2}(\l_{\sigma(p)})
 {\tilde C}_{1}(\l_{\sigma(p +1)})\ldots
 {\tilde C}_{1}(\l_{\sigma(n)})\,|0\rangle \label{eq:Omega-YR} \\
% \nonumber\\  &&
 &&=\sum_{\s\in{\cal S}_N}Y_R(\{\l_{\s(i)}\},\{\l_{\s(j)}^{(1)}\})
     \sum_{i_1<\ldots<i_{p}}\sum_{i_{p+1}<\ldots<i_n}
   2^{{p(p-1)+(n-p)(n-p-1)\over 2}}
 \prod_{l=1}^{p}~\prod_{k=p +1}^{n}a(\l_{\sigma(l)},\x_{i_{k}})
  \nonumber\\ &&\quad\times
   \mbox{det}\,{\cal B}_p(\l_{\s(1)},\ldots,\l_{\s(p)};
   \x_{i_1},\ldots,\x_{i_p})
 \nonumber\\ &&\quad\times
 \mbox{det}\,{\cal B}_{n-p}(\l_{\s(p+1)},\ldots,\l_{\s(n)};
   \x_{i_{p+1}},\ldots,\x_{i_n})
   \overrightarrow{\prod_{j=1}^{p}}E^{23}_{(i_j)}
   \overrightarrow{\prod_{j=p+1}^{n}}E^{13}_{(i_j)}|0\rangle
   \label{eq:state-proposition1}
 %  \nonumber\\
\end{eqnarray}
with the sets
$\{i_1,\ldots,i_p\}\cap\{i_{p+1},\ldots,i_n\}=\varnothing$ and the
prefactor $Y_R$ being
\begin{eqnarray}
 &&Y_R(\{\l_{\s(i)}\},\{\l_{\s(j)}^{(1)}\} ) \nonumber\\
 &&={1\over p!(n-p)!}
  c^{\sigma}_{1\ldots n}
  B^*_{n-p}\left(\l_{p +1}^{(1)},\ldots,\l_{n}^{(1)}|
     \l_{\sigma(p +1)},\ldots,\l_{\sigma(n)}\right)
%  \nonumber\\ &&\times
    \prod_{k=p +1}^{n}~\prod_{l=1}^{p}
    \left(-{2a(\l_{\sigma(k)},\l_{\sigma(l)})\over
     a(\l_{\sigma(l)},\l_{\sigma(k)})}\right).\nonumber\\
\end{eqnarray}
Here $c^\sigma_{1,\ldots,n}$ has the decomposition law
 $c^{\sigma'\sigma}_{1\ldots n}
 =c^{\sigma}_{\sigma'(1\ldots n)}
  c^{\sigma'}_{1\ldots n}$
with $c^{\sigma_i}_{1\ldots n}=c_{i\ i+1}\equiv c(v_i,v_{i+1})$
for an elementary permutation $\sigma_i$, the $c$-number $B_{p}^*$
is given by
\begin{eqnarray*}
&&B^*_{p}\left(\l_1^{(1)},\ldots,\l_{p}^{(1)}|
                \l_{1},\ldots,\l_{p}\right)
\nonumber\\
 &=&\sum_{\sigma\in{\cal S}_{p}}\prod_{k=1}^{p}
  \left(-b(\l_k^{(1)},\l_{\sigma(k)})\right)
  \prod_{j\ne \sigma(k),\ldots,\sigma(p)}
  {c(\l_k^{(1)},\l_{j})\over 2a(\l_{\s(k)},\l_{j})}
  \prod_{l=k+1}^{p} 2a(\l_k^{(1)},\l_{\sigma(l)}),
\end{eqnarray*}
and the elements of the $n\times n$ matrix ${\cal
B}_n(\{\l_i\};\{\x_{j}\})$ are
\begin{equation}
({\cal B}_n)_{\alpha\beta}
 %(\{\l_i\},\{\x_j\})
 =b(\l_\alpha,\x_\beta)
  \prod_{\gamma=1}^{\alpha-1}a(\l_\gamma,\x_\beta).
\end{equation}
\end{Proposition}
In (\ref{eq:state-proposition1}), we have used the convention
$\overrightarrow{\prod_{i=1}^n}f_i\equiv f_1\ldots f_n$. For our
later use, we also introduce the notation
$\overleftarrow{\prod_{i=1}^n}f_i\equiv f_n\ldots f_1$.

\noindent {\it Proof}. By using the exchange symmetry of the Bethe
state
\begin{eqnarray}
|\tilde\Omega_N(\{\l_{\s(j)}\})\rangle
 ={1\over c^\sigma_{1\ldots n}}
  |\tilde\Omega_N(\{\l_j\})\rangle,
  ~\sigma\in {\cal S}_{n} \label{eq:exchange}
\end{eqnarray}
and the commutation relation between $C_i(\l)$ and $C_j(\m)$
\begin{eqnarray}
 \tilde C_i(\m)\tilde C_j(\l)
 &=&-{1\over a(\l,\m)}\tilde C_j(\l)\tilde C_i(\m)
    +{b(\l,\m)\over a(\l,\m)}\tilde C_j(\m)\tilde C_i(\l),
    \label{eq:commu-CC-F}
\end{eqnarray}
we have showed in \cite{zsy04,zsy0503} that the Bethe state
(\ref{eq:Omega-F}) can be written as
\begin{eqnarray}
|{\tilde\Omega}_N(\{\l_j\}_{(p,n)})\rangle %\nonumber\\&&
 &=&\sum_{\sigma \in {\cal S}_{n}}Y_R(\{\l_{\s(i)}\},\{\l_{\s(j)}^{(1)}\} )
    \sum_{i_1<\ldots<i_{p}}\sum_{i_{p+1}<\ldots<i_n}
    \prod_{k=p +1}^{n}~\prod_{l=1}^{p}a(\l_{\sigma(l)},\x_{i_k})
      \nonumber\\
&&\times B_{p}(\l_{\sigma(1)},\ldots,
     \l_{\sigma(p)}|\x_{i_{1}},\ldots,\x_{i_{p}})
                       \nonumber\\
&&\times B_{n-p}(\l_{\sigma(p+1)},\ldots,
     \l_{\sigma(n)}|\x_{i_{p+1}},\ldots,\x_{i_{n}})
     \overrightarrow{\prod_{j=1}^{p}}E^{23}_{(i_j)}
     \overrightarrow{\prod_{j=p+1}^{n}}E^{13}_{(i_j)}|0\rangle,
\end{eqnarray}
where
\begin{eqnarray}
B_{n}(\l_1,\ldots,\l_{n}|\x_{i_1},\ldots,\x_{i_{n}})
 &=&
 \sum_{\sigma\in {\cal S}_n}\mbox{sign}(\sigma)
  \prod_{k=1}^n b(\l_k,\x_{i_{\sigma(k)}})
  \prod_{l=k+1}^n 2a(\l_k,\x_{i_{\sigma(l)}}).
\end{eqnarray}
One checks that the function
$B_{n}(\l_1,\ldots,\l_{n}|\x_{i_1},\ldots,\x_{i_{n}})$ is
equivalent to the determinant $2^{n(n-1)/2}\mbox{det}{\cal
B}_n(\{\l_k\},\{\x_j\})$, thus proving the proposition.
% $\quad\quad\quad~~~~~~~~~~~~~~~~~~~~~~~~~~~~~~~~~~\Box$
 \begin{flushright}$\Box$~~~~~\end{flushright}

By a similar procedure, one may prove the following proposition
for the dual Bethe state $\langle\t\O_N(\{\m_j\}_{(p,n)})|$
(\ref{de:dual-state}):
\begin{Proposition}
The dual Bethe state $\langle\t\O_N(\{\m_j\}_{(p,n)})|$ of the
supersymmetric t-J model can be represented by
\begin{eqnarray}
  \langle\t\O_N(\{\m_j\}_{(p,n)}|% \nonumber\\
 &=&\sum_{\s\in{\cal S}_N}Y_L(\{\m_{\s(i)}\},\{\m_{\s(j)}^{(1)}\})
 \langle0|
 {\tilde B}_{1}(\m_{\sigma(n)})\ldots {\tilde B}_{1}(\m_{\sigma(p+1)})
 \nonumber\\&&\times
 {\tilde B}_{2}(\m_{\sigma(p)})\ldots
 {\tilde B}_{2}(\m_{\sigma(1)})\, ~~~~~~\label{eq:Omega-YL} \\
% \nonumber\\  &&
 &=&\sum_{\s\in{\cal S}_N}Y_L(\{\m_{\s(i)}\},\{\m_{\s(j)}^{(1)}\})
     \sum_{i_1<\ldots<i_{p}}\sum_{i_{p+1}<\ldots<i_n}
   (-1)^n 2^{-{p(p-1)+(n-p)(n-p-1)\over 2}}
  \nonumber\\ &&\times
  \prod_{l=1}^{p}~\prod_{k=p +1}^{n}a(\m_{\sigma(l)},\x_{i_{k}})
  \prod_{l=1}^p\prod_{k\ne i_l,i_{p+1},\ldots,i_n}^N a^{-1}(\x_{k},\x_{i_l})
  \prod_{l=p+1}^n\prod_{k=1,\ne i_l}^N a^{-1}(\x_{k},\x_{i_l})
  \nonumber\\ &&\times
   \mbox{det}\,{\cal B}_p(\m_{\s(1)},\ldots,\m_{\s(p)};
   \x_{i_1},\ldots,\x_{i_p})
 \nonumber\\ &&\times
 \mbox{det}\,{\cal B}_{n-p}(\m_{\s(p+1)},\ldots,\m_{\s(n)};
   \x_{i_{p+1}},\ldots,\x_{i_n})
   \langle0|\overleftarrow{\prod_{j=p+1}^{n}}E^{31}_{(i_j)}
   \overleftarrow{\prod_{j=1}^{p}}E^{32}_{(i_j)},
 %  \nonumber\\
\end{eqnarray}
where the prefactor $Y_L$ is
\begin{eqnarray}
 &&Y_L(\{\m_{\s(i)}\},\{\m_{\s(j)}^{(1)}\} ) \nonumber\\
 &&={1\over p!(n-p)!}
  c^{\sigma}_{1\ldots n}
  B^{**}_{n-p}\left(\m_{p +1}^{(1)},\ldots,\m_{n}^{(1)}|
     \m_{\sigma(p +1)},\ldots,\m_{\sigma(n)}\right)
%  \nonumber\\ &&\times
    \prod_{k=p +1}^{n}~\prod_{l=1}^{p}
    \left(-{2a(\m_{\sigma(k)},\m_{\sigma(l)})\over
     a(\m_{\sigma(l)},\m_{\sigma(k)})}\right),\nonumber\\
\end{eqnarray}
and the $c$-number $B_{p}^{**}$ is given by
\begin{eqnarray*}
&&B^{**}_{p}\left(\m_1^{(1)},\ldots,\m_{p}^{(1)}|
                \m_{1},\ldots,\m_{p}\right)
\nonumber\\
 &&=\sum_{\sigma\in{\cal S}_{p}}\prod_{k=1}^{p}
  b(\m_k^{(1)},\m_{\sigma(k)})
  \prod_{j\ne \sigma(k),\ldots,\sigma(p)}c(\m_k^{(1)},\m_{j})
  \prod_{l=k+1}^{p}
  {a(\m_k^{(1)},\m_{\sigma(l)})\over
  2a(\m_{\s(l)},\m_{\s(k)})}\,.
\end{eqnarray*}
\end{Proposition}

%%%%%%%%%%%%%%%%%%%%%%%%%%%%%%%%%%%%%%%%%%%%%%%%%%%%%%%%%%%%%%%%%%%%
%                                                                  %
% 4. Determinant representation of the scalar product              %
%                    of the Bethe states                           %
%                                                                  %
%%%%%%%%%%%%%%%%%%%%%%%%%%%%%%%%%%%%%%%%%%%%%%%%%%%%%%%%%%%%%%%%%%%%

\sect{Determinant representation of the scalar product }
%\subsection{The determinant representation of the scalar product}

The scalar product of the Bethe states with a given quantum number
$p$ is defined by
\begin{eqnarray}
 P_n(\{\m_i\}_{(p,n)},\{\l_j\}_{(p,n)})&=&
 \langle\O_N(\{\m_j\}_{(p,n)})|\O_N(\{\l_j\}_{(p,n)})\rangle.
 \label{de:P_n}
\end{eqnarray}
The invariant property of the pseudo-vacuum state under the
$F$-transformation, i.e. $F_{1\ldots N}|0\rangle=|0\rangle$ and
$\langle0|F^{-1}_{1\ldots N}=\langle0|$, implies that in the
$F$-basis, the scalar product $P_n$ is
\begin{eqnarray}
 &&P_n(\{\m_i\}_{(p,n)},\{\l_j\}_{(p,n)})%\nonumber\\
  =
 \langle\tilde \O_N(\{\m_j\}_{(p,n)})|
 \tilde\O_N(\{\l_j\}_{(p,n)})\rangle
 \nonumber\\ &&=\sum_{\s',\s}
 Y_L(\{\m_{\sigma'(j)}\},\{\m^{(1)}_{\sigma'(k)}\})
 Y_R(\{\l_{\sigma(j)}\},\{\l^{(1)}_{\sigma(k)}\})
 \nonumber\\&& \quad\times
 \langle 0|\tilde B_1(\m_{\s'(n)})\ldots \tilde B_1(\m_{\s'(p+1)})
 \tilde B_2(\m_{\s'(p)})\ldots \tilde B_2(\m_{\s'(1)})
  \nonumber\\&&\quad \times
  \tilde C_{2}(\l_{\s(1)})\ldots \tilde C_{2}(\l_{\s(p)})
  \tilde C_{1}(\l_{\s(p+1)})\ldots \tilde C_{1}(\l_{\s(n)})|0\rangle.
  \label{de:S_n}
\end{eqnarray}

To compute the scalar product, following \cite{Kitanine98}, we
introduce the following intermediate functions
\begin{eqnarray}
&&G^{(m)}(\{\l_k\}_{(p,n)},u_1,\ldots,\m_m,i_{m+1},\ldots,i_n)\nonumber\\
 &&=\left\{\begin{array}{l}
 \langle0|\overleftarrow{\prod_{k=p+1}^n}E_{(i_k)}^{31}
    \overleftarrow{\prod_{k=m+1}^{p}}E_{(i_k)}^{32}
 \tilde B_2(\m_m)\ldots \tilde B_2(\m_1) \\ \quad\quad \times
  \tilde C_2(\l_1)\ldots \tilde C_2(\l_{p})
  \tilde C_1(\l_{p+1})\ldots \tilde C_1(\l_n)|0>\quad
    \mbox{ for }m\leq p,\\[3mm]
%\langle0|\overleftarrow{\prod_{k=m+1}^n}E_{(i_k)}^{31}
% \tilde B_2(\m_m)\ldots \tilde B_2(\m_1)\\ \quad\quad \times
%  \tilde C_2(\l_1)\ldots \tilde C_2(\l_{p})
%  \tilde C_1(\l_{p+1})\ldots \tilde C_1(\l_n)|0>\quad
%    \mbox{ for } m=p\\[3mm]
\langle0|\overleftarrow{\prod_{k=m+1}^n}E_{(i_k)}^{31}
 \tilde B_1(\m_m)\ldots \tilde B_1(\m_{p+1})
 \tilde B_2(\m_{p}) \tilde B_2(\m_1)
  \\ \quad\quad \times
  \tilde C_2(\l_1)\ldots \tilde C_2(\l_{p})
  \tilde C_1(\l_{p+1})\ldots \tilde C_1(\l_n)|0>\quad
    \mbox{ for }m\geq p+1.
  \end{array}\right.
  \label{de:G-m}
\end{eqnarray}
where the lower indices of $E^{32}_{(i_k)}$ and $E^{31}_{(i_k)}$
satisfy the relations $i_{m+1}<\ldots<i_{p}$,
$i_{p+1}<\ldots<i_{n}$ and
$\{i_1,\ldots,i_p\}\cap\{i_{p+1},\ldots,i_n\}=\varnothing$. Thus,
the scalar product can be rewritten as
\begin{eqnarray}
&&P_n(\{\m_i\}_{(p,n)},\{\l_j\}_{(p,n)}) \nonumber\\
 &&=
\sum_{\s,\s'}Y_L(\{\m_{\sigma'(j)}\},\{\m^{(1)}_{\sigma'(k)}\})
 Y_R(\{\l_{\sigma(j)}\},\{\l^{(1)}_{\sigma(k)}\})\,\
 G^{(n)}(\{\l_{\s(j)}\}_{(p,n)},\{\m_{\s'(k)}\}_{(p,n)}).
 \label{eq:Sn-Gn}
\end{eqnarray}

We now compute $G^{(m)}$ for $m\leq p$ and $m\geq p+1$ separately.

%$\bullet$ $1\leq m\leq p$
\subsection{$1\leq m\leq p$}

We first compute the function $G^{(m)}$ for $m\leq p$.

Inserting a complete set, (\ref{de:G-m}) becomes
\begin{eqnarray}
&&G^{(m)}(\{\l_k\}_{(p,n)},\m_1,\ldots,\m_m,i_{m+1},\ldots,i_n)\nonumber\\
 &&=\sum_{j\ne i_{m+1},\ldots,i_n}^N
     \langle0|\overleftarrow{\prod_{k=p+1}^n}E_{(i_k)}^{31}
    \overleftarrow{\prod_{k=m+1}^{p}}E_{(i_k)}^{32}
    \tilde B_2(\m_m)
     \overrightarrow{\prod_{k=m+1}^{m+q}}E_{(i_k)}^{23}E_{(j)}^{23}
     \overrightarrow{\prod_{k=m+q+1}^{p}}E_{(i_k)}^{23}
     \overrightarrow{\prod_{k=p+1}^n}E_{(i_k)}^{13}|0\rangle \nonumber\\
 &&\quad\times
 G^{(m-1)}(\{\l_k\}_{(p,n)},\m_1,\ldots,\m_{m-1},i_{m+1},
 \ldots,i_{m+q},j,i_{m+q+1}\ldots,i_n) \quad (0\leq q\leq p-m).
 \nonumber\\
 \label{de:Gm-Gm-1}
\end{eqnarray}
In view of (\ref{eq:B2-tilde}), we have
\begin{eqnarray}
&&\langle0|\overleftarrow{\prod_{k=p+1}^n}E_{(i_k)}^{31}
    \overleftarrow{\prod_{k=m+1}^{p}}E_{(i_k)}^{32}
    \tilde B_2(\m_m)
     \overrightarrow{\prod_{k=m+1}^{m+q}}E_{(i_k)}^{23}E_{(j)}^{23}
     \overrightarrow{\prod_{k=m+q+1}^{p}}E_{(i_k)}^{23}
     \overrightarrow{\prod_{k=p+1}^n}E_{(i_k)}^{13}|0\rangle
\nonumber\\
%&& \rightarrow -(2\cosh\eta)^{(n-m)}
% <i_{n},\ldots,i_{m+1}|\tilde B(\m_m)
%     |i_{m+1},\ldots, i_{m+p},j,i_{m+p+1},\ldots,i_n> \nonumber\\
 &&=-(-1)^{q}2^{-(p-m)}\cdot\,
 b(\m_m,\x_j)\prod_{l=m+1}^p a(\m_m,\x_{i_l})
\prod_{l=p+1}^n a(\m_m,\x_{i_l})
 \prod_{k\ne j,i_{p+1},\ldots,i_n}^Na^{-1}(\x_k,\x_j)
 . \nonumber\\ \label{eq:B-expect}
\end{eqnarray}

Substituting the expressions of $\tilde C_1$ (\ref{eq:C1-tilde})
and $\tilde C_2$ (\ref{eq:C2-tilde}) into (\ref{de:G-m}), we
obtain $G^{(0)}$:
\begin{eqnarray}
&& G^{(0)}(\{\l_k\}_{(p,n)},i_1,\ldots,i_n)%\nonumber\\
 =\langle0|\overleftarrow{\prod_{k=p+1}^n}E_{(i_k)}^{31}
    \overleftarrow{\prod_{k=1}^{p}}E_{(i_k)}^{32}
    \prod_{k=1}^{p}\tilde C_2(\l_k)
    \prod_{k=p+1}^{n}\tilde C_1(\l_k)|0> \nonumber\\
 &&=2^{{p(p-1)+(n-p)(n-p-1)\over 2}}
    \prod_{l=1}^{p}\prod_{k=p+1}^{n}
    a(\l_{l},\x_{i_k})
 \mbox{det} {\cal B}_{p}(\l_1,\ldots,\l_{p};
      \x_{i_1},\ldots,\x_{i_{p}})
 \nonumber\\ &&\quad\times
 \mbox{det} {\cal B}_{n-p}(\l_{p+1},\ldots,\l_n;
      \x_{i_{p+1}},\ldots,\x_{i_{n}}). \label{eq:G-0}
\end{eqnarray}

We compute $G^{(m)}$ by using the recursion relation
(\ref{de:Gm-Gm-1}). One sees that there are two different
determinants in $G^{(0)}$, which are labelled by different $\l$'s
and $\x_{i_{k}}$'s. For  $m\leq p$ we only focus on the first
determinant, i.e. $\mbox{det}{\cal B}_{p}$.
%The main idea to
%compute $G^{(m)}\ (m\leq p)$ is to replace $\x_{i_k} (k\leq p)$ in
%the determinant $\mbox{det}{\cal B}_{p}$ by the inhomogeneous
%parameters $\m_m\, (m\leq p)$. Therefore in our current stage, we
%constrain the condition
%\begin{equation}
%\m_m\ne \x_{i_l}-\eta\quad\quad (l=p+1,\ldots,n)
%\label{eq:mu-condition}
%\end{equation}
%to the parameters $\m_{m}\, (m\leq p)$.

To compute $G^{(1)}$, we substitute (\ref{eq:B-expect}) and
(\ref{eq:G-0}) into (\ref{de:Gm-Gm-1}) to obtain
\begin{eqnarray}
&& G^{(1)}(\{\l_k\}_{(p,n)},\m_1,i_2,\ldots,i_n)\nonumber\\
&&  =\sum^N_{j\ne
i_2,\ldots,i_n}\langle0|\overleftarrow{\prod_{k=p+1}^n}E_{(i_k)}^{31}
    \overleftarrow{\prod_{k=2}^{p}}E_{(i_k)}^{32}
    \tilde B_2(\m_1)
     \overrightarrow{\prod_{k=2}^{q+1}}E_{(i_k)}^{23}E_{(j)}^{23}
     \overrightarrow{\prod_{q+2}^{p}}E_{(i_k)}^{23}
     \overrightarrow{\prod_{k=p+1}^n}E_{(i_k)}^{13}|0\rangle \nonumber\\
&&\quad\times
G^{(0)}(\{\l_k\}_{(p,n)},i_2,\ldots,i_{q+1},j,i_{q+2},\ldots,i_n)
   \nonumber\\&&
 =-2^{{(p-1)(p-2)+(n-p)(n-p-1)\over 2}}
 \prod_{l=1}^{p}\prod_{k=p+1}^na(\l_l,\x_{i_k})
 \sum^N_{j\ne i_2,\ldots,i_n}(-1)^{q}\,
 b(\m_1,\x_j)
 \nonumber\\ &&\quad\times
 \prod_{l=2}^p a(\m_1,\x_{i_l})
 \prod_{k\ne j,i_{p+1},\ldots,i_n}^Na^{-1}(\x_k,\x_j)\,
 \mbox{det}{\cal B}_{p}(\l_1\ldots,\l_p;\x_{i_2},\ldots,\x_{i_{q+1}},
      \x_j,\x_{i_{q+2}},\ldots,\x_{i_{p}})
 \nonumber\\ && \quad\times
 \prod_{l=p+1}^n a(\m_1,\x_{i_l})
 \mbox{det}{\cal B}_{n-p}(\l_{p+1},\ldots,\l_n;
      \x_{i_{p+1}},\ldots,\x_{i_{n}})
      . \label{eq:G-1-1}
\end{eqnarray}
Let $\l_k\, (k=1,\dots,n)$ label the row and $\x_l\,
(l=i_2,\ldots,j,\ldots,i_{p}) $ label the column of the matrix
${\cal B}_{p}$. From (\ref{eq:G-0}), one sees that the column
indices in (\ref{eq:G-1-1}) satisfy the sequence
$i_2<\ldots<j<\ldots<i_{p}$. Therefore, moving the $j$th column in
the matrix ${\cal B}_{p}$ to the first column, we have
\begin{eqnarray}
&& G^{(1)}(\{\l_k\}_{(p,n)},\m_1,i_2,\ldots,i_n)\nonumber\\
&&=-2^{{(p-1)(p-2)+(n-p)(n-p-1)\over 2}}
  \prod_{l=1}^{p}\prod_{k=p+1}^na(\l_l,\x_{i_k})
 \sum^N_{j\ne i_2,\ldots,i_n}\,
 b(\m_1,\x_j)
 \nonumber\\ &&\quad\times
 \prod_{l=2}^p a(\m_1,\x_{i_l})
 \prod_{k\ne j,i_{p+1},\ldots,i_n}^Na^{-1}(\x_k,\x_j)\,
 \mbox{det}{\cal B}_{p}(\l_1,\ldots,\l_p;\x_j,\x_{i_2},\ldots,\x_{i_n})
 \nonumber\\ && \quad\times
 \mbox{det}{\cal B}_{n-p}(\l_{p+1},\ldots,\l_n;
      \x_{i_{p+1}},\ldots,\x_{i_{n}})
 \nonumber\\ &&
 =-2^{{(p-1)(p-2)+(n-p)(n-p-1)\over 2}}
 \mbox{det}{\cal B}^{(1)}_{p}(\l_1,\ldots,\l_p;\m_1,\x_{i_2},\ldots,\x_{i_{p}})
 \nonumber\\ && \quad\times
 \prod_{l=p+1}^n a(\m_1,\x_{i_l})
 \mbox{det}{\cal B}_{n-p}(\l_{p+1},\ldots,\l_n;
      \x_{i_{p+1}},\ldots,\x_{i_{n}}),
 \label{eq:G-1}
\end{eqnarray}
where the matrix $\t{\cal B}^{(1)}_{p}(\{\l_k\},\m_1,
\x_{i_2},\ldots,\x_{i_{p}})$ is given by
\begin{eqnarray}
({\cal B}^{(1)}_{p})_{\alpha\beta}
 &=&\prod_{k=p+1}^n a(\l_\a,\x_{i_k})\,a(\m_1,\x_{i_\beta})
 ({\cal B}_{p})_{\alpha\beta} \quad\quad
 (1\leq\a\leq p\,\mbox{ and }2\leq\beta\leq p), \label{eq:B-1-0}\\
({\cal B}^{(1)}_{p})_{\alpha 1}&=&
\prod_{k=p+1}^na(\l_\a,\x_{i_k}) \sum^N_{j\ne i_2,\ldots,i_n}
 b(\m_1,\x_j)b(\l_{\alpha},\x_j)
 \prod_{\gamma=1}^{\alpha-1}a(\l_\gamma,\x_j) \nonumber\\ &&\times
% \prod_{k=p+1}^n a(\m_1,\x_{i_k})
 \prod_{k\ne j,i_{p+1},\ldots,i_n}^N a^{-1}(\x_{k},\x_j)
 \quad\quad (1\leq\a\leq p). \label{eq:B-1-1}
\end{eqnarray}
Using the properties of determinant, one finds that if
$j=i_2,\ldots,i_{p}$, the corresponding terms in (\ref{eq:B-1-1})
contribute zero to the determinant. Thus, without changing the
determinant of the matrix ${\cal B}^{(1)}_{p}$, the elements
$({\cal B}^{(1)}_{p})_{\alpha 1}$ in (\ref{eq:B-1-1}) may be
replaced by
\begin{eqnarray}
({\cal B}^{(1)}_{p})_{\alpha 1}&=&
\prod_{k=p+1}^n{\l_\a-\x_{i_k}\over\l_\a-\x_{i_k}+\eta}
 \sum^N_{j\ne i_{p+1},\ldots,i_n}
 {\eta\over \m_1-\x_{j}+\eta}
 {\eta\over \l_\a-\x_{j}+\eta}
 \nonumber\\ &&\times
 \prod_{\gamma=1}^{\alpha-1}{\l_\g-\x_j\over\l_\g-\x_j+\eta}
% \prod_{k=p+1}^n {\m_1-\x_{i_k}\over\m_1-\x_{i_k}+\eta }
 \prod_{k\ne j,i_{p+1},\ldots,i_n}^N
  {\x_{k}-\x_j+\eta\over\x_{k}-\x_j }.\label{eq:B-1-2}
\end{eqnarray}
%Investigating the above formula, one sees that as a real function
%of $\m_1$, $\m_1=\x_j-\eta\, (j=1,\ldots,N)$ are all simple poles,
%and when $\m_1\rightarrow \infty$, the function will be tend to 1.
Thanks to the Bethe ansatz equation (\ref{eq:BAE}), we may
construct the function
\begin{eqnarray}
{\cal M}_{\alpha\beta}
 &=&
 {\eta\over\l_\a-\m_\b}
 \prod_{\g=1}^{\a-1}
  {\l_\g-\m_\b-\eta\over\l_\g-\m_\b}
\prod_{\epsilon=1}^{\b-1}
 {\m_\epsilon-\m_\beta-\eta\over\m_\epsilon-\m_\beta}
 % \nonumber\\&&\quad \quad\times
  \left[\prod_{j=p+1}^n
  {\m_\b-\l_j^{(1)}\over\m_\b-\l_j^{(1)}+\eta } \right.
 \nonumber\\ && \quad\quad \mbox{}\left.
 -\prod_{k=1}^N{\m_\beta-\x_k\over\m_\beta-\x_k+\eta}
  \prod_{j=p+1}^n{\m_\beta-\x_{i_j}+\eta\over\m_\beta-\x_{i_j}}
%  \prod_{l=p+1}^n
  {\l_\a-\x_{i_j}\over\l_\a-\x_{i_j}+\eta}
  \right]\nonumber\\
  &&+\sum_{j=p+1}^n\left[{\eta\over\m_\b-\l_j^{(1)}+\eta }
  {\eta\over \l_\a-\l_{j}^{(1)}+\eta}
  \prod_{\g=1}^{\a-1}{\l_\g-\l_j^{(1)}\over\l_\g-\l_j^{(1)}+\eta}
  \right. \nonumber\\ &&\quad \times\left.
  \prod_{\e=1}^{\b-1}{\m_\e-\l_j^{(1)}\over\m_\e-\l_j^{(1)}+\eta}
  \prod_{k=p+1,\ne j}^n
  {\l_k^{(1)}-\l_j^{(1)}+\eta\over\l_k^{(1)}-\l_j^{(1)}}\right]
  \nonumber\\&& \mbox{} +
  \sum_{\g=1}^{\a-1}
  {\eta\over\l_\alpha-\l_\g}{\eta\over\l_\g-\m_\b}
 \prod_{\iota=1,\ne \g}^{\alpha-1}
  {\l_\iota-\l_\g-\eta\over\l_\iota-\l_\g}
   \prod_{\e=1}^{\b-1}
  {\m_\e-\l_\g-\eta\over\m_\e-\l_\g}
% \nonumber\\&&\quad \quad\times
  \left[\prod_{j=p+1}^n
  {\l_\g-\l_j^{(1)}\over\l_\g-\l_j^{(1)}+\eta } \right.
\nonumber\\ && \quad \mbox{}\left.
 -\prod_{k=1}^N{\l_\g-\x_k\over\l_\g-\x_k+\eta}
  \prod_{j=p+1}^n {\l_\g-\x_{i_j}+\eta\over\l_\g-\x_{i_j}}
  %\prod_{l=p+1}^n
  {\l_\a-\x_{i_j}\over\l_\a-\x_{i_j}+\eta}
  \right]
    \nonumber\\&& \mbox{} +
  \sum_{\e=1}^{\b-1}
  {\eta\over\l_\a-\m_\e}{\eta\over\m_\e-\m_\b}
 \prod_{\g=1}^{\a-1}
  {\l_\g-\m_\e-\eta\over\l_\g-\m_\e}
   \prod_{\varepsilon=1,\ne\e}^{\b-1}
  {\m_\varepsilon-\m_\e-\eta\over\m_\varepsilon-\m_\e}
% \nonumber\\&&\quad \quad\times
  \left[\prod_{j=p+1}^n
  {\m_\e-\l_j^{(1)}\over\m_\e-\l_j^{(1)}+\eta } \right.
\nonumber\\ && \quad \mbox{}\left.
 -\prod_{k=1}^N{\m_\e-\x_k\over\m_\e-\x_k+\eta}
  \prod_{j=p+1}^n {\m_\e-\x_{i_j}+\eta\over\m_\e-\x_{i_j}}
%  \prod_{l=p+1}^n
  {\l_\a-\x_{i_j}\over\l_\a-\x_{i_j}+\eta}
  \right],%\nonumber\\
 \label{eq:M-ab}
\end{eqnarray}
where $\l_j^{(1)}$ ($j=p+1,\ldots,n$) satisfy the NBAE
(\ref{eq:BAE-nested}). By direct computation, one sees that the
residues of ${\cal M}_{\a 1}$ at points $\m_1=\l^{(1)}_j-\eta$,
$\m_1=\l_\g\, (\g=1,\ldots,\a-1)$ and $\m_1=\m_\e\,
(\e=1,\ldots,\b-1)$ are zero. Moreover the residues of ${\cal
M}_{\a 1}$ at the points $\m_1=\l_\a$ are also zero because
$\l_\a$ is a solution of the BAE (\ref{eq:BAE}). Then comparing
(\ref{eq:B-1-2}) with (\ref{eq:M-ab}),  one finds that as
functions of $\m_1$, the functions $({\cal B}_p^{(1)})_{\a 1}$ and
${\cal M}_{\a 1}$ have the same residues at the simple poles
$\m_1=\x_j-\eta$ $(j\ne i_{p+1},\ldots,i_n)$, and that when
$\m_1\rightarrow \infty$, they tend to zero. Therefore, according
to the properties of the rational functions, we have $({\cal
B}^{(1)}_{p})_{\alpha 1}= {\cal M}_{\a1}$.

Then, by using the function $G^{(0)},G^{(1)}$ and the intermediate
function (\ref{de:Gm-Gm-1}) repeatedly, we obtain $G^{(m)}$
$(m\leq p)$:
\begin{eqnarray}
&&G^{(m)}(\{\l_k\}_{(p,n)},\m_1,\ldots,\m_m,i_{m+1},\ldots,i_n) \nonumber\\
 &&=(-1)^m 2^{{p(p-1)-m(2p-m-1)+(n-p)(n-p-1)\over 2}}
% \prod_{1\leq j<k\leq m}a^{-1}(\m_k,\m_j)
% \nonumber\\&&\quad\quad\times
 \mbox{det}{\cal B}^{(m)}_p(\l_1,\ldots,\l_p;\m_1,\ldots,\m_m,
  i_{m+1},\ldots,i_{p})
 \nonumber\\ && \quad\times
 \prod_{l=1}^{m}\prod_{k=p+1}^na(\m_l,\x_{i_k})
 \mbox{det}{\cal B}_{n-p}(\l_{p+1},\ldots,\l_n;
      \x_{i_{p+1}},\ldots,\x_{i_{n}})
 \label{eq:G-m}
\end{eqnarray}
with the matrix elements %${\cal B}^{(m)}_{\alpha\beta}$
\begin{eqnarray}
({\cal B}^{(m)}_{p})_{\alpha\beta}&=&\prod_{\e=1}^{\b-1}
 a(\m_\e,\x_{i_\beta})({\cal B}_{p})_{\a\b},
 \quad\quad (1\leq \a\leq p,\ m<\beta\leq p),\nonumber\\
({\cal B}^{(m)}_{p})_{\alpha\beta}
 &=&{\cal M}_{\alpha\beta},
 \quad\quad\quad\quad\quad\quad\quad\quad\
 (1\leq \a\leq p,\ 1\leq\beta\leq m).
\end{eqnarray}
(\ref{eq:G-m}) can be proved by induction. Firstly from
(\ref{eq:G-1}), (\ref{eq:B-1-0}) and (\ref{eq:M-ab}),
(\ref{eq:G-m}) is true for $m=1$. Assume (\ref{eq:G-m}) for
$G^{(m-1)}$. Let us show (\ref{eq:G-m}) for general $m$.
Substituting $G^{(m-1)}$ and (\ref{eq:B-expect}) into intermediate
function (\ref{de:Gm-Gm-1}), we have
\begin{eqnarray}
&&G^{(m)}(\{\l_k\}_{(p,n)},\m_1,\ldots,\m_m,i_{m+1},\ldots,i_n)\nonumber\\
 &&=-(-1)^{q}2^{-(p-m)}
 \sum_{j\ne i_{m+1},\ldots,i_n}^N
 b(\m_m,\x_j)\prod_{l=m+1}^n a(\m_m,\x_{i_l})
 \prod_{k\ne j,p+1,\ldots,n}^Na^{-1}(\x_k,\x_j) \nonumber\\
 &&\quad\times
 G^{(m-1)}(\{\l_k\}_{(p,n)},\m_1,\ldots,\m_{m-1},i_{m+1},
 \ldots, i_{m+p},j,i_{m+p+1},\ldots,i_n)\nonumber\\
 &&=-2^{-(p-m)}\
 %\nonumber\\ &&\quad\quad \times%\nonumber\\
% &&\quad\quad\times
 \sum_{j\ne i_{m+1},\ldots,i_n}^N
 b(\m_m,\x_j)\prod_{l=m+1}^n a(\m_m,\x_{i_l})
 \prod_{k\ne j,p+1,\ldots,n}^Na^{-1}(\x_k,\x_j)\nonumber\\
 &&\quad\times
 G^{(m-1)}(\{\l_k\}_{(p,n)},\m_1,\ldots,\m_{m-1},j,i_{m+1},
 \ldots, i_n)\nonumber\\
 &&=(-1)^{m}2^{{p(p-1)-m(2p-m-1)+(n-p)(n-p+1)\over 2}}\,
% \nonumber\\ &&\quad\quad \times
% \prod_{1\leq j<k\leq m-1}a^{-1}(\m_k,\m_j)\nonumber\\
% &&\quad\times
 \mbox{det}{\cal B}^{(m)}_{p}(\l_1,\ldots,\l_p;
      \m_1\ldots,\m_m,i_{m+1},\ldots,i_{p})
 \nonumber\\ && \quad\times
 \prod_{l=1}^{m}\prod_{k=p+1}^na(\m_l,\x_{i_k})
 \mbox{det}{\cal B}_{n-p}(\l_{p+1},\ldots,\l_n;
      \x_{i_{p+1}},\ldots,\x_{i_{n}}),
% \nonumber\\ &&
% \equiv \prod_{1\leq j<k\leq m-1}a^{-1}(\m_k,\m_j)\,
% \mbox{det}{\cal B'}^{(m)}_{\alpha\beta},
 \label{eq:Gm-induction}
\end{eqnarray}
where the matrix elements $({\cal B}^{(m)}_{p})_{\alpha\beta}$ are
given by
\begin{eqnarray}
({\cal B}^{(m)}_{p})_{\alpha\beta}&=&
 \prod_{\e=1}^{\b-1} a(\m_\e,\x_{i_\beta})
 \prod_{k=p+1}^n a(\l_\a,\x_{i_k})
 ({\cal B}_{p})_{\alpha\beta}
 \quad\quad (1\leq \a\leq p,\, m<\beta\leq p),\nonumber\\
({\cal B}^{(m)}_{p})_{\alpha\beta}
 &=&{\cal M}_{\alpha\beta}
 \quad\quad\quad\quad\quad\quad\quad\quad\quad
 \quad\quad\quad\quad\quad\ \
 (1\leq \a\leq p,\, 1\leq \beta<m),
 \nonumber\\
({\cal B}^{(m)}_{p})_{\alpha m}
 &=&% \prod_{k=p+1}^na(\m_m,\x_{i_k})
 \prod_{k=p+1}^n a(\l_\a,\x_{i_k})
% \prod_{i=1}^{m-1}a(\m_i,\x_{i_m})
% \nonumber\\ &&\quad\times
 \sum_{j\ne i_{m+1},\ldots,i_n}
 b(\m_m,\x_j)b(\l_{\alpha},\x_j)
 \prod_{\g=1}^{\a-1}a(\l_\g,\x_j)
 \nonumber\\ &&\quad\times
 \prod_{\e=1}^{\b-1}a(\m_\e,\x_{j})
% \nonumber\\&&\times
% \prod_{k=p+1}^na(\m_m,\x_{i_k})
 \prod_{k\ne j,i_{p+1},\ldots,i_n}^N a^{-1}(\x_{k},\x_j)
% \nonumber\\&&
 \quad\quad\quad\quad (1\leq \a\leq p).
 %&\equiv& \prod_{k=p+1}^na(\m_m,\x_{i_k})\cdot
 %f(\m_m).
 \label{eq:Bm1}
\end{eqnarray}
By the procedure leading to $({\cal B}_{p}^{(1)})_{\alpha\beta}$,
we prove $({\cal B}^{(m)}_{p})_{\alpha m}={\cal M}_{\alpha m}$.
Therefore we have proved that the function (\ref{eq:G-m}) holds
for all $m\leq p$.

When $m=p$, we have,
\begin{eqnarray}
&&G^{(p)}(\{\l_k\}_{(p,n)},\m_1,\ldots,\m_{p},i_{p+1},\ldots,i_n) \nonumber\\
 &&=(-1)^{p} 2^{{(n-p)(n-p-1)\over 2}}
% \prod_{1\leq j<k\leq p}a^{-1}(\m_k,\m_j)%\nonumber\\
% &&\quad\quad\times
 \mbox{det}{\cal M}(\l_1,\ldots,\l_{p};\m_1,\ldots,\m_{p})
 \nonumber\\ && \quad\times
 \prod_{l=1}^p\prod_{k=p+1}^n a(\m_l,\x_{i_k})\
 \mbox{det}{\cal B}_{n-p}(\l_{p+1},\ldots,\l_n;
      \x_{i_{p+1}},\ldots,\x_{i_{n}}),
 \label{eq:G-p}
\end{eqnarray}
where the matrix elements of ${\cal M}$ are given by
(\ref{eq:M-ab}).

For later use, we rewrite the element of the matrix ${\cal
M}_{\a\b}\,(1\leq\a,\b\leq p)$ in the form
\begin{eqnarray}
 {\cal M}_{\a\b}&=&{\cal F}_{\a\b}
 +\sum_{\e=1}^\b\prod_{j=p+1}^n \left(
  a^{-1}(\m_\e,\x_{i_j})
  a(\l_\a,\x_{i_j})\right)\,{\cal G}_{\a\b}^\e
  \nonumber\\ &&\mbox{}
 +\sum_{\g=1}^{\a-1}\prod_{j=p+1}^n \left(
  a^{-1}(\l_\g,\x_{i_j})
  a(\l_\a,\x_{i_j})\right)\,
 {\cal H}^\g_{\a\b},
\end{eqnarray}
where
\begin{eqnarray}
{\cal F}_{\a\b}&=&{b(\l_\alpha,\m_\b)\over a(\l_\alpha,\m_\b)}
  \prod_{\g=1}^{\a-1}a^{-1}(\m_\b,\l_\g)
  \prod_{\e=1}^{\b-1}a^{-1}(\m_\b,\m_\e)
  \prod_{j=p+1}^{n}a(\m_\b,\l_{j}^{(1)})
  \nonumber\\ &&\mbox{}+
  \sum_{j=p+1}^n\left[
  b(\m_\b,\l_j^{(1)})b(\l_\a,\l_{j}^{(1)})
  \prod_{\g=1}^{\a-1}a(\l_\g,\l_j^{(1)})
  \prod_{\e=1}^{\b-1}a(\m_\e,\l_j^{(1)})
  \prod_{k\ne j}a^{-1}(\l_k^{(1)},\l_j^{(1)})\right]
  \nonumber\\ &&\mbox{}+\sum_{\g=1}^{\a-1}
  {b(\l_\alpha,\l_\g)\over a(\l_\alpha,\l_\g)}
  {b(\l_\g,\m_\b)\over a(\l_\g,\m_\b)}
 \prod_{\iota=1,\ne \g}^{\a-1}a^{-1}(\l_\g,\l_\iota)
 \prod_{\e=1}^{\b-1}a^{-1}(\l_\g,\m_\e)
 \prod_{j=p+1}^{n}a(\l_\g,\l_{j}^{(1)})
 \nonumber\\ &&  \mbox{}
 +\sum_{\e=1}^{\b-1}
  {b(\l_\a,\m_\e)\over a(\l_\alpha,\m_\e)}
  {b(\m_\e,\m_\b)\over a(\m_\e,\m_\b)}
 \prod_{\g=1}^{\a-1}a^{-1}(\m_\e,\l_\g)
 \prod_{\varepsilon=1,\ne \e}^{\b-1}a^{-1}(\m_\e,\m_\varepsilon)
 \prod_{j=p+1}^{n}a(\m_\e,\l_{j}^{(1)}),\nonumber \\ \\
 {\cal G}^\e_{\a\b}&=&\left\{\begin{array}{l} \displaystyle
  -{b(\l_\a,\m_\e)\over a(\l_\alpha,\m_\e)}
 \prod_{\g=1}^{\a-1}a^{-1}(\m_\e,\l_\g)
 \prod_{k=1}^N a(\m_\b,\x_k) \quad \quad (\e=\b) \\ \displaystyle
 -{b(\l_\a,\m_\e)\over a(\l_\alpha,\m_\e)}
  {b(\m_\e,\m_\b)\over a(\m_\e,\m_\b)}
 \prod_{\g=1}^{\a-1}a^{-1}(\m_\e,\l_\g)
 \prod_{\varepsilon=1,\ne \e}^{\b-1}a^{-1}(\m_\e,\m_\varepsilon)
 \prod_{k=1}^N a(\m_\e,\x_k) \quad (1\leq\e\leq \b-1)
 \end{array}\right. \\
% \prod_{l=p+1}^n a(\l_\a,\x_{i_l}),
 %
 %
  {\cal H}^\g_{\a\b}&=&-{b(\l_\alpha,\l_\g)\over a(\l_\alpha,\l_\g)}
  {b(\l_\g,\m_\b)\over a(\l_\g,\m_\b)}
 \prod_{\iota=1,\ne \g}^{\alpha-1}a^{-1}(\l_\g,\l_\iota)
 \prod_{\e=1}^{\b-1}a^{-1}(\l_\g,\m_\e)
 \prod_{k=1}^N a(\l_\g,\x_k).
% \prod_{l=p+1}^n a(\l_\a,\x_{i_l}).
\end{eqnarray}
After a tedious computation, we obtain the determinant of the
matrix $\cal M$
\begin{eqnarray}
 &&\mbox{det}{\cal M}(\{\l_\a\},\{\m_\b\})\nonumber\\
 &&=\mbox{det} {\cal F}(\{\l_\a\},\{\m_\b\})
 \nonumber\\&&\mbox{}+
 {\sum_{k,l_k,\varrho^k_{m_k}}}'\,
 \prod_{e=1}^k\prod_{g=p+1}^n
 a^{-l_e}(\m_{e},\x_{i_g})
 \sum_{\stackrel{j_1,\ldots,j_p=1}{j_1\ne\ldots\ne j_p}}^p
 (-1)^{\tau(j_1j_2\ldots j_p)}
 \nonumber\\ &&\quad\times
  \prod_{f'=1}^p\prod_{f=p+1}^n
  \prod_{t=1}^k\prod_{t'=1}^{l_t}\left[1+
  \delta_{j_{f'}\,\varrho^t_{t'}}
% \prod_{t=\{\varrho^1_{m_j}\},\ldots,\{\varrho^k_{m_k}\} }\delta_{j_{f'}t}
 \left(a(\l_{f'},\x_{i_f})-1\right)\right]
 \nonumber\\ &&\quad\times
 (A_2)_{1\,j_1}(A_2)_{2\,j_2}\ldots (A_2)_{p\,j_p}
 \nonumber\\[2mm] &&\mbox{} +
{\sum_{k,l_k,\rho^k_{m_k}}}'' \,
 \prod_{e=1}^{k}
 \prod_{g=p+1}^n a^{-l_e}(\l_{e},\x_{i_g})^{}
 \prod_{t=1}^k\prod_{t'=1}^{l_t}
% \prod_{e'=\{\rho^k_{m_k}\},\ldots,\{\rho^1_{m_1}\}}
 \prod_{g'=p+1}^n a(\l_{\rho^t_{t'}},\x_{i_{g'}})\,
  \mbox{det}\, A_3(\{\l_\a\},\{\m_\b\})
% \nonumber\\ &&\quad\times
%  \sum_{\stackrel{j_1,\ldots,j_p=1}{j_1\ne\ldots\ne j_p}}^p
% (-1)^{\tau(j_1j_2\ldots j_p)}
% (A_3)_{1\,j_1}(A_3)_{2\,j_2}\ldots (A_3)_{p\,j_p}
 \nonumber\\ &&\mbox{}+
 {\sum_{k,l_k,\varrho^k_{m_k}}}'\,
 {\sum_{k',l'_{k'},\rho^{k'}_{m'_{k'}}}}''
% \nonumber\\&& \quad\times
 \prod_{e=1}^{k}
 \prod_{g=p+1}^n a^{-l'_e}(\l_{e},\x_{i_g})
 \prod_{t=1}^{k'}\prod_{t'=1}^{l'_{t}}
% \prod_{e'=\{\rho^{k'}_{m'_{k'}}\},\ldots,\{\rho^1_{m'_1}\}}
 \prod_{g'=p+1}^n \left(a(\l_{\rho^t_{t'}},\x_{i_{g'}})\right)
 \nonumber\\ &&\quad \times
 \, \sum_{\stackrel{j_1,\ldots,j_p=1}{j_1\ne\ldots\ne j_p}}^p
 (-1)^{\tau(j_1j_2\ldots j_p)}
 \nonumber\\&& \quad\times
  \prod_{t=1}^k\prod_{f=p+1}^n
 \prod_{\stackrel{s=1}{s\ne\{\rho^1_{m_1}\},\ldots,
          \{\rho^{k}_{m_{k}}\}}}^p
 \prod_{f'=1}^p\prod_{t'=1}^{l_t}\left[1+
 \delta_{f'\,s}\,\delta_{j_{f'}\,\varrho^t_{t'}}
 \left(a^{-1}(\m_{t},\x_{i_f})-1\right)\right]
 \nonumber\\&& \quad\times
  \prod_{t=1}^k\prod_{f=p+1}^n
  \prod_{\stackrel{s=1}{s\ne\{\rho^1_{m_1}\},\ldots,
          \{\rho^{k}_{m_{k}}\}}}^p
 \prod_{f'=1}^p \prod_{t'=1}^{l_t}
 \left[1+
% \prod_{t_0=\{\varrho^{k}_{m_{k}}\},\ldots,\{\varrho^1_{m_1}\}}
 \delta_{f'\,s}\,\delta_{j_{f'}\,\varrho^t_{t'}}
 \left(a(\l_{s},\x_{i_f})-1\right)\right]
 \nonumber\\&& \quad\times
 (A_4)_{1\,j_1}(A_4)_{2\,j_2}\ldots (A_4)_{p\,j_p} \label{eq:M-t2}\\
 &\equiv& {\cal T}_1+{\cal T}_2+ {\cal T}_3+{\cal T}_4,\label{eq:M-t1t2}
\end{eqnarray}
where $\tau(x_1,\ldots,x_p)=\tau(\s)$, ($\s\in{\cal S}_p$ and
$(x_1,\ldots,x_p)=\s(1,\ldots,p)$), $\tau(\s)=0$ if $\s$ is even
and $\tau(\s)=1$ if $\s$ is odd,
\begin{eqnarray*}
{\sum_{k,l_k,\varrho^k_{m_k}}}'&=& \sum_{k=1}^{p}\left\{
 \sum_{l_k=1}^{p-k+1}{1\over l_k!}\prod_{m_k=1}^{l_k}
 \sum_{\stackrel{\varrho^k_{m_k}=k}
       {\varrho^k_{m_k}\ne\varrho^k_{m_k-1},\ldots,\varrho^k_{1} }}^{p}
 \right\}
% \nonumber\\ &&\quad\times
 \left\{
 \prod_{r=1}^{k-1}\,\,
 \sum_{l_{k-r}=0}^{p-k+r+1-\sum_{j=0}^{r-1}l_{k-j}}
 {1\over l_{k-r}!}
 \right.\nonumber\\ &&\quad\times \left.
 \prod_{m_{k-r}=1}^{l_{k-r}}
 \sum_{\stackrel{\varrho_{m_{k-r}}^{k-r}=k-r}
       {\stackrel{\varrho_{m_{k-r}}^{k-r}\ne\varrho_{m_{k-r}-1}^{k-r},
       \ldots,\varrho_{1}^{k-r}}
        {\varrho_{m_{k-r}}^{k-r}\ne\{\varrho_{\kappa}^{k-r+1}\},
        \{\varrho_{\kappa}^{k-r+2}\},\ldots,\{\varrho_{\kappa}^{k}\}}}}
       ^{p} \right\},\\
{\sum_{k,l_k,\rho^k_{m_k}}}''&=&
  \sum_{k=1}^{p-1}\left\{
 \sum_{l_k=1}^{p-k}{1\over l_k!}\prod_{m_k=1}^{l_k}
 \sum_{\stackrel{\rho^k_{m_k}=k+1}
       {\rho^k_{m_k}\ne\rho^k_{m_k-1},\ldots,\rho^k_{1} }}^{p}
  \right\}
% \nonumber\\ &&\quad\times
 \left\{
 \prod_{r=1}^{k-1}\,
 \sum_{l_{k-r}=0}^{p-k+r-\sum_{j=0}^{r-1}l_{k-j}}
 {1\over l_{k-r}!}
 \right. \nonumber\\ &&\quad\times \left.
 \prod_{m_{k-r}=1}^{l_{k-r}}
 \sum_{\stackrel{\rho_{m_{k-r}}^{k-r}=k-r+1}
       {\stackrel{\rho_{m_{k-r}}^{k-r}\ne\rho_{m_{k-r}-1}^{k-r},
       \ldots,\rho_{1}^{k-r}}
        {\rho_{m_{k-r}}^{k-r}\ne\{\rho_{\kappa}^{k-r+1}\},
        \{\rho_{\kappa}^{k-r+2}\},\ldots,\{\rho_{\kappa}^{k}\}}}}
       ^{p}
 \right\},
 % \right\}
\end{eqnarray*}
and the elements $(A_i)_{\a\b}$, $i=2,3,4$, are given by
\begin{eqnarray}
&&(A_2)_{\a\b}=\left\{\begin{array}{cl}
 {\cal F}_{\a\b}\quad\quad & \a=1,\dots,p,\,\,
 \b=1,\dots,p,\, \b \ne
 \{\varrho_{m_1}^1\},\ldots, \{\varrho_{m_{n}}^{n}\}\\[2mm]
 {\cal G}^k_{\a\b}\quad\quad &
  \a=1,\dots,p, \b=\{\varrho^k_{m_k}\} (k=1,2,\ldots,p)%\\
% {\cal H}^k_{\a\b}\quad\quad & \a=
% \{\rho_{m_k}^k\}; k=1,\ldots n-1,\,\,
% \b=1,\dots,p; \ne\nu_1,\ldots,\nu_q
 \end{array},
 \right.
\nonumber\\
\end{eqnarray}
\begin{eqnarray}
&&(A_3)_{\a\b}=\left\{\begin{array}{cl}
 {\cal F}_{\a\b}\quad\quad & \a=1,\dots,p,\,
 \a\ne \{\rho_{m_1}^1\},\ldots, \{\rho_{m_{n-1}}^{n-1}\},\,\,
 \b=1,\dots,p\\[2mm]
 \displaystyle
 {1\over\b+1}{\cal H}^k_{\a\b}\quad\quad &
  \a= \{\rho^k_{m_k}\} (k=1,2,\ldots,p-1),\,\,\b=1,\dots,p%\\
% {\cal H}^k_{\a\b}\quad\quad & \a=
% \{\rho_{m_k}^k\}; k=1,\ldots n-1,\,\,
% \b=1,\dots,p; \ne\nu_1,\ldots,\nu_q
 \end{array},
 \right.
\nonumber\\
\end{eqnarray}
and
\begin{eqnarray}
&&(A_4)_{\a\b}=\left\{\begin{array}{cl}
 {\cal F}_{\a\b}\quad\quad & \a=1,\dots,p,\, \a \ne
 \{\rho_{m_1}^1\},\ldots, \{\rho_{m_{n-1}}^{n-1}\},\\
 & \b=1,\dots,p,\,
     \b\ne\{\varrho_{m_1}^1\},\ldots, \{\varrho_{m_{n}}^{n}\}\\[2mm]
 {\cal G}^k_{\a\b}\quad\quad &
  \a=1,\dots,p,\,
  \a\ne\{\rho_{m_1}^1\},\ldots, \{\rho_{m_{n-1}}^{n-1}\},\\ &
  \b=\{\varrho_{m_k}^k\}\,(k=1,2,\ldots,n)\\[2mm]
  \displaystyle
 {1\over\b+1}{\cal H}^k_{\a\b}\quad\quad &
 \a=\{\rho_{m_k}^k\}\, (k=1,\ldots n-1),\,\,%\\ &
 \b=1,\dots,p
 \end{array}
 \right.,%\nonumber\\
\end{eqnarray}
respectively.

Thus by using (\ref{eq:M-t1t2}), the function $G^{(p)}$
(\ref{eq:G-p}) becomes
\begin{eqnarray}
 && G^{(p)}(\{\l_k\}_{(p,n)},\m_1,\ldots,\m_p,i_{p+1},\ldots,i_n)
 \nonumber\\ &&=
 (-1)^{p}2^{{(n-p)(n-p-1)\over 2}}
% \prod_{1\leq j<k\leq p}a^{-1}(\m_k,\m_j)
 \sum_{j=1}^4{\cal T}_j
  \nonumber\\ &&  \quad\times
  \prod_{l=1}^p\prod_{k=p+1}^n a(\m_l,\x_{i_k})
  \mbox{det}{\cal B}_{n-p}(\l_{p+1},\ldots,\l_n;
      \x_{i_{p+1}},\ldots,\x_{i_{n}}) %\label{eq:Gp-1234}
  \nonumber\\
% \\ && \mbox{}+ (-1)^{p}2^{{(n-p)(n-p-1)\over 2}}
% \prod_{1\leq j<k\leq p}a^{-1}(\m_k,\m_j)
% \, {\cal T}_2\,
%  \prod_{l=1}^p\prod_{k=p+1}^n a(\m_l,\x_{i_k})
%  \nonumber\\ && \times
% \mbox{det}{\cal B}_{n-p}(\{\l_{p+1},\ldots,\l_n\},
%      \{\x_{i_{p+1}},\ldots,\x_{i_{n}}\})\label{eq:Gp-2}
 &&\equiv \sum_{j=1}^4
 G^{(p)}_j(\{\l_k\}_{(p,n)},\m_1,\ldots,\m_p,i_{p+1},\ldots,i_n).
 \label{eq:Gp-G1G2}
\end{eqnarray}

%$\bullet$  $m\geq p+1$.
\subsection{$m\geq p+1$}

Then we compute the intermediate functions $G^{(m)}$ for $m\geq
p+1$. Similar to the $m\leq p$ case, inserting a complete set and
noticing (\ref{eq:Gp-G1G2}), we have
\begin{eqnarray}
&&G^{(m)}(\{\l_k\}_{(p,n)},\m_1,\ldots,\m_m,i_{m+1},\ldots,i_n)\nonumber\\
 &&=\sum_{j\ne i_{m+1},\ldots,i_n}^N
     \langle0|\overleftarrow{\prod_{k=m+1}^n}E_{(i_k)}^{31}
    \tilde B_1(\m_m)
     \overrightarrow{\prod_{k=m+1}^{m+q}}E_{(i_k)}^{13}E_{(j)}^{13}
     \overrightarrow{\prod_{m+q+1}^{n}}E_{(i_k)}^{13}|0\rangle
     \nonumber\\
 &&\quad\times
 G^{(m-1)}(\{\l_k\}_{(p,n)},\m_1,\ldots,\m_{m-1},i_{m+1},
 \ldots,i_{m+q},j,i_{m+q+1}\ldots,i_n)
 \nonumber\\
 &&=\sum_{j=1}^4G^{(m)}_j(\{\l_k\}_{(p,n)},\m_1,\ldots,\m_m,i_{m+1},\ldots,i_n),
 \label{de:Gm-Gm-1234}
\end{eqnarray}
where $G^{(m)}_j$'s correspond to $G^{(p)}_j$'s in
(\ref{eq:Gp-G1G2}), respectively.

We first compute $G^{(m)}_1$. With the help of the expression of
$\tilde B_1$ (\ref{eq:B1-tilde}), we have
\begin{eqnarray}
&&\langle0|\overleftarrow{\prod_{k=m+1}^n}E_{(i_k)}^{31}
    \tilde B_1(\m_m)
     \overrightarrow{\prod_{k=m+1}^{m+q}}E_{(i_k)}^{13}E_{(j)}^{13}
     \overrightarrow{\prod_{m+q+1}^{n}}E_{(i_k)}^{13}|0\rangle
\nonumber\\
 &&=-(-1)^{q}2^{-(n-m)}\cdot\,
 b(\m_m,\x_j)\prod_{l=m+1}^n a(\m_m,\x_{i_l})
 \prod_{k\ne j}^Na^{-1}(\x_k,\x_j)
 . \label{eq:B1-expect}
\end{eqnarray}
When $m=p+1$,  by using the expressions (\ref{eq:Gp-G1G2}) and
(\ref{eq:B1-expect}), the intermediate function $G^{(p+1)}_1$ is
given by

\begin{eqnarray}
&&G^{(p+1)}_1(\{\l_k\}_{(p,n)},\m_1,\ldots,\m_{p+1},i_{p+2},\ldots,i_n)
 \nonumber\\ &&=
 \sum_{j\ne i_{p+2},\ldots,i_n}^N
     \langle0|\overleftarrow{\prod_{k=p+2}^n}E_{(i_k)}^{31}
    \tilde B_1(\m_{p+1})
     \overrightarrow{\prod_{k=p+2}^{p+q+1}}E_{(i_k)}^{13}E_{(j)}^{13}
     \overrightarrow{\prod_{p+q+2}^{n}}E_{(i_k)}^{13}|0\rangle
     \nonumber\\
 &&\quad\times
 G^{(p)}_1(\{\l_k\}_{(p,n)},\m_1,\ldots,\m_{p},i_{p+2},
 \ldots,i_{p+q+1},j,i_{p+q+2}\ldots,i_n)
 \nonumber\\
 &&=(-1)^{m} 2^{{(n-p)(n-p-1)-2(n-p-1)\over 2}}
  \mbox{det}{\cal F}(\l_1,\ldots,\l_p;\m_1,\ldots,\m_p)
 \nonumber\\ &&\quad \times
 \mbox{det}{\cal B}^{(p+1)}_{n-p}(\l_{p+1},\ldots,\l_n;
      \m_{p+1};\x_{i_{p+2}},\ldots,\x_{i_{n}}),
 \label{eq:G-m-n-p}
\end{eqnarray}
where the matrix elements $({\cal B}^{(m)}_{n-p})_{\alpha\beta}$
$(p+1\leq\a,\b\leq n)$
\begin{eqnarray}
({\cal B}^{(p+1)}_{n-p})_{\alpha\beta}&=&
 \prod_{\e=1}^{\b-1}
 a(\m_{\e},\x_{i_\beta})({\cal B}_{n-p})_{\a\b},
 \quad\quad \mbox{for } p+1<\beta\leq n,\nonumber\\
 ({\cal B}^{(p+1)}_{n-p})_{\a\,p+1}&=&
  \sum^N_{j\ne i_{p+2},\ldots,i_n}
 b(\m_{p+1},\x_j)b(\l_{\a},\x_j)
 \prod_{\g=p+1}^{\a-1}a(\l_\g,\x_j)
 \prod_{\e=1}^{p}a(\m_\e,\x_j)
% \nonumber\\ &&\times
% \prod_{k=p+1}^n a(\m_1,\x_{i_k})
 \prod_{k\ne j}^N a^{-1}(\x_{k},\x_j). \nonumber\\
% \quad\quad (1\leq\a\leq p). ,
% \quad\quad\quad\quad\quad\quad\quad\quad\ \
% \mbox{for } p+1\leq\beta\leq p+1.
 \label{eq:B-p+1}
\end{eqnarray}
As a element of the matrix  ${\cal B}^{(p+1)}_{n-p}$, one finds if
we take $j=i_{p+2},\ldots,i_n$ in the sum of (\ref{eq:B-p+1}), the
added terms will not contribute to the determinant. Therefore we
may rewrite $({\cal B}^{(p+1)}_{n-p})_{\a\,p+1}$ as
\begin{eqnarray}
 ({\cal B}^{(p+1)}_{n-p})_{\alpha\, p+1}&=&
 \sum^N_{j=1}
 {\eta\over \m_{p+1}-\x_{j}+\eta}
 {\eta\over \l_\a-\x_{j}+\eta}
 \prod_{\g=p+1}^{\a-1}{\l_\g-\x_j\over\l_\g-\x_j+\eta}
  \nonumber\\ &&\times
  \prod_{\e=1}^{p}{\m_\e-\x_j\over\m_\e-\x_j+\eta}
% \prod_{k=p+1}^n {\m_1-\x_{i_k}\over\m_1-\x_{i_k}+\eta }
 \prod_{k\ne j}^N
  {\x_{k}-\x_j+\eta\over\x_{k}-\x_j}.
  \nonumber\\ \label{eq:B-p+1-2}
\end{eqnarray}
Then by using the properties of rational function again, we
construct the function $({\cal N}_1)_{\a\b}$ ($p+1\leq\a,\b\leq
n$)
\begin{eqnarray}
 ({\cal N}_1)_{\a\b}&=&
 {\eta\over\l_\a-\m_\b}
 \prod_{\g=p+1}^{\a-1}
  {\l_\g-\m_\b-\eta\over\l_\g-\m_\b}
 \prod_{\e=1}^{\b-1}
 {\m_\epsilon-\m_\beta-\eta\over\m_\epsilon-\m_\beta}
  \nonumber\\&&\quad \times
  \left[\prod_{j=p+1}^n
  {\m_\b-\l_j^{(1)}\over\m_\b-\l_j^{(1)}+\eta }% \right.
% \nonumber\\ && \quad\quad \mbox{}\left.
 -\prod_{k=1}^N{\m_\beta-\x_k\over\m_\beta-\x_k+\eta}
  \right]\nonumber\\
  &&+\sum_{j=p+1}^n\left[{\eta\over\m_\b-\l_j^{(1)}+\eta }
  {\eta\over \l_\a-\l_{j}^{(1)}+\eta}
  \prod_{\g=p+1}^{\a-1}{\l_\g-\l_j^{(1)}\over\l_\g-\l_j^{(1)}+\eta}
  \right. \nonumber\\ &&\quad \times\left.
  \prod_{\e=1}^{\b-1}{\m_\e-\l_j^{(1)}\over\m_\e-\l_j^{(1)}+\eta}
  \prod_{k\ne j}{\l_k^{(1)}-\l_j^{(1)}+\eta\over\l_k^{(1)}-\l_j^{(1)}}\right]
    \nonumber\\&& \mbox{} +
  \sum_{\e=1}^{\b-1}
  {\eta\over\l_\a-\m_\e}{\eta\over\m_\e-\m_\b}
 \prod_{\g=p+1}^{\a-1}
  {\l_\g-\m_\e-\eta\over\l_\g-\m_\e}
   \prod_{\varepsilon=1,\ne\e}^{\b-1}
  {\m_\varepsilon-\m_\e-\eta\over\m_\varepsilon-\m_\e}
 \nonumber\\&&\quad \quad\times
  \left[\prod_{j=p+1}^n
  {\m_\e-\l_j^{(1)}\over\m_\e-\l_j^{(1)}+\eta }% \right.
%\nonumber\\ && \quad \mbox{}\left.
 -\prod_{k=1}^N{\m_\e-\x_k\over\m_\e-\x_k+\eta}\right].
 %\nonumber\\
 \label{eq:M-ab-n}
\end{eqnarray}
Here as before, one may prove $({\cal
B}^{(p+1)}_{n-p})_{\a\,p+1}=({\cal N}_1)_{\a\, p+1}$. Moreover,
with a similar procedure, one may prove that for any $p+1\leq
m\leq n$, the function $G^{(m)}_1$ can be written as
\begin{eqnarray}
&&G^{(m)}_1(\{\l_k\}_{(p,n)},\m_1,\ldots,\m_{m},i_{m+1},\ldots,i_n) \nonumber\\
 &&=(-1)^{m} 2^{{(n-p)(n-p-1)-(m-p)(2n-m-p-1)\over 2}}
 \mbox{det}{\cal F}(\l_1,\ldots,\l_{p};\m_1,\ldots,\m_{p})
 \nonumber\\ && \quad\times
% \prod_{l=1}^{p}\prod_{k=p+1}^na(\l_l,\x_{i_k})
 \mbox{det}{\cal B}_{n-p}(\l_{p+1},\ldots,\l_n;
      \x_{i_{p+1}},\ldots,\x_{i_{n}}),
 \label{eq:G-m-n-p}
\end{eqnarray}
where the matrix elements $({\cal B}^{(m)}_{n-p})_{\alpha\beta}$
$(p+1\leq\a,\b\leq n)$
\begin{eqnarray}
({\cal B}^{(m)}_{n-p})_{\alpha\beta}&=&\prod_{\e=1}^{\b-1}
 a(\m_\e,\x_{i_\beta})({\cal B}_{p})_{\a\b},
 \quad\quad \mbox{for } m<\beta\leq n,\nonumber\\
({\cal B}^{(m)}_{n-p})_{\alpha\beta}&=&({\cal N}_1)_{\alpha\beta},
 \quad\quad\quad\quad\quad\quad\,\,\quad \mbox{for } p+1\leq\beta\leq m.
\end{eqnarray}
Therefore when $m=n$, we obtain
\begin{eqnarray}
 G_1^{(n)}(\{\l_j\}_{(p,n)},\{\m_k\}_{(p,n)})% \nonumber\\
 &=&(-1)^{n}\,\,
 \mbox{det}{\cal F}(\l_1,\ldots,\l_{p};\m_1,\ldots,\m_{p})
 \nonumber\\ && \times
% \prod_{l=1}^{p}\prod_{k=p+1}^na(\l_l,\x_{i_k})
 \mbox{det}\,{\cal N}_1(\l_{p+1},\ldots,\l_n;
      \m_{{p+1}},\ldots,\m_{{n}}).
 \label{eq:results-G1}
\end{eqnarray}

Similarly, the function $G_2^{(n)}$ is given by
\begin{eqnarray}
G_2^{(n)}(\{\l_j\}_{(p,n)},\{\m_k\}_{(p,n)})&=&(-1)^n
{\sum_{k,l_k,\varrho^k_{m_k}}}'
 \sum_{\stackrel{j_1,\ldots,j_p=1}{j_1\ne\ldots\ne j_p}}^p
 (-1)^{\tau(j_1j_2\ldots j_p)}
% \nonumber\\ &&\quad\times
 (A_2)_{1\,j_1}(A_2)_{2\,j_2}\ldots (A_2)_{p\,j_p}
 \nonumber\\ && \times
% \prod_{l=1}^{p}\prod_{k=p+1}^na(\l_l,\x_{i_k})
 \mbox{det}\,{\cal N}_2(\l_{p+1},\ldots,\l_n;
      \m_{{p+1}},\ldots,\m_{{n}})
 \label{eq:results-G2}
\end{eqnarray}
with
\begin{eqnarray}
({\cal N}_2)_{\a\b}&=&
 {\eta\over \l_\a-\m_\b}
 \prod_{e=1}^k\left({\m_e-\m_\b\over\m_e-\m_\b-\eta }
    \right)^{l_e-1}
 \prod_{\e=k+1}^{\b-1}{\m_\e-\m_\b-\eta\over\m_\e-\m_\b }
 \prod_{\g=p+1}^{\a-1}{\l_\g-\m_\b-\eta\over\l_\g-\m_\b }
 \nonumber\\ &&\quad\times
  \prod_{f'=1}^p
 \prod_{t=1}^{k}\prod_{t'=1}^{l_t}
 \left[1+\delta_{j_{f'}\,\varrho^t_{t'}}
% \prod_{t=\{\varrho^1_{m_j}\},\ldots,\{\varrho^k_{m_k}\} }\delta_{j_{f'}t}
 \left({\l_{f'}-\m_\b-\eta\over\l_{f'}-\m_\b}-1\right)\right]
 \nonumber\\ &&\quad\times
  \left[\prod_{j=p+1}^n
  {\m_\b-\l_j^{(1)}\over\m_\b-\l_j^{(1)}+\eta }
 -\prod_{l=1}^N{\m_\beta-\x_l\over\m_\beta-\x_l+\eta}\right]
 \nonumber\\ &&\mbox{}+
 \sum_{\theta=p+1}^{n}
 {\eta\over \m_\b-\l_\theta^{(1)}+\eta}
 {\eta\over \l_\a-\l_\theta^{(1)}+\eta}
 \prod_{e=1}^k\left({\m_e-\l_\theta^{(1)}+\eta\over\m_e
    -\l_\theta^{(1)}}\right)^{l_e-1}
 \nonumber\\ &&\quad\times
 \prod_{\e=k+1}^{\b-1}{\m_\e-\l_\theta^{(1)}\over\m_\e
    -\l_\theta^{(1)}+\eta }
 \prod_{\g=p+1}^{\a-1}{\l_\g-\l_\theta^{(1)}\over\l_\g
    -\l_\theta^{(1)}+\eta }
   \prod_{\vartheta=p+1,\ne\theta}^n
  {\l_\vartheta-\l_\theta^{(1)}+\eta\over\l_\vartheta-\l_\theta^{(1)} }
 \nonumber\\ &&\quad\times
  \prod_{f'=1}^p
 \prod_{t=1}^{k}\prod_{t'=1}^{l_t}
 \left[1+\delta_{j_{f'}\,\varrho^t_{t'}}
% \prod_{t=\{\varrho^1_{m_j}\},\ldots,\{\varrho^l_{m_l}\} }\delta_{j_{f'}t}
 \left({\l_{f'}-\l_\theta^{(1)}\over\l_{f'}
    -\l_\theta^{(1)}+\eta}-1\right)\right]
 \nonumber\\ &&\mbox{}+
 \sum_{\e=k+1}^{\b-1}
 {\eta\over \m_\e-\m_\b}  {\eta\over \l_\a-\m_\e}
 \prod_{e=1}^k\left({\m_e-\m_\e\over\m_e-\m_\e-\eta }
    \right)^{l_e-1}
 \prod_{\varepsilon=k+1,\ne \e}^{\b-1}
  {\m_\varepsilon-\m_\e-\eta\over\m_\varepsilon-\m_\e }
 \nonumber\\ &&\quad\times
 \prod_{\g=p+1}^{\a-1}{\l_\g-\m_\e-\eta\over\l_\g-\m_\e }
  \left[\prod_{j=p+1}^n
  {\m_\e-\l_j^{(1)}\over\m_\e-\l_j^{(1)}+\eta }
 -\prod_{l=1}^N{\m_\e-\x_l\over\m_\e-\x_l+\eta}\right]
 \nonumber\\ &&\quad\times
  \prod_{f'=1}^p
 \prod_{t=1}^{k}\prod_{t'=1}^{l_t}
 \left[1+\delta_{j_{f'}\,\varrho^t_{t'}}
% \prod_{t=\{\varrho^1_{m_j}\},\ldots,\{\varrho^k_{m_k}\} }\delta_{j_{f'}t}
 \left({\l_{f'}-\m_\e-\eta\over\l_{f'}-\m_\e}-1\right)\right]
 \nonumber\\ &&\mbox{}+
\sum_{e=1}^{k}g_2(\m_\b,l_e),
\end{eqnarray}
where the function $g_2(\m_\b,l_e)=0 $ when $l_e=1$; when $l_e$=0,
\begin{eqnarray}
g_2(\m_\b,l_e)&=& {\eta\over \m_e-\m_\b}  {\eta\over \l_\a-\m_e}
 \prod_{e'=1,\ne e}^k\left({\m_{e'}-\m_e\over\m_{e'}-\m_e-\eta }
    \right)^{l_{e'}-1}
 \prod_{\e=k+1}^{\b-1}
  {\m_\e-\m_e-\eta\over\m_\e-\m_e }
 \nonumber\\ &&\quad\times
 \prod_{\g=p+1}^{\a-1}{\l_\g-\m_e-\eta\over\l_\g-\m_e }
  \left[\prod_{j=p+1}^n
  {\m_e-\l_j^{(1)}\over\m_e-\l_j^{(1)}+\eta }
 -\prod_{l=1}^N{\m_e-\x_l\over\m_e-\x_l+\eta}\right]
 \nonumber\\ &&\quad\times
  \prod_{f'=1}^p
 \prod_{t=1}^{k}\prod_{t'=1}^{l_t}
 \left[1+\delta_{j_{f'}\,\varrho^t_{t'}}
% \prod_{t=\{\varrho^1_{m_j}\},\ldots,\{\varrho^k_{m_k}\} }\delta_{j_{f'}t}
 \left({\l_{f'}-\m_e-\eta\over\l_{f'}-\m_e}-1\right)\right]
\end{eqnarray}
and when $l_e\geq 2$,
\begin{eqnarray}
g_2(\m_\b,l_e)&=& -\sum_{k=0}^{l_e-2}{1\over k!}
 {1\over (\m_\b-\m_e+\eta)^{l_e-k-1}}
 {d^{k}\over d\chi^{k}}\left\{
 (\chi-\m_e)^{l_e-1}
 {\eta\over \l_\a-\chi}\right.
 \nonumber\\ &&\quad\times
 \prod_{e'=1,\ne e}^k\left({\m_{e'}-\chi\over\m_{e'}-\chi-\eta }
    \right)^{l_{e'}-1}
 \prod_{\e=k+1}^{\b-1}{\m_\e-\chi-\eta\over\m_\e-\chi }
 \prod_{\g=p+1}^{\a-1}{\l_\g-\chi-\eta\over\l_\g-\chi }
 \nonumber\\ &&\quad\times
  \prod_{f'=1}^p
 \prod_{t=1}^{k}\prod_{t'=1}^{l_t}
 \left[1+\delta_{j_{f'}\,\varrho^t_{t'}}
% \prod_{t=\{\varrho^1_{m_j}\},\ldots,\{\varrho^k_{m_k}\} }\delta_{j_{f'}t}
 \left({\l_{f'}-\chi-\eta\over\l_{f'}-\chi}-1\right)\right]
 \nonumber\\ &&\quad\times \left.
  \left[\prod_{j=p+1}^n
  {\chi-\l_j^{(1)}\over\chi-\l_j^{(1)}+\eta }
 -\prod_{l=1}^N{\chi-\x_l\over\chi-\x_l+\eta}\right]
  \right\}_{\chi=\m_e-\eta}.
\end{eqnarray}

The function $G_3^{(n)}$ is given by
\begin{eqnarray}
G_3^{(n)}(\{\l_j\}_{(p,n)},\{\m_k\}_{(p,n)})&=&
(-1)^n{\sum_{k,l_k,\rho^k_{m_k}}}''
 \mbox{det}\, A_3(\l_{\s(1)},\ldots,\l_{\s(p)};
         \m_{\s'(1)},\ldots,\m_{\s'(p)})
 \nonumber\\ && \times
% \prod_{l=1}^{p}\prod_{k=p+1}^na(\l_l,\x_{i_k})
 \mbox{det}\,{\cal N}_3(\l_{p+1},\ldots,\l_n;
      \m_{{p+1}},\ldots,\m_{{n}})
 \label{eq:results-G3}
\end{eqnarray}
with
\begin{eqnarray}
({\cal N}_3)_{\a\b}&=&
 {\eta\over \l_\a-\m_\b}
 \prod_{e=1}^k\left({\l_e-\m_\b\over\l_e-\m_\b-\eta }
    \right)^{l_e}
 \prod_{t=1}^k\prod_{t'=1}^{l_t}
 {\l_{\rho^t_{t'}}-\m_\b-\eta\over\l_{\rho^t_{t'}}-\m_\b }
 \prod_{\g=p+1}^{\a-1}{\l_\g-\m_\b-\eta\over\l_\g-\m_\b }
 \nonumber\\ &&\quad\times
 \prod_{\e=1}^{\b-1}{\m_\e-\m_\b-\eta\over\m_\e-\m_\b }
% \nonumber\\ &&\quad\times
  \left[\prod_{j=p+1}^n
  {\m_\b-\l_j^{(1)}\over\m_\b-\l_j^{(1)}+\eta }
 -\prod_{l=1}^N{\m_\beta-\x_l\over\m_\beta-\x_l+\eta}\right]
 \nonumber\\ &&\mbox{}+
 \sum_{\theta=p+1}^{n}
 {\eta\over \m_\b-\l_\theta^{(1)}+\eta}
 {\eta\over \l_\a-\l_\theta^{(1)}+\eta}
 \prod_{\e=1}^{\b-1}{\m_\e-\l_\theta^{(1)}\over\m_\e
    -\l_\theta^{(1)}+\eta }
 \prod_{\g=p+1}^{\a-1}{\l_\g-\l_\theta^{(1)}\over\l_\g
    -\l_\theta^{(1)}+\eta }
 \nonumber\\ &&\quad\times
 \prod_{e=1}^k\left({\m_e-\l_\theta^{(1)}+\eta\over\m_e
    -\l_\theta^{(1)}}\right)^{l_e}
  \prod_{t=1}^k\prod_{t'=1}^{l_t}
 {\l_{\rho^t_{t'}}-\l_\theta^{(1)}\over
  \l_{\rho^t_{t'}}-\l_\theta^{(1)}+\eta }
   \prod_{\vartheta=p+1,\ne\theta}^n
  {\l_\vartheta-\l_\theta^{(1)}+\eta\over\l_\vartheta-\l_\theta^{(1)} }
 \nonumber\\ &&\mbox{}+
 \sum_{\e=k+1}^{\b-1}
 {\eta\over \m_\e-\m_\b}  {\eta\over \l_\a-\m_\e}
 \prod_{e=1}^k\left({\l_e-\m_\e\over\l_e-\m_\e-\eta }
    \right)^{l_e}
 \prod_{t=1}^k\prod_{t'=1}^{l_t}
 {\l_{\rho^t_{t'}}-\m_\e-\eta\over\l_{\rho^t_{t'}}-\m_\e }
 \nonumber\\ &&\quad\times
  \prod_{\varepsilon=1,\ne \e}^{\b-1}
  {\m_\varepsilon-\m_\e-\eta\over\m_\varepsilon-\m_\e }
 \prod_{\g=p+1}^{\a-1}{\l_\g-\m_\e-\eta\over\l_\g-\m_\e }
  \left[\prod_{j=p+1}^n
  {\m_\e-\l_j^{(1)}\over\m_\e-\l_j^{(1)}+\eta }
 -\prod_{l=1}^N{\m_\e-\x_l\over\m_\e-\x_l+\eta}\right]
 \nonumber\\ &&\mbox{}+
\sum_{e=1}^{k}g_3(\m_\b,l_e),
\end{eqnarray}
where the function $g_3(\m_\b,l_e)$ is given as follows. i.) when
$\prod_{t=1}^k\prod_{t'=1}^{l_t}\delta_{e\, \rho^t_{t'}}=0$,
\begin{eqnarray}
 g_3(\m_\b,l_e)&=&
 -\sum_{k=0}^{l_e-1}{1\over k!}
 {1\over (\m_\b-\l_e+\eta)^{l_e-k}}
 {d^{k}\over d\chi^{k}}\left\{(\chi-\l_e)^{l_e}
 {\eta\over \l_\a-\chi}
 \prod_{e'=1,\ne e}^k\left({\l_e-\chi\over\l_e-\chi-\eta }
    \right)^{l_e}
 \right. \nonumber\\ && \quad \times
 \prod_{t=1}^k\prod_{t'=1}^{l_t}
 {\l_{\rho^t_{t'}}-\chi-\eta\over\l_{\rho^t_{t'}}-\chi }
 \prod_{\g=p+1}^{\a-1}{\l_\g-\chi-\eta\over\l_\g-\chi }
 \nonumber\\ &&\quad\times \left.
 \prod_{\e=1}^{\b-1}{\m_\e-\chi-\eta\over\m_\e-\chi }
% \nonumber\\ &&\quad\times
  \left[\prod_{j=p+1}^n
  {\chi-\l_j^{(1)}\over\chi-\l_j^{(1)}+\eta }
 -\prod_{l=1}^N{\chi-\x_l\over\chi-\x_l+\eta}\right]
 \right\}_{\chi=\l_e-\eta},
\end{eqnarray}
ii.) when $\prod_{t=1}^k\prod_{t'=1}^{l_t}\delta_{e\,
\rho^t_{t'}}=1$ and $l_e=1$, $g_3(\m_\b,l_e)=0 $ and  iii.) when
there is an index $\hat t$ ($\hat t\in \{1,\ldots k\}$) and $\hat
t'$ ($\hat t'\in \{1,\ldots l_{\hat t}\}$) such that $\rho^{\hat
t}_{\hat t'}=e$, and $l_e\geq 2$,
\begin{eqnarray}
 g_3(\m_\b,l_e)&=&
 -\sum_{k=0}^{l_e-2}{1\over k!}
 {1\over (\m_\b-\l_e+\eta)^{l_e-k-1}}
 {d^{k}\over d\chi^{k}}\left\{(\chi-\l_e)^{l_e-1}
 {\eta\over \l_\a-\chi}
 \prod_{e'=1,\ne e}^k\left({\l_e-\chi\over\l_e-\chi-\eta }
    \right)^{l_e}
 \right. \nonumber\\ && \quad \times
 \prod_{t=1,\ne \hat t}^k\, \prod_{t'=1,\ne \hat t'}^{l_t}
 {\l_{\rho^t_{t'}}-\chi-\eta\over\l_{\rho^t_{t'}}-\chi }
 \prod_{\g=p+1}^{\a-1}{\l_\g-\chi-\eta\over\l_\g-\chi }
 \nonumber\\ &&\quad\times \left.
 \prod_{\e=1}^{\b-1}{\m_\e-\chi-\eta\over\m_\e-\chi }
% \nonumber\\ &&\quad\times
  \left[\prod_{j=p+1}^n
  {\chi-\l_j^{(1)}\over\chi-\l_j^{(1)}+\eta }
 -\prod_{l=1}^N{\chi-\x_l\over\chi-\x_l+\eta}\right]
 \right\}_{\chi=\l_e-\eta}.
\end{eqnarray}

The function $G^{(n)}_4$ is given by
\begin{eqnarray}
 &&G_4^{(n)}(\{\l_j\}_{(p,n)},\{\m_k\}_{(p,n)})\nonumber\\
 &&=(-1)^n
 {\sum_{k,l_k,\varrho^k_{m_k}}}'
 {\sum_{k',l'_{k'},\rho^{k'}_{m'_{k'}}}}''
 \sum_{\stackrel{j_1,\ldots,j_p=1}{j_1\ne\ldots\ne j_p}}^p
 (-1)^{\tau(j_1j_2\ldots j_p)}
% \nonumber\\ &&\quad\times
 (A_4)_{1\,j_1}(A_4)_{2\,j_2}\ldots (A_4)_{p\,j_p}
 \nonumber\\ && \quad\times
% \prod_{l=1}^{p}\prod_{k=p+1}^na(\l_l,\x_{i_k})
 \mbox{det}\,{\cal N}_4(\l_{p+1},\ldots,\l_n;
      \m_{{p+1}},\ldots,\m_{{n}})
 \label{eq:results-G4}
\end{eqnarray}
with
\begin{eqnarray}
({\cal N}_4)_{\a\b}&=&
 {\eta\over \l_\a-\m_\b}
 \prod_{e=1}^k\left({\l_e-\m_\b\over\l_e-\m_\b-\eta }
    \right)^{l'_e}
 \prod_{t=1}^{k'}\prod_{t'=1}^{l'_t}
 {\l_{\rho^t_{t'}}-\m_\b-\eta\over\l_{\rho^t_{t'}}-\m_\b }
 \prod_{\g=p+1}^{\a-1}{\l_\g-\m_\b-\eta\over\l_\g-\m_\b }
 \nonumber\\ &&\quad\times
 \prod_{\e=1}^{\b-1}{\m_\e-\m_\b-\eta\over\m_\e-\m_\b }
% \nonumber\\ &&\quad\times
  \left[\prod_{j=p+1}^n
  {\m_\b-\l_j^{(1)}\over\m_\b-\l_j^{(1)}+\eta }
 -\prod_{l=1}^N{\m_\beta-\x_l\over\m_\beta-\x_l+\eta}\right]
 \nonumber\\ &&\quad\times
  \prod_{t=1}^k
 \prod_{\stackrel{s=1}{s\ne\{\rho^1_{m_1}\},\ldots,
          \{\rho^{k}_{m_{k}}\}}}^p
 \prod_{f'=1}^p\prod_{t'=1}^{l_t}
 \left[1+
 \delta_{f'\,s}\,\delta_{j_{f'}\,\varrho^t_{t'}}
 \left({\m_{t}-\m_\b\over\m_{t}-\m_\b-\eta }
 -1\right)\right]
 \nonumber\\&& \quad\times
  \prod_{t=1}^k
 \prod_{\stackrel{s=1}{s\ne\{\rho^1_{m_1}\},\ldots,
          \{\rho^{k}_{m_{k}}\}}}^p
 \prod_{f'=1}^p \prod_{t'=1}^{l_t}
 \left[1+
% \prod_{t_0=\{\varrho^{k}_{m_{k}}\},\ldots,\{\varrho^1_{m_1}\}}
 \delta_{f'\,s}\,\delta_{j_{f'}\,\varrho^t_{t'}}
 \left({\l_{s}-\m_\b-\eta\over\l_{s}-\m_\b }
 -1\right)\right]
 \nonumber\\ &&\mbox{}+
 \sum_{\theta=p+1}^{n}
 {\eta\over \m_\b-\l_\theta^{(1)}+\eta}
 {\eta\over \l_\a-\l_\theta^{(1)}+\eta}
 \prod_{\e=1}^{\b-1}{\m_\e-\l_\theta^{(1)}\over\m_\e
    -\l_\theta^{(1)}+\eta }
 \prod_{\g=p+1}^{\a-1}{\l_\g-\l_\theta^{(1)}\over\l_\g
    -\l_\theta^{(1)}+\eta }
 \nonumber\\ &&\quad\times
 \prod_{e=1}^k\left({\m_e-\l_\theta^{(1)}+\eta\over\m_e
    -\l_\theta^{(1)}}\right)^{l'_e}
  \prod_{t=1}^{k'}\prod_{t'=1}^{l'_t}
 {\l_{\rho^t_{t'}}-\l_\theta^{(1)}\over
  \l_{\rho^t_{t'}}-\l_\theta^{(1)}+\eta }
   \prod_{\vartheta=p+1,\ne\theta}^n
  {\l_\vartheta-\l_\theta^{(1)}+\eta\over\l_\vartheta-\l_\theta^{(1)} }
   \nonumber\\ &&\quad\times
  \prod_{t=1}^k
 \prod_{\stackrel{s=1}{s\ne\{\rho^1_{m_1}\},\ldots,
          \{\rho^{k}_{m_{k}}\}}}^p
 \prod_{f'=1}^p\prod_{t'=1}^{l_t}\left[1+
 \delta_{f'\,s}\,\delta_{j_{f'}\,\varrho^t_{t'}}
 \left({\m_{t}-\l_\theta^{(1)}+\eta\over\m_{t}-\l_\theta^{(1)}}
 -1\right)\right]
 \nonumber\\&& \quad\times
  \prod_{t=1}^k
 \prod_{\stackrel{s=1}{s\ne\{\rho^1_{m_1}\},\ldots,
          \{\rho^{k}_{m_{k}}\}}}^p
 \prod_{f'=1}^p \prod_{t'=1}^{l_t}\left[1+
% \prod_{t_0=\{\varrho^{k}_{m_{k}}\},\ldots,\{\varrho^1_{m_1}\}}
 \delta_{f'\,s}\,\delta_{j_{f'}\,\varrho^t_{t'}}
 \left({\l_{s}-\l_\theta^{(1)}\over\l_{s}-\l_\theta^{(1)}+\eta }
 -1\right)\right]
 \nonumber\\ &&\mbox{}+
 \sum_{\e=k+1}^{\b-1}
 {\eta\over \m_\e-\m_\b}  {\eta\over \l_\a-\m_\e}
 \prod_{e=1}^k\left({\l_e-\m_\e\over\l_e-\m_\e-\eta }
    \right)^{l'_e}
 \prod_{t=1}^{k'}\prod_{t'=1}^{l'_t}
 {\l_{\rho^t_{t'}}-\m_\e-\eta\over\l_{\rho^t_{t'}}-\m_\e }
 \nonumber\\ &&\quad\times
  \prod_{\varepsilon=1,\ne \e}^{\b-1}
  {\m_\varepsilon-\m_\e-\eta\over\m_\varepsilon-\m_\e }
 \prod_{\g=p+1}^{\a-1}{\l_\g-\m_\e-\eta\over\l_\g-\m_\e }
 \nonumber\\ &&\quad\times
  \left[\prod_{j=p+1}^n
  {\m_\e-\l_j^{(1)}\over\m_\e-\l_j^{(1)}+\eta }
 -\prod_{l=1}^N{\m_\e-\x_l\over\m_\e-\x_l+\eta}\right]
   \nonumber\\ &&\quad\times
  \prod_{t=1}^k\
 \prod_{\stackrel{s=1}{s\ne\{\rho^1_{m_1}\},\ldots,
          \{\rho^{k}_{m_{k}}\}}}^p
 \prod_{f'=1}^p\prod_{t'=1}^{l_t}\left[1+
 \delta_{f'\,s}\,\delta_{j_{f'}\,\varrho^t_{t'}}
 \left({\m_{t}-\m_\e\over\m_{t}-\m_\e-\eta}
 -1\right)\right]
 \nonumber\\&& \quad\times
  \prod_{t=1}^k
 \prod_{\stackrel{s=1}{s\ne\{\rho^1_{m_1}\},\ldots,
          \{\rho^{k}_{m_{k}}\}}}^p
 \prod_{f'=1}^p \prod_{t'=1}^{l_t}\left[1+
% \prod_{t_0=\{\varrho^{k}_{m_{k}}\},\ldots,\{\varrho^1_{m_1}\}}
 \delta_{f'\,s}\,\delta_{j_{f'}\,\varrho^t_{t'}}
 \left({\l_{s}-\m_\e-\eta\over\l_{s}-\m_\e }
 -1\right)\right]
 \nonumber\\ &&\mbox{}+
 \sum_{e=1}^{k'}g_4(\m_\b,l'_e) +
 \sum_{t=1}^{k}g'_4(\m_\b,l_t),
\end{eqnarray}
where $g_4(\m_\b,l'_e)$ is given as follows. i.) when
$\prod_{t=1}^{k'}\prod_{t'=1}^{l'_t}\delta_{e\, \rho^t_{t'}}=0$,
\begin{eqnarray}
 g_4(\m_\b,l'_e)&=&
 -\sum_{k=0}^{l'_e-1}{1\over k!}
 {1\over (\m_\b-\l_e+\eta)^{l'_e-k}}
 {d^{k}\over d\chi^{k}}\left\{(\chi-\l_e)^{l'_e}
 {\eta\over \l_\a-\chi}
 \prod_{e'=1,\ne e}^{k'}\left({\l_e-\chi\over\l_e-\chi-\eta }
    \right)^{l'_e}
 \right. \nonumber\\ && \quad \times
 \prod_{t=1}^{k'}\prod_{t'=1}^{l'_t}
 {\l_{\rho^t_{t'}}-\chi-\eta\over\l_{\rho^t_{t'}}-\chi }
 \prod_{\g=p+1}^{\a-1}{\l_\g-\chi-\eta\over\l_\g-\chi }
 \nonumber\\ &&\quad\times
 \prod_{\e=1}^{\b-1}{\m_\e-\chi-\eta\over\m_\e-\chi }
% \nonumber\\ &&\quad\times
  \left[\prod_{j=p+1}^n
  {\chi-\l_j^{(1)}\over\chi-\l_j^{(1)}+\eta }
 -\prod_{l=1}^N{\chi-\x_l\over\chi-\x_l+\eta}\right]
\nonumber\\ &&\quad\times
  \prod_{t=1}^k
 \prod_{\stackrel{s=1}{s\ne\{\rho^1_{m_1}\},\ldots,
          \{\rho^{k}_{m_{k}}\}}}^p
 \prod_{f'=1}^p\prod_{t'=1}^{l_t}\left[1+
 \delta_{f'\,s}\,\delta_{j_{f'}\,\varrho^t_{t'}}
 \left({\m_{t}-\chi\over\m_{t}-\chi-\eta }
 -1\right)\right]
 \nonumber\\&& \quad\times \left.
  \prod_{t=1}^k
 \prod_{\stackrel{s=1}{s\ne\{\rho^1_{m_1}\},\ldots,
          \{\rho^{k}_{m_{k}}\}}}^p
 \prod_{f'=1}^p \prod_{t'=1}^{l_t}\left[1+
% \prod_{t_0=\{\varrho^{k}_{m_{k}}\},\ldots,\{\varrho^1_{m_1}\}}
 \delta_{f'\,s}\,\delta_{j_{f'}\,\varrho^t_{t'}}
 \left({\l_{s}-\chi-\eta\over\l_{s}-\chi }
 -1\right)\right] \right\}_{\chi=\l_e-\eta}, \nonumber\\
\end{eqnarray}
ii.) when $\prod_{t=1}^k\prod_{t'=1}^{l'_t}\delta_{e\,
\rho^t_{t'}}=1$ and $l'_e=1$, $g_4(\m_\b,l_e)=0 $ and  iii.) when
there are indices $\hat t$ ($\hat t\in \{1,\ldots k\}$) and $\hat
t'$ ($\hat t'\in \{1,\ldots l_{\hat t}\}$) such that $\rho^{\hat
t}_{\hat t'}=e$, and $l_e\geq 2$,
\begin{eqnarray}
 g_4(\m_\b,l_e)&=&
 -\sum_{k=0}^{l_e-2}{1\over k!}
 {1\over (\m_\b-\l_e+\eta)^{l_e-k-1}}
 {d^{k}\over d\chi^{k}}\left\{(\chi-\l_e)^{l_e-1}
 {\eta\over \l_\a-\chi}
 \right. \nonumber\\ && \quad \times
 \prod_{e'=1,\ne e}^{k'}
 \left({\l_e-\chi\over\l_e-\chi-\eta }\right)^{l'_e}
 \prod_{t=1,\ne \hat t}^{k'}\,\prod_{t'=1,\ne \hat t'}^{l'_t}
 {\l_{\rho^t_{t'}}-\chi-\eta\over\l_{\rho^t_{t'}}-\chi }
 \prod_{\g=p+1}^{\a-1}{\l_\g-\chi-\eta\over\l_\g-\chi }
  \nonumber\\ &&\quad\times
 \prod_{\e=1}^{\b-1}{\m_\e-\chi-\eta\over\m_\e-\chi }
% \nonumber\\ &&\quad\times
  \left[\prod_{j=p+1}^n
  {\chi-\l_j^{(1)}\over\chi-\l_j^{(1)}+\eta }
 -\prod_{l=1}^N{\chi-\x_l\over\chi-\x_l+\eta}\right]
\nonumber\\ &&\quad\times
  \prod_{t=1}^k
 \prod_{\stackrel{s=1}{s\ne\{\rho^1_{m_1}\},\ldots,
          \{\rho^{k}_{m_{k}}\}}}^p
 \prod_{f'=1}^p\prod_{t'=1}^{l_t}\left[1+
 \delta_{f'\,s}\,\delta_{j_{f'}\,\varrho^t_{t'}}
 \left({\m_{t}-\chi\over\m_{t}-\chi-\eta }
 -1\right)\right]
 \nonumber\\&& \quad\times \left.
  \prod_{t=1}^k
 \prod_{\stackrel{s=1}{s\ne\{\rho^1_{m_1}\},\ldots,
          \{\rho^{k}_{m_{k}}\}}}^p
 \prod_{f'=1}^p \prod_{t'=1}^{l_t}\left[1+
% \prod_{t_0=\{\varrho^{k}_{m_{k}}\},\ldots,\{\varrho^1_{m_1}\}}
 \delta_{f'\,s}\,\delta_{j_{f'}\,\varrho^t_{t'}}
 \left({\l_{s}-\chi-\eta\over\l_{s}-\chi }
 -1\right)\right] \right\}_{\chi=\l_e-\eta}; \nonumber\\
\end{eqnarray}
for the function $g'_4(\m_\b,l_t)$, one has : i.)
$g'_4(\m_\b,l_t)=0$ when
$$n_t\equiv \sum_{\stackrel{s=1}{s\ne\{\rho^1_{m_1}\},\ldots,
  \{\rho^{k}_{m_{k}}\}}}^p \sum_{f'=1}^p\sum_{t'=1}^{l_t}
 \delta_{f'\,s}\,\delta_{j_{f'}\,\varrho^t_{t'}}=1,$$
ii.) when $n_t=0$,
\begin{eqnarray}
g'_4(\m_\b,l_t)&=&{\eta\over \m_t-\m_\b}  {\eta\over \l_\a-\m_t}
 \prod_{e=1}^k\left({\l_e-\m_t\over\l_e-\m_t-\eta }
    \right)^{l'_e}
 \prod_{t=1}^{k'}\prod_{t'=1}^{l'_t}
 {\l_{\rho^t_{t'}}-\m_t-\eta\over\l_{\rho^t_{t'}}-\m_t }
 \nonumber\\ &&\quad\times
  \prod_{\e=1,\ne t}^{\b-1}
  {\m_\e-\m_t-\eta\over\m_\e-\m_t }
 \prod_{\g=p+1}^{\a-1}{\l_\g-\m_t-\eta\over\l_\g-\m_t }
 \nonumber\\ &&\quad\times
  \left[\prod_{j=p+1}^n
  {\m_t-\l_j^{(1)}\over\m_t-\l_j^{(1)}+\eta }
 -\prod_{l=1}^N{\m_t-\x_l\over\m_t-\x_l+\eta}\right]
   \nonumber\\ &&\quad\times
  \prod_{\tau=1,\ne t}^k
 \prod_{\stackrel{s=1}{s\ne\{\rho^1_{m_1}\},\ldots,
          \{\rho^{k}_{m_{k}}\}}}^p
 \prod_{f'=1}^p\prod_{\tau'=1}^{l_\tau}\left[1+
 \delta_{f'\,s}\,\delta_{j_{f'}\,\varrho^\tau_{\tau'}}
 \left({\m_{\tau}-\m_t\over\m_{\tau}-\m_t-\eta}
 -1\right)\right]
 \nonumber\\&& \quad\times
  \prod_{\tau=1}^k
 \prod_{\stackrel{s=1}{s\ne\{\rho^1_{m_1}\},\ldots,
          \{\rho^{k}_{m_{k}}\}}}^p
 \prod_{f'=1}^p \prod_{\tau'=1}^{l_\tau}\left[1+
% \prod_{t_0=\{\varrho^{k}_{m_{k}}\},\ldots,\{\varrho^1_{m_1}\}}
 \delta_{f'\,s}\,\delta_{j_{f'}\,\varrho^\tau_{\tau'}}
 \left({\l_{s}-\m_t-\eta\over\l_{s}-\m_t }
 -1\right)\right],
\end{eqnarray}
iii.) when $n_t\geq 2$,
\begin{eqnarray}
g'_4(\m_\b,n_t)&=& -\sum_{k=0}^{n_t-2}{1\over k!}
 {1\over (\m_\b-\m_e+\eta)^{n_t-k-1}}
% \nonumber\\ && \quad\times
 {d^{k}\over d\chi^{k}}
 \left\{(\m_t-\chi)^{-1}
% {1\over p\cdot(\m_{t}-\chi)
% (\m_{t}-\chi+\eta)^{p\cdot l_t(p-{l'}_1-\ldots-l_{k'})-n_t}}
 \right.
 \nonumber\\&&\quad \times%\left.
 \prod_{\stackrel{s=1}{s\ne\{\rho^1_{m_1}\},\ldots,
          \{\rho^{k}_{m_{k}}\}}}^p
 \prod_{f'=1}^p\prod_{t'=1}^{l_t}
 \left[(\m_{t}-\chi -\eta)+
% \nonumber\\ &&\quad\quad\mbox{}\left.\left. +
 \eta\cdot\delta_{f'\,s}\,\delta_{j_{f'}\,\varrho^t_{t'}}
 \right]
 \nonumber\\&&\quad \times
  {\eta\over \l_\a-\chi}
 \prod_{e=1}^k\left({\l_e-\m_\b\over\l_e-\chi-\eta }
    \right)^{l'_e}
 \prod_{t=1}^{k'}\prod_{t'=1}^{l'_t}
 {\l_{\rho^t_{t'}}-\chi-\eta\over\l_{\rho^t_{t'}}-\chi }
 \prod_{\g=p+1}^{\a-1}{\l_\g-\chi-\eta\over\l_\g-\chi }
 \nonumber\\ &&\quad\times
 \prod_{\e=1,\ne t}^{\b-1}{\m_\e-\chi-\eta\over\m_\e-\chi }
% \nonumber\\ &&\quad\times
  \left[\prod_{j=p+1}^n
  {\chi-\l_j^{(1)}\over\chi-\l_j^{(1)}+\eta }
 -\prod_{l=1}^N{\chi-\x_l\over\chi-\x_l+\eta}\right]
 \nonumber\\ &&\quad\times
  \prod_{\tau=1,\ne t}^k
 \prod_{\stackrel{s=1}{s\ne\{\rho^1_{m_1}\},\ldots,
          \{\rho^{k}_{m_{k}}\}}}^p
 \prod_{f'=1}^p\prod_{\tau'=1}^{l_\tau}
 \left[1+
 \delta_{f'\,s}\,\delta_{j_{f'}\,\varrho^\tau_{\tau'}}
 \left({\m_{\tau}-\chi\over\m_{\tau}-\chi-\eta }
 -1\right)\right]
 \nonumber\\&& \quad\times \left.
  \prod_{\tau=1}^k
 \prod_{\stackrel{s=1}{s\ne\{\rho^1_{m_1}\},\ldots,
          \{\rho^{k}_{m_{k}}\}}}^p
 \prod_{f'=1}^p \prod_{\tau'=1}^{l_\tau}
 \left[1+
% \prod_{t_0=\{\varrho^{k}_{m_{k}}\},\ldots,\{\varrho^1_{m_1}\}}
 \delta_{f'\,s}\,\delta_{j_{f'}\,\varrho^\tau_{\tau'}}
 \left({\l_{s}-\chi-\eta\over\l_{s}-\chi }
 -1\right)\right]
  \right\}_{\chi=\m_e-\eta}. \nonumber\\
\end{eqnarray}

From (\ref{eq:Sn-Gn}) and (\ref{de:Gm-Gm-1}), we have the
following theorem:
\begin{Theorem}
Let the spectral parameters $\{\l_k\}$ of the Bethe state
$|\O_N(\{\l_k\}_{(p,n)})\rangle $ be solutions of the $BAE$
(\ref{eq:BAE}). The scalar products
$P_n(\{\m_k\}_{(p,n)},\{\l_k\}_{(p,n)})$ defined by (\ref{de:P_n})
are represented by
\begin{eqnarray}
 && P_n(\{\m_k\}_{(p,n)},\{\l_k\}_{(p,n)})
 \nonumber\\
 &&=(-1)^n\sum_{\s,\s'\in{\cal S}_n}
 Y_L(\{\m_{\sigma'(j)}\},\{\m^{(1)}_{\sigma'(k)}\})
 Y_R(\{\l_{\sigma(j)}\},\{\l^{(1)}_{\sigma(k)}\})\,
 \nonumber\\&& \quad \times \left\{
 \mbox{det}\, {\cal F}(\l_{\s(1)},\ldots,\l_{\s(p)};
         \m_{\s'(1)},\ldots,\m_{\s'(p)})
 \right.
 \nonumber\\ && \quad\quad\quad \times
% \prod_{l=1}^{p}\prod_{k=p+1}^na(\l_l,\x_{i_k})
 \mbox{det}\,{\cal N}_1(\l_{\s(p+1)},\ldots,\l_{\s(n)};
      \m_{{\s'(p+1)}},\ldots,\m_{{\s'(n)}})
 \nonumber\\ &&\quad\quad \mbox +
 {\sum_{k,l_k,\varrho^k_{m_k}}}'
 \sum_{\stackrel{j_1,\ldots,j_p=1}{j_1\ne\ldots\ne j_p}}^p
 (-1)^{\tau(\s'(j_1)\ldots \s'(j_p))}
% \nonumber\\ &&\quad\times
 (A_2)_{\s(1)\,\s'(j_1)}\ldots (A_2)_{\s(p)\,\s'(j_p)}
 \nonumber\\ && \quad\quad\quad\times
% \prod_{l=1}^{p}\prod_{k=p+1}^na(\l_l,\x_{i_k})
 \mbox{det}\,{\cal N}_2(\l_{\s(p+1)},\ldots,\l_{\s(n)};
      \m_{{\s'(p+1)}},\ldots,\m_{{\s'(n)}})
 \nonumber\\ &&\quad\quad \mbox +
 {\sum_{k,l_k,\rho^k_{m_k}}}''
 \mbox{det}\, A_3(\l_{\s(1)},\ldots,\l_{\s(p)};
         \m_{\s'(1)},\ldots,\m_{\s'(p)})
% \sum_{\stackrel{j_1,\ldots,j_p=1}{j_1\ne\ldots\ne j_p}}^p
% (-1)^{\tau(\s'(j_1)\ldots \s'(j_p))}
%% \nonumber\\ &&\quad\times
% (A_3)_{\s(1)\,\s'(j_1)}\ldots (A_3)_{\s(p)\,\s'(j_p)}
 \nonumber\\ &&\quad\quad\quad \times
% \prod_{l=1}^{p}\prod_{k=p+1}^na(\l_l,\x_{i_k})
 \mbox{det}\,{\cal N}_3(\l_{\s(p+1)},\ldots,\l_{\s(n)};
      \m_{{\s'(p+1)}},\ldots,\m_{{\s'(n)}})
 \nonumber\\ &&\quad\quad  \mbox +
 {\sum_{k,l_k,\varrho^k_{m_k}}}'
  {\sum_{k',l'_{k'},\rho^{k'}_{m'_{k'}}}}''
 \sum_{\stackrel{j_1,\ldots,j_p=1}{j_1\ne\ldots\ne j_p}}^p
 (-1)^{\tau(\s'(j_1)\ldots \s'(j_p))}
% \nonumber\\ &&\quad\times
 (A_4)_{\s(1)\,\s'(j_1)}\ldots (A_2)_{\s(p)\,\s'(j_p)}
 \nonumber\\ &&\quad \quad\quad\times \left.
% \prod_{l=1}^{p}\prod_{k=p+1}^na(\l_l,\x_{i_k})
 \mbox{det}\,{\cal N}_4(\l_{\s(p+1)},\ldots,\l_{\s(n)};
      \m_{{\s'(p+1)}},\ldots,\m_{{\s'(n)}})\right\}.
      \label{eq:theorem}
\end{eqnarray}
\end{Theorem}

%obtain the final results for the scalar products
%\begin{eqnarray}
%&&P_n(\{\m_j\},\{\l_k\})=(-1)^n\sum_{\s,\s'\in{\cal S}_n}
% X(\{\m_{\sigma'(j)}\},\{\m^{(1)}_{\sigma'(k)}\})
% Y(\{\l_{\sigma(j)}\},\{\l^{(1)}_{\sigma(k)}\})\,
% \nonumber\\&& \times
%\sum_{i=1}^4 G^{(n)}_i(\{\l_{\s(1)}\},\ldots,\{\l_{\s(n)}\};
% \{\m_{\s'(1)}\},\ldots,\{\m_{\s'(n)}\}),
% \label{eq:Pn-final}
%\end{eqnarray}
%where the functions $G^{(n)}_i(\{\l_j\};
% \{\m_{(k)}\})$ are given by (\ref{eq:results-G1}), (\ref{eq:results-G2}),
% (\ref{eq:results-G3}) and  (\ref{eq:results-G4}), respectively.
% \\[3mm]

{\bf Remark:} In the derivation of (\ref{eq:theorem}), the
spectral parameters $\{\l_i\}$ in the state
$|\O_N(\{\l_j\}_{(p,n)})\rangle$ are required to satisfy the BAE
(\ref{eq:BAE}). However, the parameters $\m_j$ $(j=1,\ldots,n)$ in
the dual state $\langle\O_N(\{\m_j\}_{(p,n)})|$ do not need to
satisfy the BAE.

On the other hand, if we compute the scalar product by starting
from the dual state $\langle\O_N(\{\l_j\}_{(p,n)})|$, then by
using the same procedure, we have
\begin{eqnarray}
G^{(n)}(\{\l_k\}_{(p,n)},\{\m_j\}_{(p,n)})
 =G^{(n)}(\{\m_j\}_{(p,n)},\{\l_k\}_{(p,n)}).
\end{eqnarray}
Therefore, the corresponding scalar product
$P_{n}^L(\{\l_k\}_{(p,n)},\{\m_j\}_{(p,n)})$ is given by
\begin{eqnarray}
 &&P_{n}^L(\{\l_k\}_{(p,n)},\{\m_j\}_{(p,n)})\nonumber\\ &&
 =\sum_{\s,\s'\in{\cal S}_n}Y_L(\{\l_{\s(j)}\},\{\l^{(1)}_{\s(k)}\})
 Y_R(\{\m_{\s'(j)}\,\{\m^{(1)}_{\s'(k)}\})\,\
 G^{(n)}(\{\m_{\s'(j)}\}_{(p,n)},\{\l_{\s(k)}\}_{(p,n)}). %\nonumber\\
 \label{eq:Pn-L}
\end{eqnarray}

In (\ref{eq:Pn-L}), we have also assumed that any element of the
spectral parameter set $\{\l_i\}$ satisfy the BAE.

%%%%%%%%%%%%%%%%%%%%%%%%%%%%%%%%%%%%%%%%%%%%%%%%%%%%%%%%%%%%%%%%%%%%%
%                                                                   %
%   5. Correlation functions                                        %
%                                                                   %
%%%%%%%%%%%%%%%%%%%%%%%%%%%%%%%%%%%%%%%%%%%%%%%%%%%%%%%%%%%%%%%%%%%%%

\sect{Correlation  functions}

Having obtained the scalar product and the norm, we are now in the
position to compute the k-point correlation functions of the
model. In general, a k-point correlation function is defined by
\begin{eqnarray}
F^{\epsilon^1,\ldots,\epsilon^k}_n
 =\langle\O_N(\{\m_j\})| \epsilon^1_{i_1}\ldots\epsilon^k_{i_k}
  |\O_N(\{\l_j\})\rangle, \label{de:cf-general}
\end{eqnarray}
where $\epsilon^j_{i_j}$ stand for the local fermion
representations, (\ref{de:repre-fermi}), of the generators of the
superalgebra $gl(2|1)$, and the lower indices $i_j$ indicate the
positions of the fermion operators.

The authors in \cite{Korepin99} proved that the local spin and
field operators of the fundamental graded models can be
represented in terms of monodromy matrix. Specializing to the
current system, we obtain
\begin{eqnarray}
     (1 - n_{\kappa, \up}) c_{\kappa, \down} &=
    & \prod_{j=1}^{\kappa-1} t(\x_j) \cdot B_1 (\x_\kappa)
          \cdot \prod_{j = \kappa+1}^N t(\x_j) \quad, \\
             \label{tj2}
    (1 - n_{\kappa, \down}) c_{\kappa, \up}  &=
    & \prod_{j=1}^{\kappa-1} t(\x_j) \cdot B_2 (\x_\kappa)
          \cdot \prod_{j = \kappa+1}^N t(\x_j) \quad, \\
             \label{tj3}
     (1 - n_{\kappa, \up}) c_{\kappa, \down}^\dag &=
    & \prod_{j=1}^{\kappa-1} t(\x_j) \cdot C_1 (\x_\kappa)
          \cdot \prod_{j = \kappa+1}^N t(\x_j) \quad, \\
             \label{tj4}
     (1 - n_{\kappa,\down})c_{\kappa, \up}^\dag  &=
    & \prod_{j=1}^{n-1}t(\x_j) \cdot C_2 (\x_\kappa)
          \cdot \prod_{j = \kappa+1}^N t(\x_j) \quad,\\
             \label{tj5}
                          \label{tj8}
     (1 - n_{\kappa, \down}) (1 - n_{\kappa, \up}) &=
    & \prod_{j=1}^{\kappa-1} t(\x_j) \cdot D (\x_\kappa)
          \cdot \prod_{j = \kappa+1}^N t(\x_j) \quad,\\
     S_\kappa^\dag &=
    &-  \prod_{j=1}^{\kappa-1} t(\x_j) \cdot A_{21} (\x_\kappa)
          \cdot \prod_{j = \kappa+1}^N t(\x_j) \quad, \\
             \label{tj6}
     S_\kappa &=
    &-  \prod_{j=1}^{\kappa-1} t(\x_j) \cdot A_{12} (\x_\kappa)
          \cdot \prod_{j = \kappa+1}^N t(\x_j) \quad, \\
             \label{tj7}
     S_\kappa^z &=&
    - \frac{1}{2}  \prod_{j=1}^{\kappa-1} t(\x_j) \cdot
          \bigl( A_{11} (\x_\kappa) - A_{22} (\x_\kappa) \bigr)
          \cdot \prod_{j = \kappa+1}^N t(\x_j) \quad,
          \label{tj-Sz}
\end{eqnarray}
where in the left of (\ref{tj8}), $n_{\kappa\down}n_{\kappa\up}=0$
since the double occupancy of lattice sites on the restriced
Hilbert space of the supersymmetric $t$-$J$ model is excluded.

\subsection{One-point functions}

We first calculate one point correlation functions for the local
fermion operators $(1-n_{\kappa,\down})c_{\kappa,\up}$ and
$(1-n_{\kappa,\down})c_{\kappa,\up}^{\dag}$. According to
(\ref{de:cf-general}), the correlation functions are given by
\begin{eqnarray}
&&F_n^{\up}(\{\m_j\}_{(p,n)},\x_\k,\{\l_k\}_{(p+1,n+1)})\nonumber\\
&=&\langle\O_N(\{\m_j\}_{(p,n)})|\cdot
   (1-n_{\kappa,\down})c_{\kappa,\up}\cdot
   |\O_N(\{\l_j\}_{(p+1,n+1)})\rangle
   \nonumber\\
&=& \sum_{\s\in {\cal S}_{n+1}}\sum_{\s'\in{\cal S}_{n}}
    Y_L(\{\m_{\s'(j)}\},\{\m_{\s'(k)}^{(1)}\})
    Y_R(\{\l_{\s(j)}\},\{\l_{\s(k)}^{(1)}\})\nonumber\\ && \times
    \langle0|\overleftarrow{\prod_{i=p+1}^{n}} B_1(\m_{\s'(i)})
    \overleftarrow{\prod_{i=1}^{p}}B_2(\m_{\s'(i)})\cdot
    (1-n_{\kappa,\down})c_{\kappa,\up}^{\dag}\cdot%\nonumber\\ &&\times
    \overrightarrow{\prod_{i=1}^{p+1}}C_2(\l_{\s(i)})
    \overrightarrow{\prod_{i=p+2}^{n+1}}
    C_1(\l_{\s(i)})|0\rangle,
    \nonumber\\&& \\\label{de:cf-up-dag}
&&F_n^{\up\dag}(\{\l_j\}_{(p+1,n+1)},\x_\k,\{\m_k\}_{(p,n)})\nonumber\\
&=&\langle\O_N(\{\l_j\}_{(p+1,n+1)})|\cdot
   (1-n_{\kappa,\down})c_{\kappa,\up}^{\dag}\cdot
   |\O_N(\{\m_j\}_{(p,n)})\rangle\nonumber\\
&=& \sum_{\s'\in {\cal S}_n}\sum_{\s\in{\cal S}_{n+1}}
    Y_L(\{\l_{\s(j)}\},\{\l_{\s(k)}^{(1)}\})
    Y_R(\{\m_{\s'(j)}\},\{\m_{\s'(k)}^{(1)}\})\nonumber\\ && \times
    \langle0|\overleftarrow{\prod_{i=p+2}^{n+1}} B_1(\l_{\s(i)})
    \overleftarrow{\prod_{i=1}^{p+1}} B_2(\l_{\s(i)})\cdot
    (1-n_{\kappa,\down})c_{\kappa,\up}^{\dag}\cdot%\nonumber\\ &&\times
    \overrightarrow{\prod_{i=1}^{p}} C_2(\m_{\s'(i)})
    \overrightarrow{\prod_{i=p+1}^{n}}
    C_1(\m_{\s'(i)})|0\rangle,
    \nonumber \label{de:cf-up-dag}\\
    \end{eqnarray}
where $\{\m_j\},\{\l_k\} $ are solutions of BAE, $p$ and $p+1$ are
quantum numbers of the corresponding states. For the
representations of the correlation functions, we prove the
following proposition:
\begin{Proposition}
If both the Bethe state $|\O_N(\{u_j\})\rangle$ and the dual Bethe
state $\langle\O_N(\{u_j\})|$ ($u_j=\l_j,\m_j$) are eigenstates of
the transfer matrix, then the correlation functions corresponding
to the local fermion operators
$(1-n_{\kappa,\down})c_{\kappa,\up}$ and
$(1-n_{\kappa,\down})c_{\kappa,\up}^{\dag}$ can be represented by
\begin{eqnarray}
 &&F^{\up}_n(\{\m_j\}_{(p,n)},\x_\kappa,\{\l_k\}_{(p+1,n+1)})\nonumber\\
 &&=(-1)^{n+1}
 \sum_{\s\in{\cal S}_{n+1}}\sum_{\s'\in{\cal S}_n}
 \phi_{\kappa-1}(\{\m_{j}\})
 \phi^{-1}_\kappa(\{\l_{k}\})
 \nonumber\\ &&\quad\times
  P_{n+1}\left(
    \x_\k,\m_{\s'(1)},\ldots, \m_{\s'(n)}; \{\l_{\s(j)}\}_{(p+1,n+1)}
  \right), \label{eq:F-up}\\[2mm]
 &&F^{\up\dag}_n(\{\l_j\}_{(p+1,n+1)},\kappa,\{\m_k\}_{(p,n)})\nonumber\\
 &&=(-1)^{n+1}
 \sum_{\s\in{\cal S}_{n+1}}\sum_{\s'\in{\cal S}_n}
 \phi_{\kappa-1}(\{\l_{j}\})
 \phi^{-1}_\kappa(\{\m_{k}\})
 \nonumber\\ &&\quad\times
  P_{n+1}^L\left(%\{\l_{\s(k)}\},\{\l^{(1)}_{\s(k)}\}
     \{\l_{\s(j)}\}_{(p+1,n+1)};
    \x_\k,\m_{\s'(1)},\ldots, \m_{\s'(n)}\right), \label{eq:F-up-dag}
\end{eqnarray}
respectively, where $\phi_i(\{\m_j\})=\prod_{k=1}^i\prod_{l=1}^n
a^{-1}(\m_l,\x_k)$.

\end{Proposition}

\noindent {\it Proof}. We first prove (\ref{eq:F-up}). From the
definition of $F^{\up}_n$, we have
\begin{eqnarray}
 &&F_{n+1}^{\up}(\{\m_j\}_{(p,n)},\x_\kappa,\{\l_k\}_{(p+1,n+1)})\nonumber\\
 &=&\langle\O_N(\{\m_j\}_{(p,n)})|\cdot
   (1-n_{\kappa,\down})c_{\kappa,\up}\cdot
   |\O_N(\{\l_j\}_{(p+1,n+1)})\rangle
   \nonumber\\
 &=&\langle\O_N(\{\m_j\}_{(p,n)})|
   \prod_{j=1}^{\kappa-1}t(\x_j) \cdot B_2 (\x_\kappa)
          \cdot \prod_{j = \kappa+1}^N t(\x_j)
   |\O_N(\{\l_j\}_{(p+1,n+1)})\rangle
 \nonumber\\
 &=&(-1)^{n+1}\sum_{\s\in{\cal S}_{n+1}}\sum_{\s'\in{\cal S}_n}
%
% \phi_{\kappa-1}(\{\m_{\s'(j)}\})\phi^{-1}_\kappa(\{\l_{\s(k)}\})
 \prod_{j=1}^n\prod_{k=1}^{\k-1}a^{-1}(\m_{j},\x_k)
 \prod_{j=1}^{n+1}\prod_{k=\k+1}^{N}a^{-1}(\l_{j},\x_k)
 \nonumber\\ &&\quad\times
  P_{n+1}\left(%\{\l_{\s(k)}\},\{\l^{(1)}_{\s(k)}\}
    \x_\k,\m_{\s'(1)},\ldots, \m_{\s'(n)};
     \{\l_{\s(j)}\}_{(p+1,n+1)}\right).
\end{eqnarray}
Then by using the relation
\begin{eqnarray}
\prod_{j=1}^{n}\prod_{k=1}^Na^{-1}(\l_j,\x_k)=1,
\end{eqnarray}
which is from the BAE and the NBAE, we prove (\ref{eq:F-up}). The
proof of (\ref{eq:F-up-dag}) is similar.
\begin{flushright} $\square$ \end{flushright}

For the local operators $(1-n_{\k,\up})c_{\k,\down}$ and
$(1-n_{\k\up})c_{\k\down}^\dag$, the calculation of their
correlation functions, which are defined by
\begin{eqnarray}
 F^{\down}_{n+1}(\{\m_j\}_{(p,n)},\x_\k,\{\l_k\}_{(p,n+1)})
  &=&\langle\O_N(\{\m_j\}_{(p,n)})|\cdot
 (1-n_{\k,\up})c_{\k,\down}\cdot
 |\O_N(\{\l_k\}_{(p,n+1)})\rangle, \nonumber\\ \\
 F^{\down\dag}_{n+1}(\{\l_j\}_{(p,n+1)},\x_\k,\{\m_k\}_{(p,n)})
  &=&\langle\t\O_N(\{\l_k\}_{(p,n+1)})|\cdot
 (1-n_{\k,\up})c_{\k,\down}^\dag\cdot
 |\O_N(\{\m_j\}_{(p,n)})\rangle, \nonumber\\
\end{eqnarray}
respectively, leads to the following proposition:
\begin{Proposition}
If both the Bethe state $|\O_N(\{u_j\})\rangle$ and the dual Bethe
state $\langle\O_N(\{u_j\})|$ ($u_j=\l_j,\m_j$) are eigenstates of
the transfer matrix, then the correlation functions corresponding
to the local fermion operators
$(1-n_{\kappa,\up})c_{\kappa,\down}$ and
$(1-n_{\kappa,\up})c_{\kappa,\down}^{\dag}$ can be represented by
\begin{eqnarray}
 &&F^{\down}_{n+1}(\{\m_j\}_{(p,n)},\x_\kappa,\{\l_k\}_{(p,n+1)})\nonumber\\
 &&=(-1)^{n+1}
 \sum_{\s\in{\cal S}_{n+1}}\sum_{\s'\in{\cal S}_n}
 \phi_{\kappa-1}(\{\m_{j}\})
 \phi^{-1}_\kappa(\{\l_{k}\})
 \nonumber\\ &&\quad\times \left\{
 (-1)^p\prod_{j=1}^p{1\over a(\m_{\s'(j)},\x_\k)}
  P_{n+1}\left(
    \m_{\s'(1)},\ldots,\m_{\s'(p)},\x_\k,
    \m_{\s'(p+1)},\ldots, \m_{\s'(n)};
    \{\l_{\s(j)}\}_{(p,n+1)}\right)\right.
\nonumber\\ &&\quad \mbox{}
  +(-1)^{p-1}\sum_{j=1}^p
  {b(\m_{\s'(j)},\x_\k)\over a(\m_{\s'(j)},\x_\k)}
   \prod_{k=1}^{j-1}
   {c(\m_{\s'(k)},\m_{\s'(j)})\over c(\m_{\s'(k)},\x_\k)}
   \prod_{l=1,\ne j}^{p}{1\over a(\m_{\s'(l)},\m_{\s'(j)})}
 \nonumber\\ &&\quad\times\left.
  P_{n+1}\left(
    \m_{\s'(1)},\ldots,\m_{\s'(j-1)},\x_\k,\m_{\s'(j+1)},
    \ldots,\m_{\s'(p)},\right.\right.
 \nonumber\\ &&\quad\quad~~~~~~~\left.\left.
    \m_{\s'(j)},\m_{\s'(p+1)},\ldots, \m_{\s'(n)} ;
    \{\l_{\s(j)}\}_{(p,n+1)}\right)
    {\Huge \mbox{}}\right\},\nonumber\\
   \label{eq:F-down}\\%[2mm]
 &&F^{\down\dag}_{n+1}(\{\l_j\}_{(p,n+1)},\x_\kappa,\{\m_k\}_{(p,n)})\nonumber\\
 &&=(-1)^{n+1}
 \sum_{\s\in{\cal S}_{n+1}}\sum_{\s'\in{\cal S}_n}
 \phi_{\kappa-1}(\{\l_{j}\})
 \phi^{-1}_\kappa(\{\m_{k}\})
 \nonumber\\ &&\quad\times \left\{
 (-1)^p\prod_{j=1}^p{1\over a(\m_{\s'(j)},\x_\k)}
  P_{n+1}^L\left(
     \{\l_{\s(j)}\}_{(p+1,n+1)};
    \m_{\s'(1)},\ldots,\m_{\s'(p)},\x_\k,
    \m_{\s'(p+1)},\ldots, \m_{\s'(n)}\right)\right.
\nonumber\\ &&\quad \mbox{}
  +(-1)^{p-1}\sum_{j=1}^p
  {b(\m_{\s'(j)},\x_\k)\over a(\m_{\s'(j)},\x_\k)}
   \prod_{k=1}^{j-1}
   {c(\m_{\s'(k)},\m_{\s'(j)})\over c(\m_{\s'(k)},\x_\k)}
   \prod_{l=1,\ne j}^{p}{1\over a(\m_{\s'(l)},\m_{\s'(j)})}
 \nonumber\\ &&\quad\times\left.
  P_{n+1}^L\left(
     \{\l_{\s(j)}\}_{(p,n+1)};
    \m_{\s'(1)},\ldots,\m_{\s'(j-1)},\x_\k,\m_{\s'(j+1)},
    \ldots,\m_{\s'(p)},\right.\right.
 \nonumber\\ &&\quad\quad~~~~~~~\left.\left.
 \m_{\s'(j)},\m_{\s'(p+1)},\ldots, \m_{\s'(n)}\right)
    {\Huge \mbox{}}\right\},\nonumber\\
   \label{eq:F-down-dag}
\end{eqnarray}
respectively.
\end{Proposition}

\noindent {\it Proof}. We first prove (\ref{eq:F-down}).
Considering the definition of $F^{\down}_n$ the representation of
the local fermion operator (\ref{tj2}), we have
\begin{eqnarray}
 &&
 F^{\down}_{n+1}(\{\m_j\}_{(p,n)},\x_\kappa,\{\l_k\}_{(p,n+1)})\nonumber\\
 &&=\langle\O_N(\{\m_j\}_{(p,n)})|\cdot
 \prod_{j=1}^{\k-1}t(\x_j)\cdot B_1(\x_\k)\cdot
 \prod_{j=\k+1}^{N}t(\x_j)
 |\O_N(\{\l_k\}_{(p,n+1)})\rangle, \nonumber\\
 &&= \sum_{\s\in {\cal S}_{n+1}}\sum_{\s'\in{\cal S}_{n}}
 \phi_{\kappa-1}(\{\l_{\s(j)}\})
 \phi^{-1}_\kappa(\{\m_{\s'(k)}\})
    Y_L(\{\m_{\s'(j)}\},\{\m_{\s'(k)}^{(1)}\})
    Y_R(\{\l_{\s(j)}\},\{\l_{\s(k)}^{(1)}\})\nonumber\\ && \times
    \langle0|\overleftarrow{\prod_{i=p+1}^{n}}B_1(\m_{\s'(i)})
    \overleftarrow{\prod_{i=1}^{p}}B_2(\m_{\s'(i)})\cdot
    B_1(\x_\k)\cdot%\nonumber\\ &&\times
    \overrightarrow{\prod_{i=1}^{p+1}} C_2(\l_{\s(i)})
    \overrightarrow{\prod_{i=p+2}^{n+1}}
    C_1(\l_{\s(i)})|0\rangle.
    \label{eq:F-down-1}
\end{eqnarray}
From the GYBE (\ref{eq:GYBE}), we have the following commutation
relations
\begin{eqnarray}
 && B_a(\m)B_a(\l)=-c(\l,\m)B_a(\l)B_a(\m) \label{eq:commu-BaBa} \\
 && B_a(\m)B_b(\l)=-{1\over a(\m,\l)}B_b(\l)B_a(\m)
 + {b(\m,\l)\over a(\m,\l)}B_b(\m)B_a(\l)\quad (a\ne b).
 \label{eq:commu-BaBb}
\end{eqnarray}
Then we have
\begin{eqnarray}
&&B_2(\m_p)\ldots B_2(\m_1)B_1(\x_\k) \nonumber\\
 &&=(-1)^p\prod_{j=1}^p{1\over a(\m_j,\x_\k)}
 B_1(\x_\k)B_2(\m_p)\ldots B_2(\m_1)
 \nonumber\\ &&\mbox{} \quad
 +(-1)^{p-1}\sum_{j=1}^p {b(\m_j,\x_\k)\over a(\m_j,\x_\k)}
 \prod_{k=1}^{j-1}{c(\m_k,\m_j)\over c(\m_k,\x_\k)}
 \prod_{l=1,\ne j}^p{1\over a(\m_l,\m_j)}
 \nonumber\\ &&\quad\quad\times
 B_1(\m_j)B_2(\m_p)\ldots B_2(\m_{j+1})B_2(\x_\k)B_2(\m_{j-1})
 \ldots B_2(\m_1).
\end{eqnarray}
Substituting the about relation into (\ref{eq:F-down-1}), we
obtain (\ref{eq:F-down}).

Similarly, by using the commutation relations
\begin{eqnarray}
 && C_a(\l)C_a(\m)=-c(\l,\m)C_a(\m)C_a(\l) \label{eq:commu-CaCa} \\
 && C_a(\l)C_b(\m)=-{1\over a(\m,\l)}C_b(\m)C_a(\l)
 + {b(\m,\l)\over a(\m,\l)}C_b(\l)C_a(\m)\quad (a\ne b).
 \label{eq:commu-CaCb}
\end{eqnarray}
one may prove (\ref{eq:F-down-dag}).
% ~~~~~~~~~~~~~~~~~~~~~~~~~~~~~~~~~~~~~~~~~~~~~~~~~~~~~~~~~~~~~
% ~~~~~~~~~~~~~~~~~~~~~~~~~~~~$\Box$
\begin{flushright} $\square$ \end{flushright}

For correlation functions associated with the local fermion
operators $S_\k$ and $S^\dag_\k$ defined by
\begin{eqnarray}
  F^S_n(\{\m_j\}_{(p,n)},\x_\k,\{\l_k\}_{(p-1,n)})
 &=&\langle\O_N(\{\m_j\}_{(p,n)})|\cdot
 S\cdot
 |\O_N(\{\l_k\}_{(p-1,n)})\rangle,  \\
 F^{S^\dag}_{n}(\{\l_j\}_{(p-1,n)},\x_\k,\{\m_k\}_{(p,n)})
  &=&\langle\O_N(\{\l_k\}_{(p-1,n)})|\cdot
 S^\dag\cdot
 |\O_N(\{\m_j\}_{(p,n)})\rangle,
\end{eqnarray}
we have the following proposition:
\begin{Proposition}
If both the Bethe state $|\O_N(\{u_j\})\rangle$ and the dual Bethe
state $\langle\O_N(\{u_j\})|$ ($u_j=\l_j,\m_j$) are eigenstates of
the transfer matrix, then the correlation functions corresponding
to the local fermion operators $S$ and $S^{\dag}$ can be
represented by
\begin{eqnarray}
 &&F^{S}_n(\{\m_j\}_{(p,n)},\x_\kappa,\{\l_k\}_{(p-1,n)})\nonumber\\
 &&=(-1)^{n+1}
 \sum_{\s\in{\cal S}_{n}}\sum_{\s'\in{\cal S}_n}
 \phi_{\kappa-1}(\{\m_{j}\})
 \phi^{-1}_\kappa(\{\l_{k}\})
 \sum_{i=1}^p{b(\x_\k,\m_{\s'(i)})\over a(\x_\k,\m_{\s'(i)})}
 \nonumber\\ &&\quad\times \left\{
 (-1)^{p-i}\prod_{j=i+1}^p
 {c(\m_{\s'(i)},\m_{\s'(j)})\over a(\m_{\s'(i)},\m_{\s'(j)})}
 \prod_{k=i+1}^p{1\over a(\m_{\s'(k)},\x_\k)}
 \prod_{\a=1}^N a(\m_{\s'(i)},\x_\a) \right.
 \nonumber\\ && \quad\quad\times%\right.
  P_{n}\left(\{
    \m_{\s'(1)},\ldots,\m_{\s'(i-1)},\m_{\s'(i+1)},\ldots,
    \m_{\s'(p)},\x_\k,
    \m_{\s'(p+1)},\ldots, \m_{\s'(n)}\}_{(p-1,n)}; \right.
  \nonumber\\ && \quad\quad\quad\,\,\quad \left.
    \{\l_{\s(j)}\}_{(p-1,n)}\right)
\nonumber\\ &&\quad \mbox{}% \left.
  +(-1)^{p-i-1}\prod_{j=i+1}^p
  {c(\m_{\s'(i)},\m_{\s'(j)})\over a(\m_{\s'(i)},\m_{\s'(j)})}
  \sum_{k=i+1}^p{b(\m_{\s'(k)},\x_\k)\over a(\m_{\s'(k)},\x_\k)}
   \prod_{l=i+1}^{k-1}
   {c(\m_{\s'(l)},\m_{\s'(k)})\over c(\m_{\s'(l)},\x_\k)}
  \nonumber\\ && \quad\quad\times
   \prod_{m=i+1,\ne k}^{p}{1\over a(\m_{\s'(m)},\m_{\s'(k)})}
   \prod_{\a=1}^N a(\m_{\s'(i)},\x_\a)\,
  P_{n}\left(\{
    \m_{\s'(1)},\ldots,\m_{\s'(i-1)},\m_{\s'(i+1)},\ldots
    \right.
  \nonumber\\ && \quad\quad\quad\,\,\quad \left.
    \m_{\s'(k-1)},\x_\k,\m_{\s'(k+1)},\ldots,\m_{\s'(p)},\m_{\s'(k)},
    \m_{\s'(p+1)},\ldots, \m_{\s'(n)}\}_{(p-1,n)};
    \{\l_{\s(j)}\}_{(p-1,n)}\right),
  \nonumber\\&& \quad\mbox{}
  +\sum_{j=i+1}^p
  {b(\m_{\s'(i)},\m_{\s'(j)})\over a(\m_{\s'(i)},\m_{\s'(j)})}
  \prod_{k=i+1}^{j-1}
  {c(\m_{\s'(k)},\m_{\s'(j)})\over c(\m_{\s'(k)},\m_{\s'(i)})}
  \prod_{l=i+1,\ne j}^p
  {c(\m_{\s'(j)},\m_{\s'(l)})\over a(\m_{\s'(j)},\m_{\s'(l)})}
  \prod_{\a=1}^N a(\m_{\s'(j)},\x_\a)
  \nonumber\\ &&\quad\quad \times \left[
  (-1)^{p-i}\prod_{m=i+1}^p{1\over a(\m'_{\s'(m)},\x_\k)}
  \,\,  P_{n}\left(\{
    \m_{\s'(1)},\ldots,\m_{\s'(i-1)},\m'_{\s'(i+1)},\ldots,
    \m'_{\s'(p)},\right.\right.
  \nonumber\\ && \quad\quad\quad\,\,\quad \left.
   \x_\k,\m_{\s'(p+1)},\ldots, \m_{\s'(n)}\}_{(p-1,n)};
    \{\l_{\s(j)}\}_{(p-1,n)}\right)
  \nonumber\\ && \quad\quad\quad\mbox{}
  +(-1)^{p-i-1}\sum_{m=i+1}^{p}
  {b(\m'_{\s'(m)},\x_\k)\over a(\m'_{\s'(m)},\x_\k)}
  \prod_{n=i+1}^{m-1}
  {c(\m'_{\s'(n)},\m'_{\s'(m)})\over c(\m'_{\s'(n)},\x_\k)}
  \prod_{s=i+1,\ne m}^p
  {1\over a(\m'_{\s'(s)},\m'_{\s'(m)})}
  \nonumber\\ &&\quad\quad\quad \times
  P_{n}\left(\{
    \m_{\s'(1)},\ldots,\m_{\s'(i-1)},\m'_{\s'(i+1)},\ldots
    \m'_{\s'(m-1)},\x_\k,\m'_{\s'(m+1)},\ldots,\m'_{\s'(p)},\m'_{\s'(m)},
    \right.
  \nonumber\\ && \quad\quad\quad\,\,\quad \left.\left.
    \m_{\s'(p+1)},\ldots, \m_{\s'(n)}\}_{(p-1,n)};
    \{\l_{\s(j)}\}_{(p-1,n)}\right) \right]
  \nonumber\\&& \quad\mbox{}
  -\sum_{j=i+1}^p
  {b(\x_\k,\m_{\s'(j)})\over a(\x_\k,\m_{\s'(j)})}
  \prod_{k=i+1}^{j-1}
  {c(\m_{\s'(k)},\m_{\s'(j)})\over c(\m_{\s'(k)},\x_\k)}
  \prod_{l=i+1,\ne j}^p
  {c(\m_{\s'(j)},\m_{\s'(l)})\over a(\m_{\s'(j)},\m_{\s'(l)})}
  \prod_{\a=1}^N a(\m_{\s'(j)},\x_\a)
  \nonumber\\ &&\quad\quad \times \left[
  (-1)^{p-i}\prod_{m=i+1}^p{1\over a(\m''_{\s'(m)},\m_{\s'(i)}))}
  \,\,  P_{n}\left(\{
    \m_{\s'(1)},\ldots,\m_{\s'(i-1)},\m''_{\s'(i+1)},\ldots,
    \m''_{\s'(p)},\right.\right.
  \nonumber\\ && \quad\quad\quad\,\,\quad \left.
   \m_{\s'(i)},\m_{\s'(p+1)},\ldots, \m_{\s'(n)}\}_{(p-1,n)};
    \{\l_{\s(j)}\}_{(p-1,n)}\right)
  \nonumber\\ && \quad\quad\quad\mbox{}
  +(-1)^{p-i-1}\sum_{m=i+1}^{p}
  {b(\m''_{\s'(m)},\m_{\s'(i)})\over a(\m''_{\s'(m)},\m_{\s'(i)})}
  \prod_{n=i+1}^{m-1}
  {c(\m''_{\s'(n)},\m''_{\s'(m)})\over c(\m''_{\s'(n)},\m_{\s'(i)})}
  \prod_{s=i+1,\ne m}^p
  {1\over a(\m''_{\s'(s)},\m''_{\s'(m)})}
  \nonumber\\ &&\quad\quad\quad \times
  P_{n}\left(\{
    \m_{\s'(1)},\ldots,\m_{\s'(i-1)},\m''_{\s'(i+1)},\ldots
    \m''_{\s'(m-1)},\m_{\s'(i)},\m''_{\s'(m+1)},\ldots,\m''_{\s'(p)},\m''_{\s'(m)},
    \right.
  \nonumber\\ && \quad\quad\quad\,\,\quad \left.\left.\left.
    \m_{\s'(p+1)},\ldots, \m_{\s'(n)}\}_{(p-1,n)};
    \{\l_{\s(j)}\}_{(p-1,n)}\right) \right]\right\},
   \label{eq:F-S}\\%[2mm]
 &&F^{S^\dag}_n(\{\l_k\}_{(p-1,n)},\x_\kappa,\{\m_j\}_{(p,n)})\nonumber\\
 &&=(-1)^{n+1}
 \sum_{\s\in{\cal S}_{n}}\sum_{\s'\in{\cal S}_n}
 \phi_{\kappa-1}(\{\l_{j}\})
 \phi^{-1}_\kappa(\{\m_{k}\})
 \sum_{i=1}^p{b(\x_\k,\m_{\s'(i)})\over a(\x_\k,\m_{\s'(i)})}
 \nonumber\\ &&\quad\times \left\{
 (-1)^{p-i}\prod_{j=i+1}^p
 {c(\m_{\s'(i)},\m_{\s'(j)})\over a(\m_{\s'(i)},\m_{\s'(j)})}
 \prod_{k=i+1}^p{1\over a(\m_{\s'(k)},\x_\k)}
 \prod_{\a=1}^N a(\m_{\s'(i)},\x_\a) \right.
 \nonumber\\ && \quad\quad\times%\right.
  P^L_{n}\left(
    \{\l_{\s(d)}\}_{(p-1,n)},\{\m_{\s'(1)},\ldots,\m_{\s'(i-1)},\m_{\s'(i+1)},\ldots,
    \m_{\s'(p)},\x_\k, \right.
  \nonumber\\ && \quad\quad\quad\,\,\quad \left.
    \m_{\s'(p+1)},\ldots, \m_{\s'(n)}\}_{(p-1,n)}\right)
\nonumber\\ &&\quad \mbox{}% \left.
  +(-1)^{p-i-1}\prod_{j=i+1}^p
  {c(\m_{\s'(i)},\m_{\s'(j)})\over a(\m_{\s'(i)},\m_{\s'(j)})}
  \sum_{k=i+1}^p{b(\m_{\s'(k)},\x_\k)\over a(\m_{\s'(k)},\x_\k)}
   \prod_{l=i+1}^{k-1}
   {c(\m_{\s'(l)},\m_{\s'(k)})\over c(\m_{\s'(l)},\x_\k)}
  \nonumber\\ && \quad\quad\times
   \prod_{m=i+1,\ne k}^{p}{1\over a(\m_{\s'(m)},\m_{\s'(k)})}
   \prod_{\a=1}^N a(\m_{\s'(i)},\x_\a)\,
  P^L_{n}\left(
    \{\l_{\s(d)}\}_{(p-1,n)},
    \{\m_{\s'(1)},\ldots,\m_{\s'(i-1)},
    \right.
  \nonumber\\ && \quad\quad\quad\,\,\quad \left.
    \m_{\s'(i+1)},\ldots,
    \m_{\s'(k-1)},\x_\k,\m_{\s'(k+1)},\ldots,\m_{\s'(p)},\m_{\s'(k)},
    \m_{\s'(p+1)},\ldots, \m_{\s'(n)}\}_{(p-1,n)}
    \right),
  \nonumber\\&& \quad\mbox{}
  +\sum_{j=i+1}^p
  {b(\m_{\s'(i)},\m_{\s'(j)})\over a(\m_{\s'(i)},\m_{\s'(j)})}
  \prod_{k=i+1}^{j-1}
  {c(\m_{\s'(k)},\m_{\s'(j)})\over c(\m_{\s'(k)},\m_{\s'(i)})}
  \prod_{l=i+1,\ne j}^p
  {c(\m_{\s'(j)},\m_{\s'(l)})\over a(\m_{\s'(j)},\m_{\s'(l)})}
  \prod_{\a=1}^N a(\m_{\s'(j)},\x_\a)
  \nonumber\\ &&\quad\quad \times \left[
  (-1)^{p-i}\prod_{m=i+1}^p{1\over a(\m'_{\s'(m)},\x_\k)}
  \,\,  P^L_{n}\left(
    \{\l_{\s(d)}\}_{(p-1,n)};
    \{\m_{\s'(1)},\ldots,\m_{\s'(i-1)},
    \right.\right.
  \nonumber\\ && \quad\quad\quad\,\,\quad \left.
   \m'_{\s'(i+1)},\ldots,\m'_{\s'(p)},\x_\k,\m_{\s'(p+1)},\ldots,
   \m_{\s'(n)}\}_{(p-1,n)}
   \right)
  \nonumber\\ && \quad\quad\quad\mbox{}
  +(-1)^{p-i-1}\sum_{m=i+1}^{p}
  {b(\m'_{\s'(m)},\x_\k)\over a(\m'_{\s'(m)},\x_\k)}
  \prod_{n=i+1}^{m-1}
  {c(\m'_{\s'(n)},\m'_{\s'(m)})\over c(\m'_{\s'(n)},\x_\k)}
  \prod_{s=i+1,\ne m}^p
  {1\over a(\m'_{\s'(s)},\m'_{\s'(m)})}
  \nonumber\\ &&\quad\quad\quad \times
  P^L_{n}\left(
    \{\l_{\s(d)}\}_{(p-1,n)};\{\m_{\s'(1)},\ldots,\m_{\s'(i-1)},
    \m'_{\s'(i+1)},\ldots
    \m'_{\s'(m-1)},\x_\k,\m'_{\s'(m+1)},\ldots,
    \right.
  \nonumber\\ && \quad\quad\quad\,\,\quad \left.\left.
    \m'_{\s'(p)},\m'_{\s'(m)},\m_{\s'(p+1)},\ldots, \m_{\s'(n)}\}_{(p-1,n)}
    \right) \right]
  \nonumber\\&& \quad\mbox{}
  -\sum_{j=i+1}^p
  {b(\x_\k,\m_{\s'(j)})\over a(\x_\k,\m_{\s'(j)})}
  \prod_{k=i+1}^{j-1}
  {c(\m_{\s'(k)},\m_{\s'(j)})\over c(\m_{\s'(k)},\x_\k)}
  \prod_{l=i+1,\ne j}^p
  {c(\m_{\s'(j)},\m_{\s'(l)})\over a(\m_{\s'(j)},\m_{\s'(l)})}
  \prod_{\a=1}^N a(\m_{\s'(j)},\x_\a)
  \nonumber\\ &&\quad\quad \times \left[
  (-1)^{p-i}\prod_{m=i+1}^p{1\over a(\m''_{\s'(m)},\m_{\s'(i)}))}
  \,\,  P^L_{n}\left(
    \{\l_{\s(j)}\}_{(p-1,n)};
    \{\m_{\s'(1)},\ldots,\m_{\s'(i-1)},\right.\right.
  \nonumber\\ && \quad\quad\quad\,\,\quad \left.
   \m''_{\s'(i+1)},\ldots,
    \m''_{\s'(p)},\m_{\s'(i)},\m_{\s'(p+1)},\ldots, \m_{\s'(n)}\}_{(p-1,n)}
    \right)
  \nonumber\\ && \quad\quad\quad\mbox{}
  +(-1)^{p-i-1}\sum_{m=i+1}^{p}
  {b(\m''_{\s'(m)},\m_{\s'(i)})\over a(\m''_{\s'(m)},\m_{\s'(i)})}
  \prod_{n=i+1}^{m-1}
  {c(\m''_{\s'(n)},\m''_{\s'(m)})\over c(\m''_{\s'(n)},\m_{\s'(i)})}
  \prod_{s=i+1,\ne m}^p
  {1\over a(\m''_{\s'(s)},\m''_{\s'(m)})}
  \nonumber\\ &&\quad\quad\quad \times
  P^L_{n}\left(
    \{\l_{\s(d)}\}_{(p-1,n)};
    \{\m_{\s'(1)},\ldots,\m_{\s'(i-1)},\m''_{\s'(i+1)},\ldots
    \m''_{\s'(m-1)},\m_{\s'(i)},
    \right.
  \nonumber\\ && \quad\quad\quad\,\,\quad \left.\left.\left.
    \m''_{\s'(m+1)},\ldots,\m''_{\s'(p)},\m''_{\s'(m)},
    \m_{\s'(p+1)},\ldots, \m_{\s'(n)}\}_{(p-1,n)}
    \right) \right]\right\},
   \label{eq:F-S}%[2mm]
\end{eqnarray}
respectively, where $\m'_{\s'(k)}=\m_{\s'(k)}$, for
$k=i+1,\ldots,j-1,j+1,\ldots,p$, $\m'_{\s'(k)}=\m_{\s'(i)}$, for
$k=j$, $\m''_{\s'(k)}=\m_{\s'(k)}$, for
$k=i+1,\ldots,j-1,j+1,\ldots,p$ and $\m''_{\s'(k)}=\x_\k$, for
$k=j$.
\end{Proposition}

In proving this proposition, we have used the commutation
relations
\begin{eqnarray}
 A_{ab}(\l)C_c(\m)&=&
 {r(\l-\m)^{bc}_{de}\over a(\l-\m)}C_e(\m)A_{ad}(\l)
 +{b(\l-\m)\over a(\l-\m)}C_b(\m)A_{ac}(\l),\label{eq:commu-AC}\\
 B_{c}(\m)A_{ab}(\l)&=&
 {r(\l-\m)_{bc}^{de}\over a(\l-\m)}A_{db}(\l)B_e(\m)
 +{b(\l-\m)\over a(\l-\m)}A_{cb}(\m)B_a(\l).\label{eq:commu-BA}
\end{eqnarray}
We do not write down the detailed proof here, since the procedure
is similar to that of the previous propositions.

For the one-point correlation function associated with the fermion
operators $(1-n_{\k,\down})(1-n_{\k,\up})$ and $S^z_\k$
\begin{eqnarray}
 F^{n_{\k}}_{n}(\{\m_{j}\},\x_\k,\{\l_{k}\})
 &=&\langle\O_N(\{\m_j\}_{(p,n)})|\cdot(1-n_{\k,\down})(1-n_{\k,\up})
 \cdot |\O_N(\{\l_k\}_{(p,n)})\rangle, \\
F^{S^z}_{n}(\{\m_{j}\},\x_\k,\{\l_{k}\})
 &=&\langle\O_N^p(\{\m_j\}_{(p,n)})|\cdot S^z
 \cdot |\O_N(\{\l_k\}_{(p,n)})\rangle,
 \label{de:cf-Sz}
\end{eqnarray}
we have the following proposition:
\begin{Proposition}
If both the Bethe state $|\O_N(\{u_j\})\rangle$ and the dual Bethe
state $\langle\O_N(\{u_j\})|$ ($u_j=\l_j,\m_j$) are eigenstates of
the transfer matrix, then the correlation functions corresponding
to the local fermion operators $(1-n_{\k,\down})(1-n_{\k,\up})$
and $S^z$ can be represented by
\begin{eqnarray}
 &&F^{n_\k}_n(\{\m_j\}_{(p,n)},\x_\kappa,\{\l_k\}_{(p,n)})
\nonumber\\&&
 =(-1)^{n}
 \sum_{\s\in{\cal S}_{n}}\sum_{\s'\in{\cal S}_n}
 \phi_{\kappa-1}(\{\m_{j}\})
 \phi^{-1}_\kappa(\{\l_{k}\})\,
 {\cal P}(1;\x_\k;\{\m_{\s'(j)}\};\{\l_{\s(k)}\})
 \nonumber\\&&
  \label{eq:nk} \\
 &&F^{S^z}_n(\{\m_j\}_{(p,n)},\x_\kappa,\{\l_k\}_{(p,n)})\nonumber\\
 &&=(-1)^{n}
 \sum_{\s\in{\cal S}_{n}}\sum_{\s'\in{\cal S}_n}
 \phi_{\kappa-1}(\{\m_{j}\})
 \phi^{-1}_\kappa(\{\l_{k}\})
 \nonumber\\ && \quad\quad\times
 \sum_{i=1}^p
  {b(\x_\k,\l_{\s(i)})\over a(\x_\k, \l_{\s(i)})}
 \prod_{k=p+1}^{i-1}
  {c(\l_{\s(k)},\l_{\s(i)})\over c( \l_{\s(k)},\x_\k)}
 \prod_{j=1,\ne i}^{n}
 {c(\l_{\s(i)},\l_{\s(j)})\over a(\l_{\s(i)},\l_{\s(j)})}
 \prod_{\a=1}^N a(\l_{\s(i)},\x_\a)
 \nonumber\\ && \quad\quad\quad\times
 P^L_n\left(\{\m_{\s'(d)}\}_{(p,n)};\{\l_{\s(1)},\ldots,
  \l_{\s(i-1)},\x_\k,\l_{\s(i+1)},\ldots,\l_{\s(n)}\}_{(p,n)}\right)
 \nonumber\\ && \quad\mbox{}
 +(-1)^{n}
 \sum_{\s\in{\cal S}_{n}}\sum_{\s'\in{\cal S}_n}
 {1\over 2}
 \phi_{\kappa}(\{\m_{\s'(j)}\})
 \phi^{-1}_\kappa(\{\l_{\s(k)}\})
% \prod_{j=1}^n a^{-1}(\l_{\s(j)},\x_\k)
%
% \nonumber\\ && \quad\mbox{}
-{1\over2}F^{n_\k}_n(\{\m_j\}_{(p,n)},\x_\k,\{\l_k\}_{(p,n)}),
  \nonumber\\ \label{eq:Sz}
\end{eqnarray}
respectively, where
\begin{eqnarray}
 &&{\cal P}(e;\delta;\{\m_{\s'(j)}\};\{\l_{\s(k)}\}) \nonumber\\
 &&= \prod_{i=e}^{n} {1\over a(\l_{\s(i)},\delta)}
 P^L_n\left(\{\m_{\s'(j)}\}_{(p,n)};\{\l_{\s(k)}\}_{(p,n)}\right)
 \nonumber\\ && \quad\quad\mbox{}
 -\sum_{j=p+1}^n
  {b(\l_{\s(j)},\delta)\over a(\l_{\s(j)}, \delta)}
 \prod_{k=p+1}^{j-1}
  {c(\l_{\s(k)},\l_{\s(j)})\over c( \l_{\s(k)},\delta)}
 \prod_{l=e,\ne j}^{n} {1\over a(\l_{\s(l)},\l_{\s(j)})}
 \nonumber\\ && \quad\quad\quad\times
 P^L_n\left(\{\m_{\s'(d)}\}_{(p,n)};\{\l_{\s(1)},\ldots,\l_{\s(e)},\ldots,
  \l_{\s(j-1)},\delta,\l_{\s(j+1)},\ldots,\l_{\s(n)}\}_{(p,n)}\right)
 \nonumber\\ && \quad\quad\mbox{}
 -\sum_{i=e}^p{b(\l_{\s(i)},\delta)\over a(\l_{\s(i)},\delta)}
  \prod_{k=1}^{i-1}
   {c(\l_{\s(k)},\l_{\s(i)})\over c( \l_{\s(k)},\delta)}
 \prod_{j=1,\ne i}^{p} {1\over a(\l_{\s(j)},\l_{\s(i)})}
 \left[
 \prod_{l=p+1}^{n} {1\over a(\l_{\s(l)},\l_{\s(i)})}\right.
 \nonumber\\ && \quad\quad\quad\times
 P^L_n\left(\{\m_{\s'(d)}\}_{(p,n)};\{\l_{\s(1)},\ldots,\l_{\s(e)},\ldots,
  \l_{\s(i-1)},\delta,\l_{\s(i+1)},\ldots,\l_{\s(n)}\}_{(p,n)}\right)
 \nonumber\\ && \quad\quad\quad\mbox{}
 -\sum_{l=p+1}^n
  {b(\l_{\s(l)},\l_{\s(i)})\over a(\l_{\s(l)},\l_{\s(i)})}
 \prod_{m=p+1}^{l-1}
  {c(\l_{\s(m)},\l_{\s(l)})\over c( \l_{\s(m)},\l_{\s(i)})}
 \prod_{q=p+1,\ne l}^{n} {1\over a(\l_{\s(q)},\l_{\s(l)})}
 \nonumber\\ && \quad\quad\quad\times % \left.\left.
 P^L_n\left(\{\m_{\s'(d)}\}_{(p,n)};\{\l_{\s(1)},\ldots,\l_{\s(e)},\ldots,
  \l_{\s(i-1)},\delta,\l_{\s(i+1)},\ldots, \right.
 \nonumber\\ && \quad\quad\quad\quad\quad\quad\left.\left.
  \l_{\s(l-1)},\l_{\s(i)},\l_{\s(l+1),\ldots,\l_{\s(n)}}\}_{(p,n)}
  \right)\right]
  \label{de:P-cal}
\end{eqnarray}
\end{Proposition}

\noindent {\it Proof}. With the help of the commutation relations
(\ref{eq:commu-CaCb}) and
\begin{eqnarray}
D(\l)C_c(\m)={1\over a(\m,\l)} C_c(\m)D(\l)-
 {b(\m,\l)\over a(\m,\l)} C_c(\l)D(\m), \label{eq:commu-DC}
\end{eqnarray}
by using the similar approach as before, one easily proves
(\ref{eq:nk}).

From (\ref{tj-Sz}), we have
\begin{eqnarray}
 S_\kappa^z &=&
    - \frac{1}{2}  \prod_{j=1}^{\kappa-1} t(\x_j) \cdot
          \bigl( A_{11} (\x_\kappa) - A_{22} (\x_\kappa) \bigr)
          \cdot \prod_{j = \kappa+1}^N t(\x_j) \nonumber\\
&=&\frac{1}{2}  \prod_{j=1}^{\kappa-1} t(\x_j) \cdot
          \bigl(t(\x_\k)-D(\x_\k)+2A_{22}(\x_\k)\bigr)
          \cdot \prod_{j = \kappa+1}^N t(\x_j).
          \label{eq:tj-Sz}
\end{eqnarray}
Substituting (\ref{eq:tj-Sz}) into (\ref{de:cf-Sz}), we obtain
\begin{eqnarray}
&&F^{S^z}_n(\{\m_j\}_{(p,n)},\x_\kappa,\{\l_k\}_{(p,n)})\nonumber\\
 &&=\sum_{\s\in{\cal S}_{n}}\sum_{\s'\in{\cal S}_n}
 \phi_{\kappa-1}(\{\m_{j}\})
 \phi^{-1}_\kappa(\{\l_{k}\})
 \nonumber\\ && \quad\quad\times
\langle \O_N(\{\m_j\}_{(p,n)})| \cdot \left[
 A_{22}(\x_\k)+{1\over 2}(t(\x_\k)-D(\x_\k))\right]\cdot
 |\O_N(\{\m_j\}_{(p,n)})\rangle. \label{eq:cf-Sz}
\end{eqnarray}
Then by using the commutation relations (\ref{eq:commu-AC}) and
(\ref{eq:commu-CaCb}), one may prove that (\ref{eq:cf-Sz}) gives
rise to (\ref{eq:Sz}).
\begin{flushright} $\Box$ \end{flushright}

\subsection{two-point functions}
In principle, by equations (\ref{de:cf-general})-(\ref{tj-Sz})
with proper commutation relations derived from the GYBE, and
similar method as that in the previous subsection, we may obtain
any correlation function defined by (\ref{de:cf-general}).

As an example, in this subsection, we compute the correlation
function associated with two adjacent fermion operators
$(1-n_{\k,\down})c_{\k,\up}^\dag$ and $(1-n_{\k+1,
\down})c_{\k+1,\up}$. Considering the representations of the
fermion operators (\ref{tj5}) and (\ref{tj3}),  the correlation
function is defined by
\begin{eqnarray}
 &&F^{\up\dag,\up}_n(\{\m_j\}_{(p,n)},\x_\k,\x_{\k+1},\{\l_k\}_{(p,n)})
 \nonumber\\ &&
 =\langle \O^p_N(\{\m_j\})|(1-n_{\k,\down})c_{\k,\up}^\dag
 \,
 (1-n_{\k+1,\down})c_{\k+1,\up}|\O^p_N(\{\l_k\})\rangle
 \nonumber\\ &&
 =\langle \O^p_N(\{\m_j\})| \prod_{j=1}^{\k-1}t(\x_j)\cdot
  C_2(\x_\k)\,B_2(\x_{\k+1})\cdot \prod_{j=k+2}^{N}t(\x_j)
 |\O^p_N(\{\l_k\})\rangle.
\end{eqnarray}
Here we have used the following property: for the supersymmetric
$t$-$J$ model with periodic boundary condition, the transfer
matrices satisfy the relation
%\begin{equation}
$\prod_{i=1}^N t(\l_i)=1.$
%\end{equation}
Then with the help of the commutation relations
(\ref{eq:commu-CaCa}), (\ref{eq:commu-AC}), (\ref{eq:commu-DC}),
and
\begin{eqnarray}
 B_a(\l)C_b(\m)=-C_b(\m)B_a(\l)+{b(\l,\m)\over a(\l,\m)}[
 D(\m)A_{ab}(\l)-D(\l)A_{ab}(\m)],
\end{eqnarray}
one proves the following proposition
\begin{Proposition}
If both the Bethe state $|\O_N(\{u_j\})\rangle$ and the dual Bethe
state $\langle\O_N(\{u_j\})|$ ($u_j=\l_j,\m_j$) are eigenstates of
the transfer matrix, then the two-point correlation functions
associated with the local fermion operators
$(1-n_{\k,\down})c_{\k,\up}^\dag$ and
$(1-n_{\k+1,\down})c_{\k+1,\up}$ can be represented by
\begin{eqnarray}
 &&F^{\up\dag,\up}_n(\{\m_j\}_{(p,n)},\x_\k,\x_{\k+1},\{\l_k\}_{(p,n)})
 \nonumber\\ &&
 =\sum_{\s\in{\cal S}_{n}}\sum_{\s'\in{\cal S}_n}
 \phi_{\kappa-1}(\{\m_{j}\})
 \phi^{-1}_{\kappa+1}(\{\l_{k}\})
% \nonumber\\ && \quad\quad\times
 \sum_{i=1}^p (-1)^{i-1}
  {b(\x_{\k+1},\l_{\s(i)})\over a(\x_{\k+1},\l_{\s(i)})}
 \nonumber\\ && \quad\times \left\{
 \sum_{j=i+1}^p{b(\x_{\k+1},\l_{\s(j)})\over a(\x_{\k+1},\l_{\s(j)})}
 \prod_{l=i+1}^{j-1}
  {c(\l_{\s(l)},\l_{\s(j)})\over a(\l_{\s(l)},\x_{\k})}
 \prod_{k=i+1,\ne j}^{p}
  {c(\l_{\s(j)},\l_{\s(k)})\over a(\l_{\s(j)},\l_{\s(k)})}
 \prod_{\a=1}^N a(\l_{\s(j)},\x_\a) \right.
 \nonumber\\ && \quad\quad\times
 {\cal P}\left(i+1;\l_{\s(i)};\{\m_{\s'(d)}\}_{(p,n)};\{\l'_{\s(f)}\}_{(p,n)}\right)
 \nonumber\\ && \quad\quad \mbox{}
 -\prod_{j=i+1}^p
  {c(\l_{\s(i)},\l_{\s(j)})\over a(\l_{\s(i)},\l_{\s(j)})}
 \prod_{\a=1}^N a(\l_{\s(i)},\x_\a)\,
 {\cal P}\left(i+1;\x_{\k+1};\{\m_{\s'(d)}\}_{(p,n)};\{\l^*_{\s(f)}\}_{(p,n)}\right)
 \nonumber\\ && \quad\quad\mbox{}
 -\sum_{j=i+1}^p{b(\l_{\s(i)},\l_{\s(j)})\over a(\l_{\s(i)},\l_{\s(j)})}
 \prod_{l=i+1}^{j-1}
  {c(\l_{\s(l)},\l_{\s(j)})\over a(\l_{\s(l)},\l_{\s(i)})}
 \prod_{k=i+1,\ne j}^{p}
  {c(\l_{\s(j)},\l_{\s(k)})\over a(\l_{\s(j)},\l_{\s(k)})}
 \prod_{\a=1}^N a(\l_{\s(j)},\x_\a)
 \nonumber\\ && \quad\quad\, \times \left.
 {\cal P}\left(i+1;\x_{\k+1};\{\m_{\s'(d)}\}_{(p,n)};
    \{\l''_{\s(f)}\}_{(p,n)}
  \right)\right\},
\end{eqnarray}
where ${\cal P}$ is given by (\ref{de:P-cal}) and the spectral
parameters $\l',\l^*$ and $\l''$ are given by
\begin{eqnarray*}
 &&\l'_{\s(k)}=\left\{\begin{array}{cl}
 \x_\k & (k=1)\\
 \l_{\s(k-1)} &(2\leq k\leq i)\\
 \l_{\s(k)} & (i+1\leq k\leq n \mbox{ and } k\ne j)\\
 \x_{\k+1} &(k=j)\end{array} \right.,
 \l^*_{\s(k)}=\left\{\begin{array}{cl}
 \x_\k & (k=1)\\
 \l_{\s(k-1)} &(k=2,\ldots,i)\\
 \l_{\s(k)} & (i+1\leq k\leq n)\end{array} \right.,\quad
 \\ && \mbox{ and }  \quad\quad
 \l''_{\s(k)}=\left\{\begin{array}{cl}
 \x_\k & (k=1)\\
 \l_{\s(k-1)} &(2\leq k\leq i)\\
 \l_{\s(k)} & (i+1\leq k\leq n \mbox{ and } k\ne j)\\
 \l_{\s(i)} &(k=j)\end{array} \right.,
\end{eqnarray*}
respectively.
\end{Proposition}

%%%%%%%%%%%%%%%%%%%%%%%%%%%%%%%%%%%%%%%%%%%%%%%%%%%%%%%%%%%%%%%%%%
%                                                                %
%         Conclusion and discussion                              %
%                                                                %
%%%%%%%%%%%%%%%%%%%%%%%%%%%%%%%%%%%%%%%%%%%%%%%%%%%%%%%%%%%%%%%%%%

\sect{Conclusion and discussion}

In this paper, we constructed the determinant representations of
scalar products and the correlation functions of the
supersymmetric $t$-$J$ model with the help of the factorizing
$F$-matrix. Because  the $t$-$J$ model is an important model in
the realm of the high $T_c$ superconductivity, we hope our results
may enlarge the range of applications in this field. We also hope
our results may offer a new understanding of the mathematical
structures of the model. An interesting problem is to extend the
results in this paper to the $t$-$J$ model with open boundary
condition. This is under consideration and results will be
reported elsewhere.

\vspace{4mm}

%%%%%%%%%%%%%%%%%%%%%%%%%%%%%%%%%%%%%%%%%%%%%%%%%%%%%%%%%%%%%%%%%%
%                                                                %
%         Acknowledgements                                       %
%                                                                %
%%%%%%%%%%%%%%%%%%%%%%%%%%%%%%%%%%%%%%%%%%%%%%%%%%%%%%%%%%%%%%%%%%

\noindent{\small{\em Acknowledgements.} This work was financially
supported by the Australian Research Council. S.-Y. Zhao was
supported by the UQ Postdoctoral Research Fellowship.}

%%%%%%%%%%%%%%%%%%%%%%%%%%%%%%%%%%%%%%%%%%%%%%%%%%%%%%%%%%%%%%%%%%
%                                                                %
%         References                                             %
%                                                                %
%%%%%%%%%%%%%%%%%%%%%%%%%%%%%%%%%%%%%%%%%%%%%%%%%%%%%%%%%%%%%%%%%%

\end{document}